\pdfoutput=1

\documentclass[11pt,twoside,a4paper,cmspaper,final,collab]{cms-tdr}
\def\svnVersion{338608}\def\cmsCernNoTag{CERN-PH-EP/2015-234}\def\cmsCernDate{\today}\def\cmsMessage{Published in Physical Review D as \href{http://dx.doi.org/10.1103/PhysRevD.93.072004}{\doi{10.1103/PhysRevD.93.072004.}}}

\begin{document}\cmsNoteHeader{TOP-14-022}

\hyphenation{had-ron-i-za-tion}
\hyphenation{cal-or-i-me-ter}
\hyphenation{de-vices}
\RCS$Revision: 318018 $
\RCS$HeadURL: svn+ssh://svn.cern.ch/reps/tdr2/papers/TOP-14-022/trunk/TOP-14-022.tex $
\RCS$Id: TOP-14-022.tex 318018 2016-01-17 21:03:15Z alverson $
\newlength\cmsFigWidth
\ifthenelse{\boolean{cms@external}}{\setlength\cmsFigWidth{0.98\columnwidth}}{\setlength\cmsFigWidth{0.65\textwidth}}
\ifthenelse{\boolean{cms@external}}{\providecommand{\cmsLeft}{top\xspace}}{\providecommand{\cmsLeft}{left\xspace}}
\ifthenelse{\boolean{cms@external}}{\providecommand{\cmsRight}{bottom\xspace}}{\providecommand{\cmsRight}{right\xspace}}

\newcommand{\mtop}{\ensuremath{m_{\PQt}}\xspace}
\providecommand{\qqbar}{\ensuremath{\PQq\PAQq}\xspace}

\newcommand{\figdiff}[3]{
  \begin{figure}[h]
    \includegraphics[width=0.49\textwidth]{diff_#1_MTH_cal.pdf}
    \caption{\diffcaption{#2}{#3} }
    \label{fig:#1}
  \end{figure}
}

\newcommand{\diffcaption}{The statistical uncertainty of the data is displayed by the inner error bars. For the outer error bars, the systematic uncertainties are added in quadrature.}

\cmsNoteHeader{TOP-14-022}
\title{Measurement of the top quark mass using proton-proton data at \texorpdfstring{$\sqrt{s} = 7$ and 8\TeV}{sqrt(s) = 7 and 8 TeV}}

\date{\today}

\abstract{
A new set of measurements of the top quark mass are presented, based on the proton-proton data recorded by the CMS experiment at the LHC at $\sqrt{s} = 8\TeV$ corresponding to a luminosity of 19.7\fbinv. The top quark mass is measured using the lepton+jets, all-jets and dilepton decay channels, giving values of
$172.35 \pm 0.16\stat\pm0.48\syst\GeV$,
$172.32\pm0.25\stat\pm0.59\syst\GeV$,
and $172.82\pm0.19\stat\pm1.22\syst\GeV$, respectively.
When combined with the published CMS results at $\sqrt{s} = 7\TeV$, they provide a top quark mass measurement of $172.44\pm 0.13\stat\pm 0.47\syst\GeV$.
The top quark mass is also studied as a function of the event kinematical properties in the lepton+jets decay channel. No indications of a kinematic bias are observed and the collision data are consistent with a range of predictions from current theoretical models of \ttbar production.
}

\hypersetup{%
pdfauthor={CMS Collaboration},%
pdftitle={Measurement of the top quark mass using proton-proton data at sqrt(s) = 7 and 8 TeV},%
pdfsubject={CMS},%
pdfkeywords={CMS, physics, top quark}}

\maketitle

\section{Introduction}

The mass of the top quark (\mtop) is one of the fundamental parameters of the standard model (SM). A precise measurement of its value provides a key input to global electroweak fits and to tests of the internal consistency of the SM~\cite{EWfit,GFitter}. Its value
leads to constraints on the stability of the electroweak vacuum~\cite{HiggsStab,HiggsStab1} and affects models with broader cosmological implications~\cite{HiggsStab2,HiggsStab3}.

The most precise measurements of \mtop have been derived from combinations of the results from the CDF and D0 experiments at the Tevatron, and ATLAS and CMS at the CERN LHC. The current combination from the four experiments gives a top quark mass of
$173.34\pm0.76\GeV$~\cite{World2013}, while the latest combination from the Tevatron experiments gives a mass of
$174.34\pm0.64\GeV$~\cite{Tevatron2014}. The Tevatron combination is currently the most precise measurement and it includes all of the current Tevatron measurements. In contrast, the current four experiment combination has not been updated since 2013 and does not include the latest Tevatron and LHC measurements, in particular the measurement from ATLAS using a combination of the lepton+jets and dilepton channels~\cite{Aad:2015nba}.

Beyond the leading order (LO) in quantum chromodynamics (QCD), the numerical value of  \mtop depends on the renormalization scheme~\cite{Renorm1,Renorm2}. The available Monte Carlo (MC) generators contain matrix elements at LO or next-to-leading order (NLO), while higher orders are approximated by applying parton showering. Each of the measurements used in the combinations has been calibrated against the mass implemented in a MC program. Given the precision of the experimental results, a detailed understanding of the relationship
between the measurements and the value of \mtop\ in different theoretical schemes is needed.
Current indications are that the present measurements based on the kinematic reconstruction of the top quark mass correspond approximately to the pole (``on-shell'') mass to within a precision of about 1\GeV~\cite{Massdef}.

At the LHC, top quarks are predominantly produced in quark-antiquark pairs ($\ttbar$) and top quark events are characterized by the decays of the daughter W bosons. This leads to experimental signatures with two jets associated with the hadronization of the bottom quarks and either a single lepton (\Pe, \PGm), one undetected neutrino and two light quark jets (lepton+jets channel), or four light quark jets (all-jets channel), or two leptons ($\Pe\Pe$, $\Pe\PGm$, $\PGm\PGm$) and two undetected neutrinos (dilepton channel).
While the events which contain leptonic \PGt decays are included in the analysis samples, they contribute very little to the mass measurements as their yields are negligible.
The results presented in this paper focus on the analysis of data in these three channels recorded by the CMS experiment in the 2012 part of what is commonly referred to as Run 1 of the LHC.

The paper is organized as follows. The main features of the detector and the data are discussed in Sections~\ref{sec:det} and \ref{sec:ds}. Section~\ref{sec:analysistech} is a discussion of the analysis techniques, which lead to the measurements of Section~\ref{sec:mass}. The categorization of the systematic uncertainties is presented in Section~\ref{sec:uncert}, followed by the full results for the three decay channels in Section~\ref{sec:channels}. Section~\ref{sec:kine} presents a study of \mtop as a function of the kinematical properties of the \ttbar system in the lepton+jets channel. This is followed in Sections~\ref{sec:combo} and \ref{sec:results}, which discuss the combination of the measurements and the final result for \mtop, respectively.
\section{The CMS detector\label{sec:det}}
The central feature of the CMS apparatus is a superconducting solenoid of 6\unit{m} internal diameter, providing a magnetic field of 3.8\unit{T}. Within the solenoid volume are a silicon pixel and strip tracker, a lead tungstate crystal electromagnetic calorimeter (ECAL), and a brass and scintillator hadron calorimeter (HCAL),  each composed of a barrel and two endcap sections. The tracker has a track-finding efficiency of more than 99\% for muons with transverse momentum $\pt>1\GeV$ and pseudorapidity $\abs{\eta} <2.5$.
The ECAL is a fine-grained hermetic calorimeter with quasi-projective geometry, and is distributed in the barrel region of $\abs{\eta} <1.48$ and in two endcaps that extend up to $\abs{\eta} < 3.0$. The HCAL barrel and endcaps similarly cover the region $\abs{\eta} < 3.0$. In addition to the barrel and endcap detectors, CMS has extensive forward calorimetry. Muons are measured in gas-ionization detectors, which are embedded in the steel flux-return yoke outside of the solenoid. A more detailed description of the CMS detector, together with a definition of the coordinate system used, and the relevant kinematic variables can be found in Ref.~\cite{CMSDET}.

\section{Data sets\label{sec:ds}}
The measurements
presented in this paper are based on the data recorded
at a center-of-mass energy of 8\TeV during 2012, and correspond to an integrated luminosity of 19.7\fbinv.

\subsection{Event simulation and reconstruction}
\label{sec:simul}

Simulated \ttbar signal events are generated with the \MADGRAPH~5.1.5.11 LO matrix element generator with up to three additional partons~\cite{Alwall:2011uj}. \textsc{MadSpin}~\cite{Artoisenet:2012st} is used for the decay of heavy resonances, \PYTHIA~6.426 for parton showering~\cite{Sjostrand:2006za} using the Z2* tune,
and \TAUOLA~\cite{Jadach:1993hs} for decays of $\tau$ leptons. The most recent \PYTHIA~Z2* tune is derived from the Z1 tune~\cite{Z1_tune}, which uses the CTEQ5L parton distribution function (PDF) set, whereas Z2* uses CTEQ6L~\cite{CTEQ6}. A full simulation of the CMS detector based on \GEANTfour~\cite{Agostinelli:2002hh} is used.
The \ttbar signal events are generated for seven different values of \mtop ranging from 166.5 to 178.5\GeV.
The \PW/\Z{}+jets background events are generated with \MADGRAPH~5.1.3.30.
The diboson background ($\PW\PW$, $\PW\Z$, $\Z\Z$) is simulated using \PYTHIA~6.426 using the Z2* tune.
The single top quark background is simulated using \POWHEG~1.380~\cite{Alioli:2009je,Re:2010bp,Nason:2004rx,Frixione:2007vw,Alioli:2010xd}, assuming an \mtop of 172.5\GeV.
The \ttbar, \PW/\Z{}+jets, and single top quark samples are normalized to the theoretical predictions described in Refs.~\cite{Kidonakis:2010dk,Kidonakis:2012db,Melnikov:2006kv,Campbell:2010ff,TopXS}.
The simulation includes the effects of additional proton-proton collisions (pileup) by overlapping minimum bias events with the same multiplicity distribution and location as in data.

Events are reconstructed using a particle-flow (PF) algorithm~\cite{PF1, PF2}. This proceeds by reconstructing and identifying each final-state particle using an optimized combination of all of the subdetector information. Each event is required to have at least one reconstructed collision vertex. The primary vertex is chosen as the vertex with the largest value of $\sum \pt^{2}$ of the tracks associated with that vertex.
Additional criteria are applied to each event to reject events with features consistent with detector noise and beam-gas interactions.

The energy of electrons is determined from a combination of the track momentum at the primary vertex, the corresponding ECAL energy cluster, and the sum of the reconstructed bremsstrahlung photons associated with the track~\cite{elec_rec}. The momentum of muons is obtained from the track momentum determined in a combined fit to information from the silicon trackers and the muon detectors~\cite{muon_rec}. The energy of charged hadrons is determined from a combination of the track momentum and the corresponding ECAL and HCAL energies, corrected for the suppression of small signals and calibrated for the nonlinear response of the calorimeters. Finally, the energy of the neutral hadrons is obtained from remaining calibrated HCAL and ECAL energies. As the charged leptons originating from top quark decays are typically isolated from other particles, a relative isolation variable ($I_{\text{rel}}$) is constructed to select lepton candidates. This is defined as the scalar sum of the \pt values of the additional particles reconstructed within an angle $\Delta R = \sqrt{\smash[b]{ (\Delta \eta)^{2} + ( \Delta \phi)^{2} }}$ of the lepton direction, divided by the \pt of the lepton. Here $\Delta \eta$ and $\Delta \phi$ are the differences in the pseudorapidity and azimuthal angles between the lepton direction and other tracks and energy depositions. A muon candidate is rejected if $I_{\text{rel}} \ge 0.12$ for $\Delta R = 0.4$, and an electron candidate is rejected if $I_{\text{rel}} \ge0.10$ for $\Delta R =0.3$.

Jets are clustered from the reconstructed PF candidates using the anti-\kt algorithm~\cite{ktalg} with a distance parameter of 0.5, as implemented in the \FASTJET package~\cite{fastjet}. The jet momentum is determined from the vector sum of the momenta of the particles in each jet, and is found in simulation to be within 5 to 10\% of the jet momentum at hadron level for the full \pt range~\cite{Chatrchyan:2011ds}. Corrections to the jet energy scale (JES) and the jet energy resolution (JER) are obtained from the simulation and through in-situ measurements of the energy balance of exclusive dijet, photon+jet, and \Z{}+jet events.
Muons, electrons, and charged hadrons originating from pileup interactions are not included in the jet reconstruction. Missing transverse energy (\MET) is defined as the magnitude of the negative vector \pt sum of all selected PF candidates in the event. Charged hadrons originating from pileup interactions are not included in the calculation of \MET. Jets are classified as \PQb jets through their probability of originating from the hadronization of bottom quarks, using the combined secondary vertex (CSV) \PQb tagging algorithm, which combines information from the significance of the track impact parameters, the kinematical properties of the jets, and the presence of tracks that form vertices within the jet.
Three different minimum thresholds are used for the CSV discriminator to define the loose (CSVL), medium (CSVM), and tight (CSVT) working points. These have \PQb tagging efficiencies of approximately 85\%, 67\%, and 50\%, and misidentification probabilities for light-parton jets of 10\%, 1\%, and 0.1\%, respectively~\cite{CSVtag}.
\subsection{Event selection}
\begin{figure*}
\centering
\includegraphics[width=0.49\textwidth]{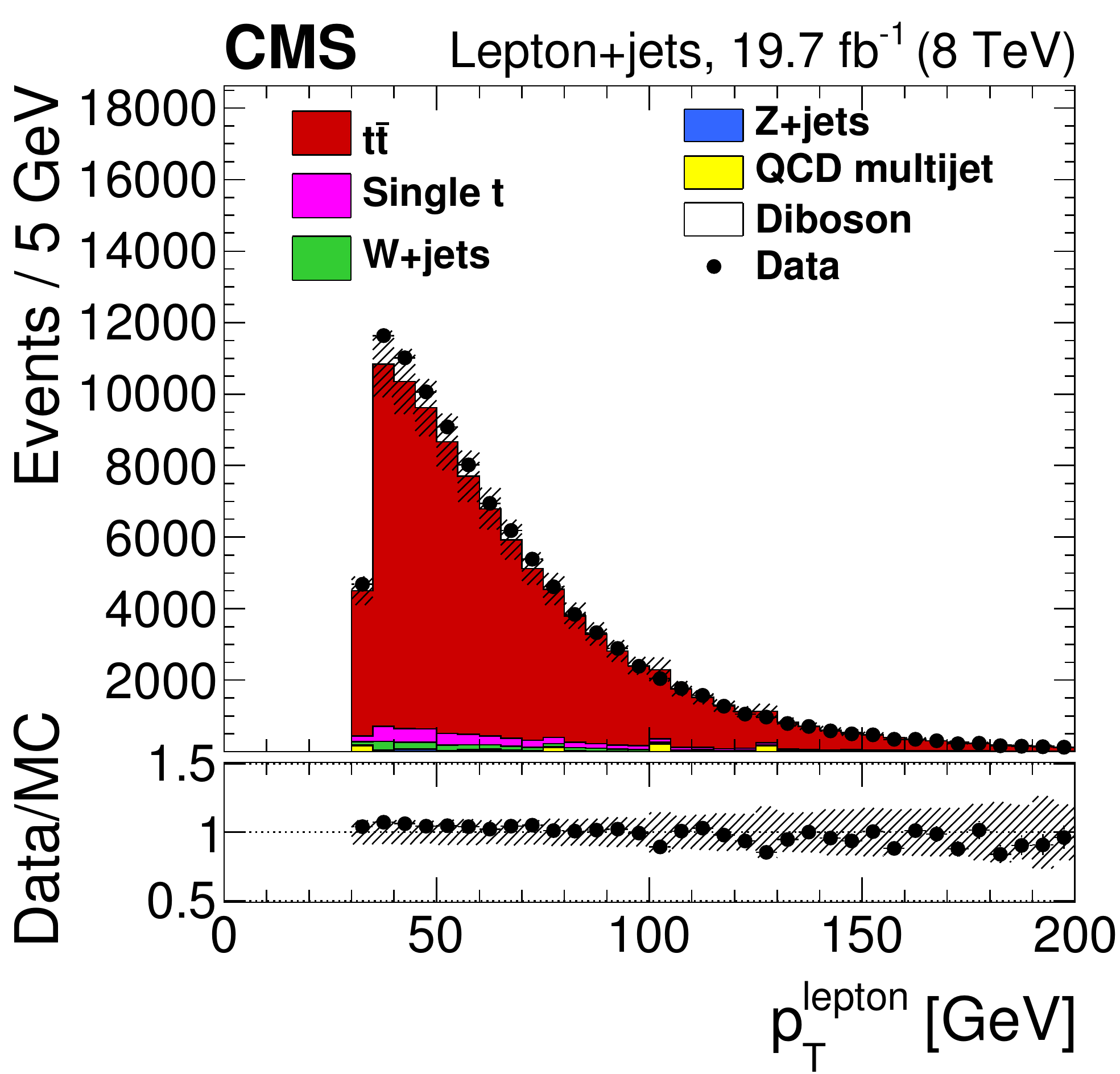}
\includegraphics[width=0.49\textwidth]{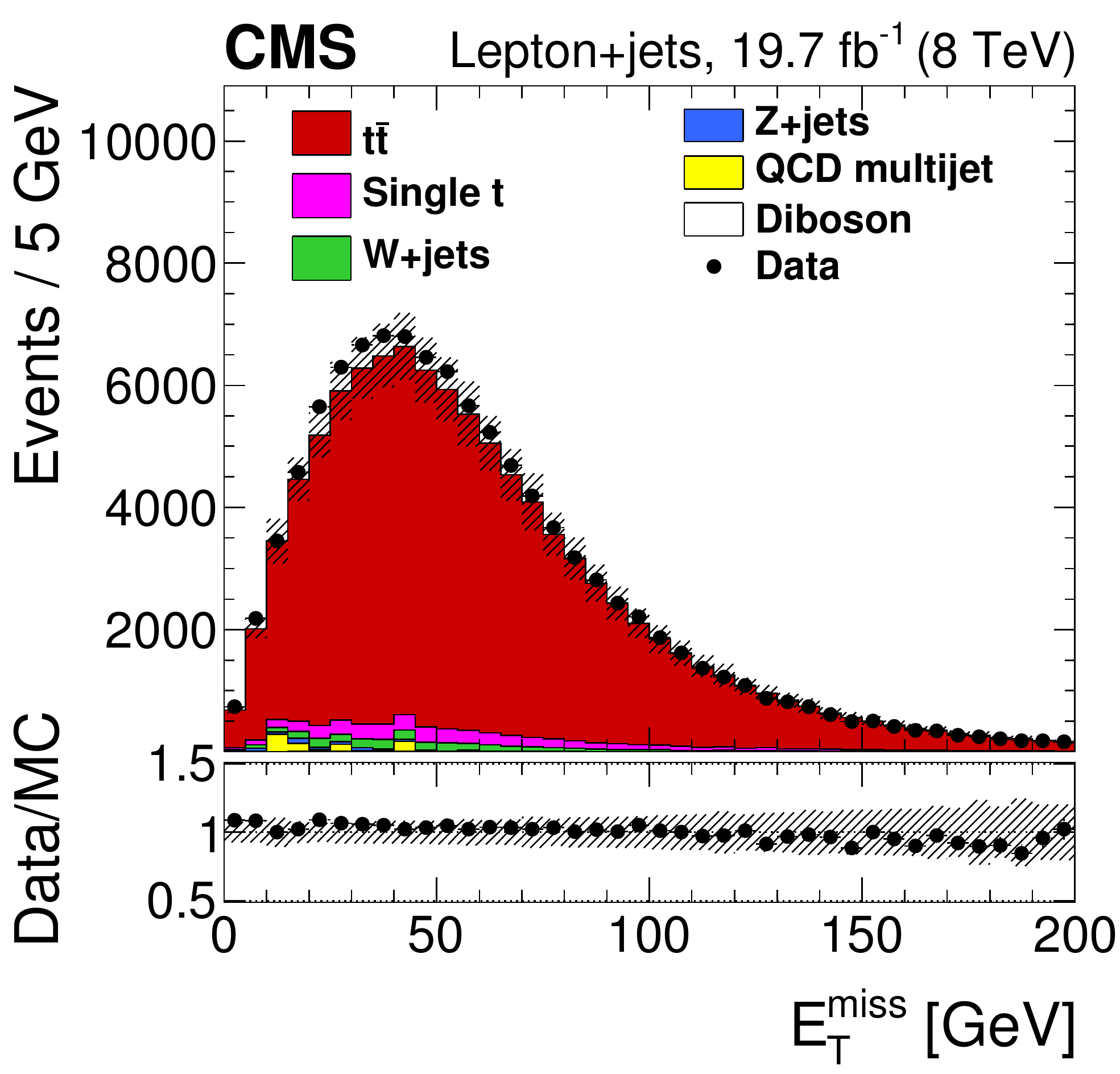}
\includegraphics[width=0.49\textwidth]{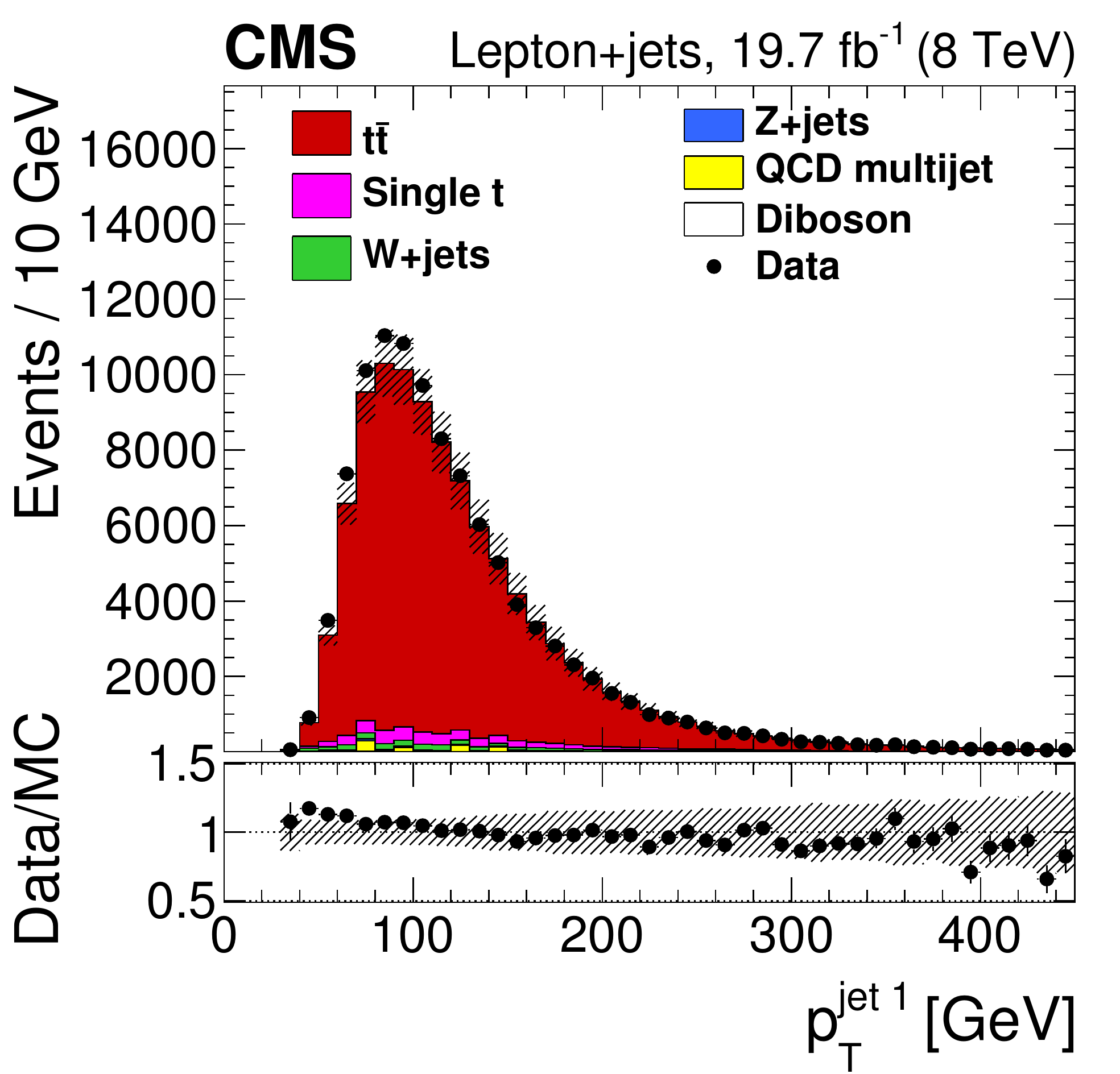}
\includegraphics[width=0.49\textwidth]{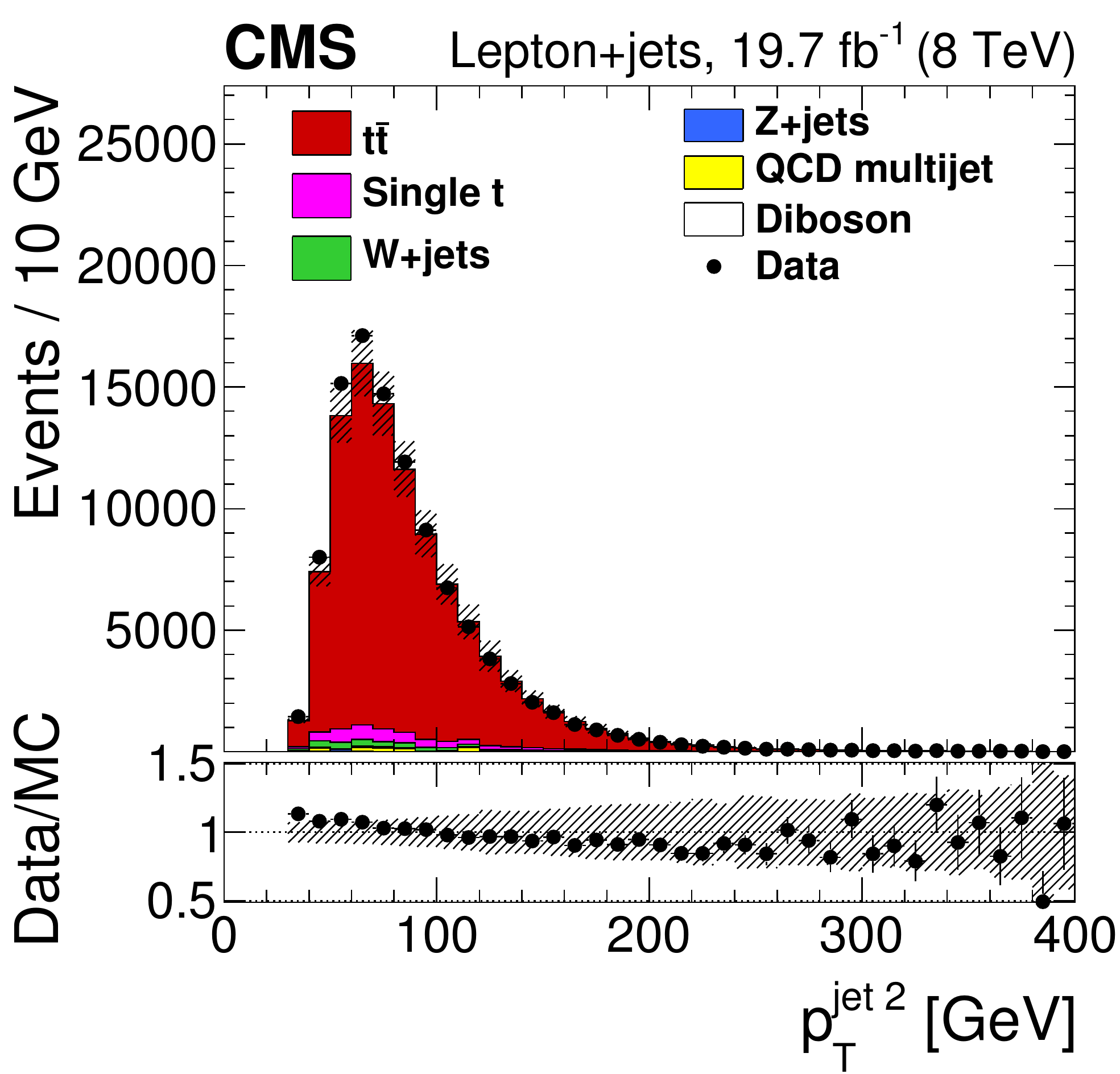}
\caption{\label{fig:LepJetsDataMC} Distributions for the lepton+jets channel of (upper left) lepton \pt, (upper right) missing transverse energy, (lower left) leading jet \pt, (lower right) second-leading jet \pt
 for data and simulation,  summed over all channels and normalized by luminosity. The vertical bars show the statistical uncertainty and the hatched bands show the statistical and systematic uncertainties added in quadrature. The lower portion of each panel shows the ratio of the yields between the collision data and the simulation.}
\end{figure*}

For the lepton+jets channel we use the data collected using a single-muon or single-electron trigger with a minimum trigger \pt threshold for an isolated muon (electron) of 24\GeV (27\GeV), corresponding to an integrated luminosity of 19.7\fbinv.
We then select events that have exactly one isolated muon or electron, with $\pt>33\GeV$ and $\abs{\eta}<2.1$. In addition, at least four jets with $\pt>30\GeV$ and  $\abs{\eta}<2.4$ are required.
Jets originating from \PQb quarks (denoted as \PQb jets) are identified using the CSV algorithm at the medium working point~\cite{CSVtag}.
With the requirement of exactly two \PQb-tagged jets among the four jets with the highest \pt, 104\,746 \ttbar candidate events are selected in data.
From simulation, the sample composition is expected to be  93\% \ttbar, 4\% single top quark, 2\% \PW{}+jets, and 1\% other processes.
Figure~\ref{fig:LepJetsDataMC} shows the comparison of the data and simulation for the selected events in some representative distributions.
The simulation shown is not corrected for the uncertainty in the shape of the top quark \pt distribution~\cite{Top_ptunc}, which accounts for almost all of the slope visible in the data/MC ratio plots. However, even without making a correction, the data and simulation are consistent within the quoted uncertainties.

For the all-jets channel we use the data collected using
a multijet trigger, corresponding to an integrated luminosity of 18.2\fbinv. The trigger requires the presence of at least four jets, reconstructed from the energies deposited in the calorimeters, with transverse momenta $\pt > 50$\GeV.
Since fully hadronically decaying top quark pairs lead to six partons in the final state, events are required to have at least four jets with $\pt > 60$\GeV and a fifth and sixth jet with $\pt > 30$\GeV. Jets originating from \PQb quarks are identified using the CSV \PQb algorithm at the tight working point~\cite{CSVtag}.
With the requirement of exactly two \PQb-tagged jets among the six leading ones,  356\,231 candidate events are selected.
From simulation, the sample is expected to be dominated by the QCD multijet background and to have a signal fraction of about 13\%.
The QCD multijet background cannot be reliably simulated and we determine its kinematic dependence from a control sample in the data. The background normalization is determined as a part of the fit process, which is discussed in Section~\ref{alljets}.

\begin{figure*}
\centering
\includegraphics[width=0.49\textwidth]{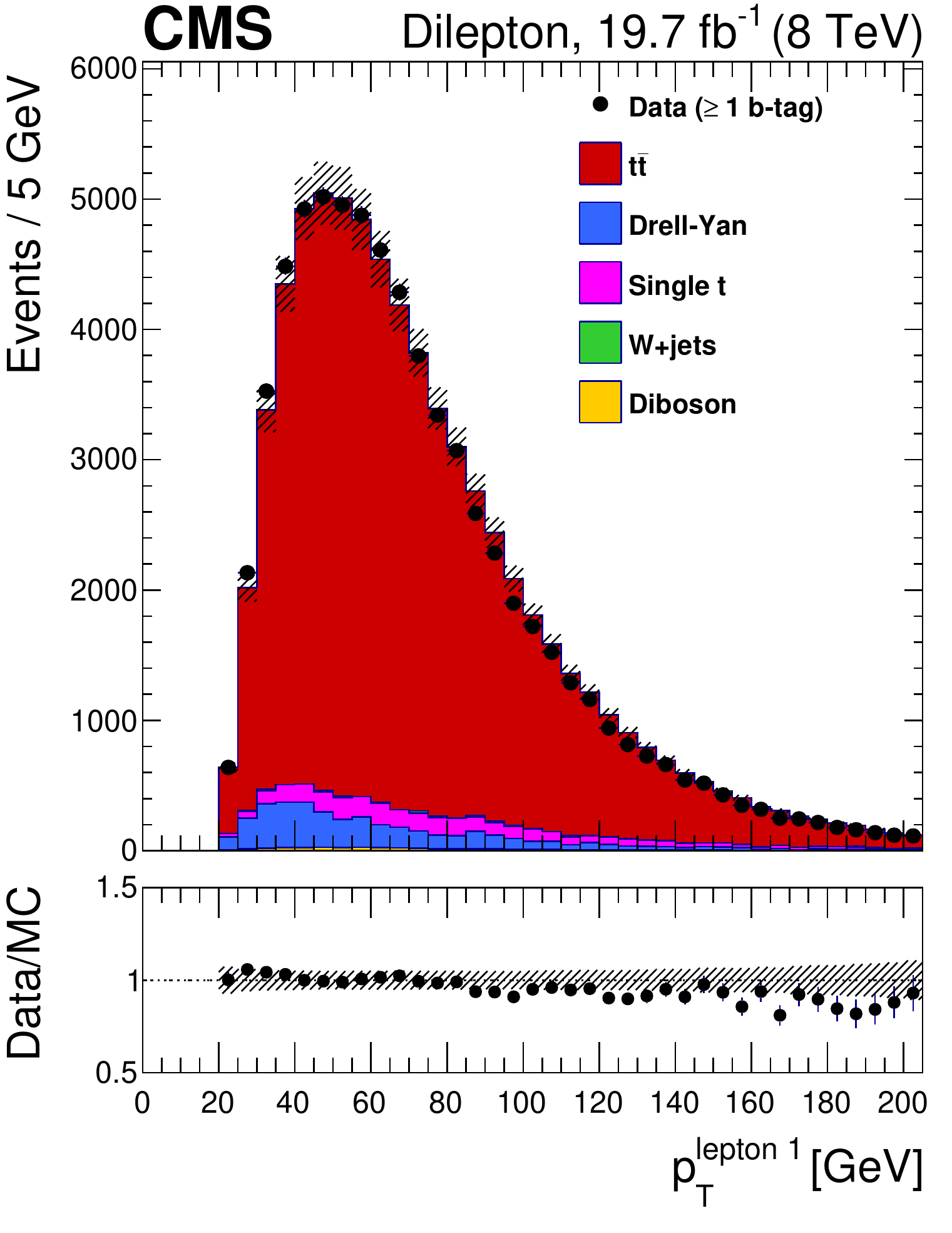}
\includegraphics[width=0.49\textwidth]{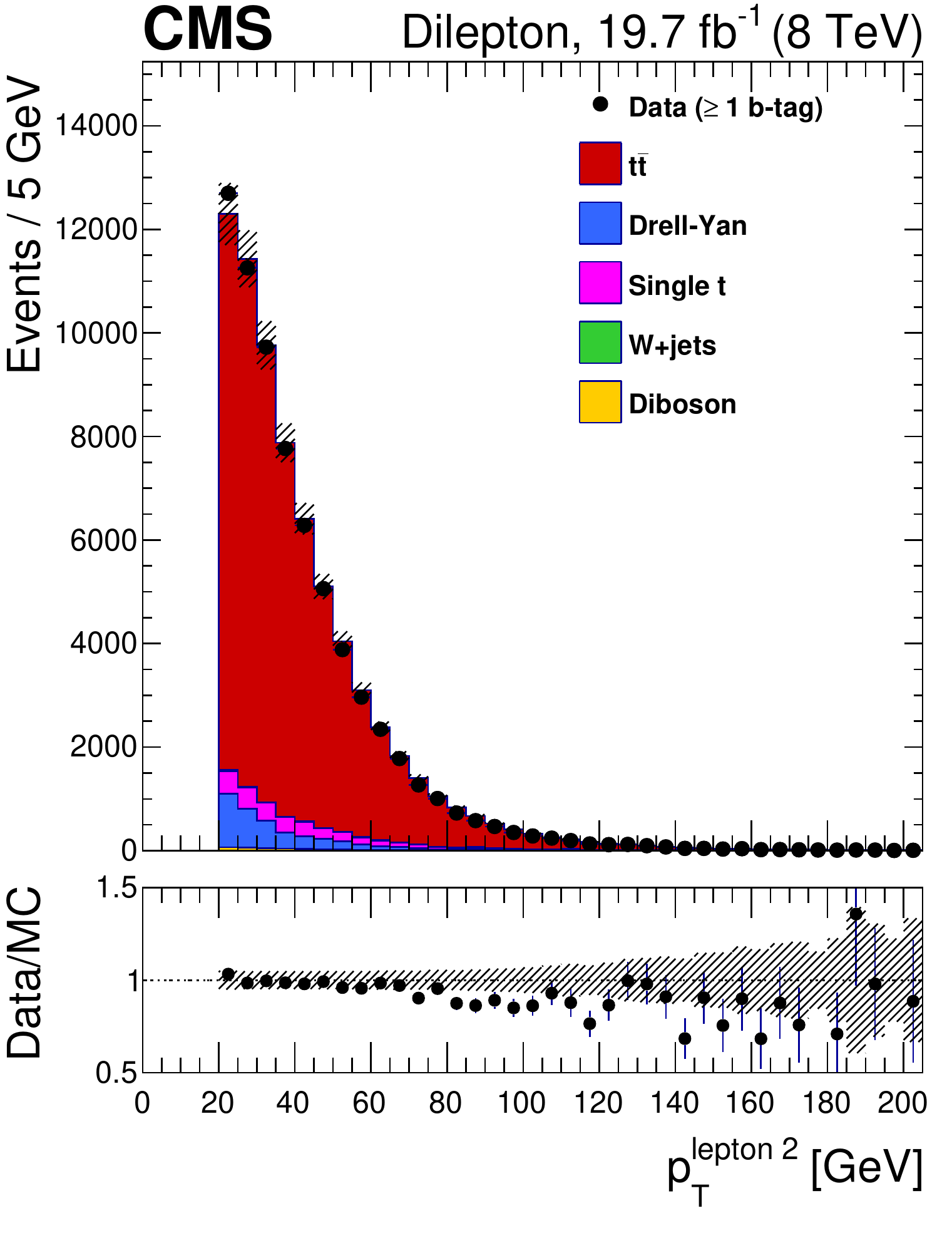}
\includegraphics[width=0.49\textwidth]{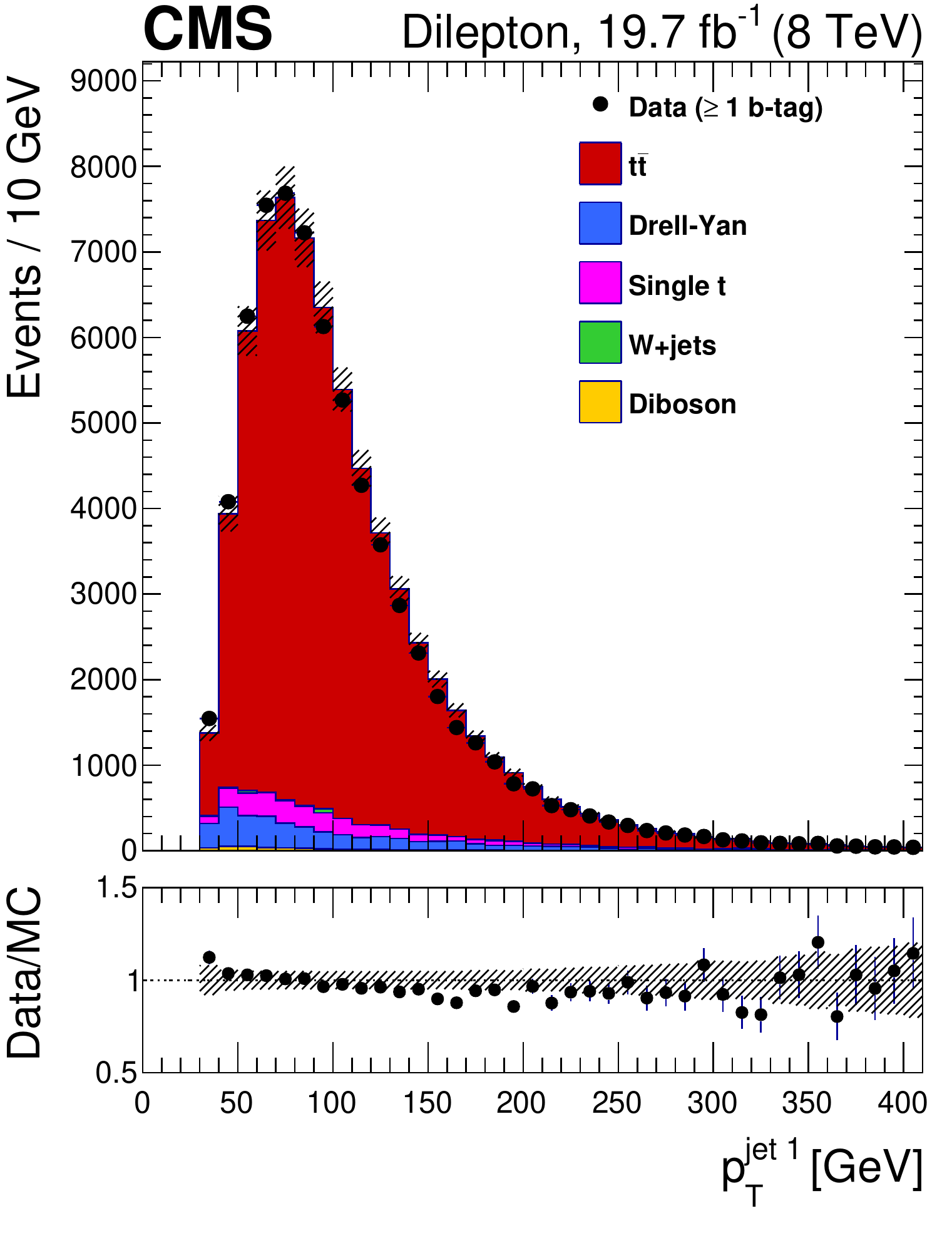}
\includegraphics[width=0.49\textwidth]{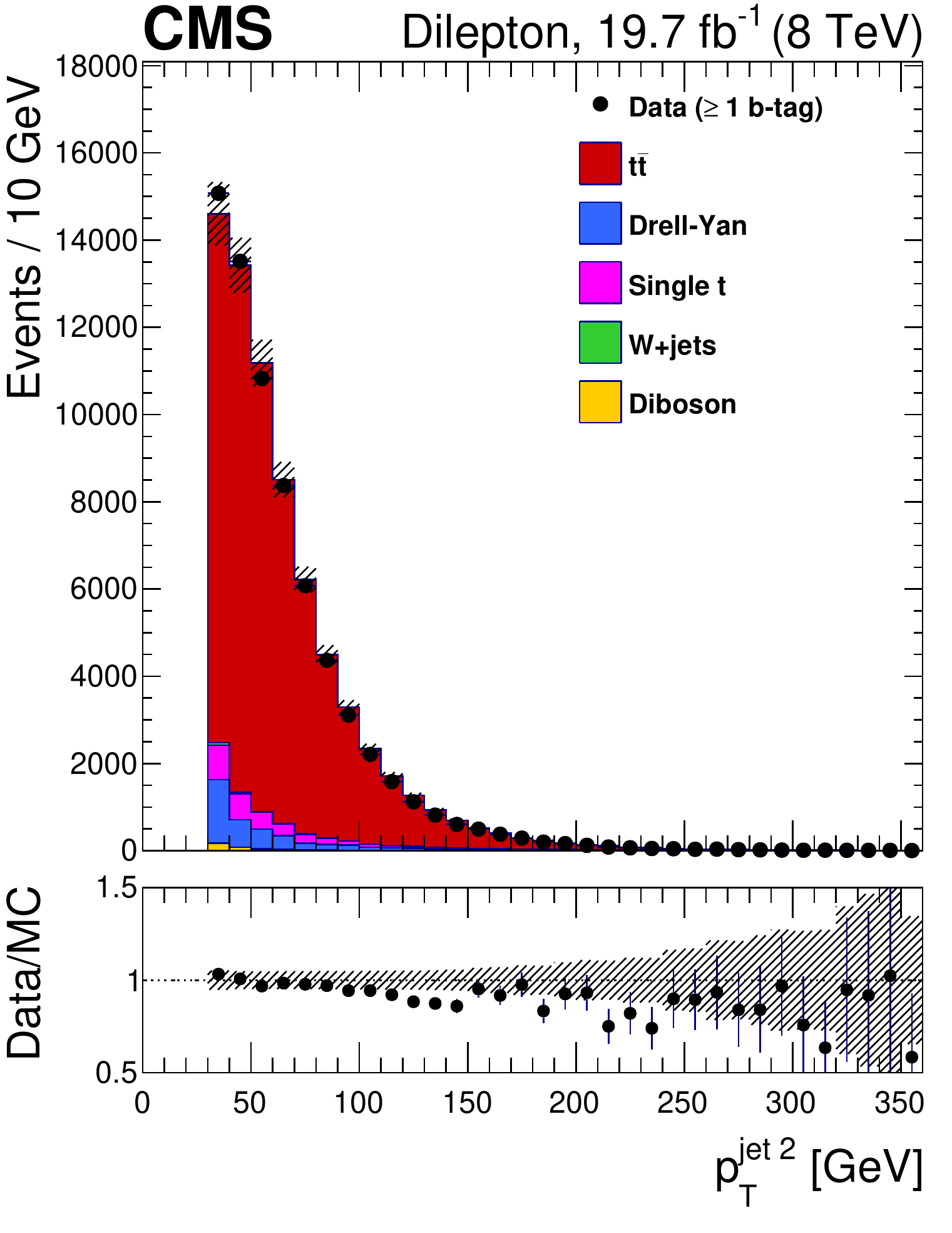}
\caption{\label{fig:AMWTDataMC} Distributions for the dilepton channel: (upper left) leading lepton \pt, (upper right) second-leading lepton \pt, (lower left) leading jet \pt, (lower right) second-leading jet \pt
 for data and simulation, summed over all channels and normalized by luminosity. The vertical bars show the statistical uncertainty and the hatched bands show the statistical and systematic uncertainties added in quadrature. The lower portion of each panel shows the ratio of the yields between the collision data and the simulation.}
\end{figure*}

For the dilepton channel, events are required to pass the triggers appropriate for each of the three channels. The $\Pe\PGm$ channel uses a logical OR of two triggers that require a muon of $\pt> 17$ (or 8)\GeV and an isolated electron of 8 (or 17)\GeV. Dimuon events must pass a trigger which requires a $\pt> 17$\GeV for the muon with the highest (``leading") \pt and 8\GeV for the second-leading muon. Similarly, dielectron events must satisfy a trigger with a threshold of $\pt>17$\GeV for the leading electron and 8\GeV for the second-leading electron.
In this case both electrons are required to be isolated. In all three cases the amount data corresponds to a luminosity of 19.7\fbinv. We select events for analysis if they have two isolated opposite-sign leptons with $\pt>20$\GeV and $\abs{\eta}< 2.4\,(2.5)$ for muons (electrons). Jets originating from \PQb quarks are identified with the CSV algorithm at the loose working point~\cite{CSVtag}. Events are retained if they have at least two \PQb-tagged jets. Background contamination from low-mass resonances is reduced by demanding a dilepton pair invariant mass, $m_{\ell\ell}$, of at least 10\GeV. To suppress the background from \Z boson decays, events with $\Pe\Pe$ and $\PGm\PGm$ signatures are required to  have $\MET> 40$\GeV, and to fall outside of the dilepton invariant mass window $76<m_{\ell\ell}<106\GeV$.
The remaining Drell--Yan background is estimated from the data using the ratio of the event yield inside vs. outside the invariant mass window~\cite{CSVtag}.
After all of the requirements, we find 41\,125 candidate events in data for which the sample compositions is expected to be 95\% \ttbar, 3\% single top quark, 2\% Drell--Yan, and $<$0.3\% other processes. Figure~\ref{fig:AMWTDataMC} shows a comparison of the data and simulation for events with at least one \PQb jet for some representative distributions. As with the lepton+jets plot (Fig.~\ref{fig:LepJetsDataMC}) the simulation is not corrected for the discrepancy in the top quark \pt distribution, leading to the slopes visible in the data/MC plots.

\section{Analysis techniques\label{sec:analysistech}}
The measurements discussed in the following sections use analysis techniques in which either \mtop alone is determined
or \mtop and the overall jet energy scale factor (JSF)
are determined simultaneously. For the lepton+jets and the all-jets channels we use analyses based on the ideogram technique (Section~\ref{ideogram}). While the ideogram technique provides the most precise measurements, it is not suitable for dilepton events where the presence of more than one neutrino introduces uncertainties in the use of the measured \MET. Instead, for the dilepton channel,  we use the Analytical Matrix Weighting Technique (AMWT) method (Section~\ref{amwt}).

\subsection{One- and two-dimensional ideogram analyses}
\label{sec:Ideogram}
\label{ideogram}

The ideogram method is a joint maximum likelihood fit that determines \mtop and, optionally, the JSF from a sample of selected \ttbar candidate events in the lepton+jets or all-jets channels.
The observable used for measuring $\mtop$ is the mass $\mtop^\text{fit}$ estimated by a kinematic fit~\cite{Abbott:1998dc}.
The kinematic fit constrains the candidates for the \ttbar decay products to the hypothesis of the production of two heavy particles of equal masses, each one decaying to a $\PW$~boson and a bottom quark, where the $\PW$~boson invariant mass is constrained to 80.4\GeV~\cite{PDG}.
The JSF is defined as a multiplicative factor to be applied in addition to the standard jet energy corrections (JEC)~\cite{Chatrchyan:2011ds} to the four-momenta of the jets.
The JSF is determined from the invariant masses of the jet pairs, $m_{\PW}^\text{reco}$, associated with the $\PW$ bosons before the jet momenta are constrained by the kinematic fit.
For the case of a simultaneous fit to both \mtop\ and the JSF (2D approach), no prior knowledge of the JSF is assumed.
If only \mtop\ is fitted (1D approach), the jet energy scale determined from the JEC is taken as the JSF prior, fixing it to unity.
A third category of fits (hybrid approach) incorporates the prior knowledge about the jet energy scale by using a Gaussian constraint, $P(\mathrm{JSF})$, centered at 1 with a variance depending on the total
JEC uncertainty. For the hybrid analysis in the lepton+jets channel, the JSF determined from the \PW boson decays and the jet energy scale from the JEC are given equal weight in the fit. In contrast, for the hybrid fit in the all-jets channel, the jet energy scale from the JEC contributes 80\% of the information, because of the larger uncertainty on the JSF from the 2D fit.

The distributions of $\mtop^\text{fit}$ and  $m_\PW^\text{reco}$ are obtained from simulation for three to seven different $\mtop$ and three to five different JSF values for the \ttbar signal, and from simulated background events (lepton+jets) or the control sample for the multijet background (all-jets).
From these distributions, probability density functions are derived separately for different cases of jet-parton assignments for the signal, and for the background contribution.
The signal functions depend on  $\mtop$ and JSF, and are labeled $P(m_{\PQt}^\text{fit}|\mtop,\mathrm{JSF})$ and  $P(m_{\PW}^\text{reco}|\mtop,\mathrm{JSF})$, respectively, for an event in the final likelihood.

The likelihood for measuring \mtop and the JSF in an observed data sample can be expressed as:
\begin{equation}
\mathcal{L}\left(\text{sample}|\mtop,\mathrm{JSF}\right) = \prod_\text{events} \mathcal{L}\left(\text{event}|\mtop,\mathrm{JSF}\right)^{w_\text{event}},
\end{equation}
where the event weight $w_\text{event}=c\,\sum_{i=1}^{n}P_\mathrm{gof}\left(i\right)$ is used in the lepton+jets analysis
to reduce the impact of events for which the chosen permutation of the jets is incorrect.
Here, $c$ is a normalization constant and the remaining quantities are defined as in Eq.~(\ref{eq:ideogram}).
For the all-jets channel, $w_\text{event}=1$ is used.
The event likelihoods (or \emph{ideograms}) are given by
\ifthenelse{\boolean{cms@external}}{
\begin{multline}
\mathcal{L}\left(\text{event}|\mtop,\mathrm{JSF}\right) = \sum_{i=1}^{n}P_{\mathrm{gof}}\left(i\right)\Biggl\{\\ f_{\text{sig}}P_{\text{sig}}\left(m_{\text{t},i}^{\text{fit}},m_{\text{W},i}^{\text{reco}}|\mtop,\mathrm{JSF}\right)\\+\left(1-f_{\text{sig}}\right)P_{\text{bkg}}\left(m_{\text{t},i}^{\text{fit}},m_{\text{W},i}^{\text{reco}}\right)\Biggr\},
\label{eq:ideogram}
\end{multline}
}{
\begin{equation}
\mathcal{L}\left(\text{event}|\mtop,\mathrm{JSF}\right) = \sum_{i=1}^{n}P_{\mathrm{gof}}\left(i\right)\Biggl\{ f_{\text{sig}}P_{\text{sig}}\left(m_{\text{t},i}^{\text{fit}},m_{\text{W},i}^{\text{reco}}|\mtop,\mathrm{JSF}\right)+\left(1-f_{\text{sig}}\right)P_{\text{bkg}}\left(m_{\text{t},i}^{\text{fit}},m_{\text{W},i}^{\text{reco}}\right)\Biggr\},
\label{eq:ideogram}
\end{equation}
}
where the index $i$ runs over the $n$ selected permutations of an event that each have a goodness-of-fit probability $P_{\mathrm{gof}}$ assigned from the kinematic fit.
The signal fraction $f_{\text{sig}}$ is assumed to be 1 for the lepton+jets channel
and is left as a free parameter of the fit for the all-jets channel.
The background term $P_{\text{bkg}}$ is independent of both \mtop and the JSF for backgrounds determined from the collision data.

As the W boson mass is fixed to 80.4\GeV in the fit, the observables $\mtop^{\text{fit}}$ and $m_{\text{W}}^{\text{reco}}$ have a low correlation coefficient (less than 5\%) and
the probability density $P$ can be factorized into one-dimensional expressions,
\ifthenelse{\boolean{cms@external}}{
\begin{multline}
P\Big(m_{\text{t}}^{\text{fit}},m_{\text{W}}^{\text{reco}}|\mtop,\mathrm{JSF}\Big)	=	\sum_{j}f_{j}P_{j}\Big(m_{\text{t}}^{\text{fit}}|\mtop,\mathrm{JSF}\Big)\\
\times P_{j}\Big(m_{\text{W}}^{\text{reco}}|\mtop,\mathrm{JSF}\Big),
\end{multline}
}{
\begin{equation}
P\Big(m_{\text{t}}^{\text{fit}},m_{\text{W}}^{\text{reco}}|\mtop,\mathrm{JSF}\Big)	=	\sum_{j}f_{j}P_{j}\Big(m_{\text{t}}^{\text{fit}}|\mtop,\mathrm{JSF}\Big)\, P_{j}\Big(m_{\text{W}}^{\text{reco}}|\mtop,\mathrm{JSF}\Big),
\end{equation}
}
where the index $j$ denotes the different jet-parton permutation classes defined for the measurement.
Their relative fraction $f_{j}$ is either determined from the simulated sample with $m_{\text{t,gen}}=172.5\GeV$ or by the fit.

The most likely $\mtop$ and JSF values are obtained by minimizing $-2\ln \mathcal{L}\left(\text{sample}|\mtop,\mathrm{JSF}\right)$ for the 2D and hybrid analyses. For the 1D analyses only $\mtop$ is determined and the JSF is set to unity during the minimization.

\subsection{Analytical matrix weighting technique (AMWT)}
\label{amwt}

The measurement of $\mtop$ for the dileptonic \ttbar decays is performed using the analytical matrix weighting technique (AMWT).
This is based on a matrix weighting technique used by the D0 collaboration~\cite{D0amwt}, combined with an analytical algorithm to find solutions of the kinematic equations~\cite{DiLep2011}. The method allows the determination of $m_{\PQt}$ with the assumption of $\mathrm{JSF} = 1$, and in this sense, the results are comparable to the 1D fits performed in either the lepton+jets or all-jets channels (see Section~\ref{ideogram}).

In dileptonic \ttbar decays, the final state consists of two charged leptons, two neutrinos, and two \PQb quarks, resulting in 18 unknowns: three momentum components for each of the six final state particles.
Of these, we observe the momenta of the two charged leptons, the momenta of the two jets, and the momenta of all of the other charged particles and jets. If there are more than two jets in an event we have to select the jets to assign to the \PQb quarks from the decay of the top quark pair. We preferentially assign \PQb-tagged jets to these.
Hence, after physics object reconstruction, we measure the following observables for each event:
\begin{itemize}
\item the momenta $\vec{p}_{\ell^+}$ and $\vec{p}_{\ell^-}$ of the charged leptons from the \PWp and \PWm decays,
\item the momenta $\vec{p}_\PQb$ and $\vec{p}_{\PAQb}$ of the \PQb and \PAQb quarks produced by the \PQt and \PAQt quark decays,
\item the total transverse momentum $\vec{p}_{\mathrm{T}}^{\kern1pt\ttbar}$ of the $\ttbar$  pair.
\end{itemize}

This leaves four unknowns that must be solved analytically.  Conservation of four-momentum provides the following four constraints on the kinematics, if a hypothetical value for the top-quark mass is assumed:
\begin{itemize}
\item the masses $m_{\ell^+\nu}$ and $m_{\ell^-\bar{\nu}}$
of the lepton-neutrino pairs from the \PWp and \PWm decays are constrained to be 80.4\GeV~\cite{PDG},
\item the masses of the systems of particles from the $\PQt$ and $\cPaqt$ decays must equal the hypothesized mass of the top quark.
\end{itemize}

Hence, the system of equations is appropriately constrained. However, there is not a unique solution, because the equations are nonlinear. For a given assignment of reconstructed momenta to final-state particles there can be up to four solutions for the neutrino momenta such that the event satisfies all of the constraints. There is a twofold ambiguity of assigning jet momenta to the \PQb and \PAQb jets, which doubles this to eight possible solutions. We follow the algorithm given in Refs.~\cite{39,37} to find these solutions. In rare cases, a latent singularity in the equations used to find these solutions can prohibit the calculation of the longitudinal momenta. In such events, a numerical method is employed to find the incalculable variables \cite{39}.

For each event, we find all solutions of neutrino momenta for hypothesized top quark masses between 100 and 600\GeV, in 1\GeV increments. In general, we expect solutions to be found for a large range of mass hypotheses.
To each solution we assign a weight $w$ given by \cite{DG}:
\ifthenelse{\boolean{cms@external}}{
\begin{multline}
w(\vec{X}|\mtop) = \left[\sum_{\text{initial partons}} F\left(x_1\right)F\left(x_2\right)\right]\\
\times p\left(E_{\ell^+}|\mtop\right)
 p\left(E_{\ell^-}|\mtop\right),
\end{multline}
}{
\begin{equation}
w(\vec{X}|\mtop) = \left[\sum_{\text{initial partons}} F\left(x_1\right)F\left(x_2\right)\right]p\left(E_{\ell^+}|\mtop\right)p\left(E_{\ell^-}|\mtop\right),
\end{equation}
}
where $\vec X$ represents the momentum vectors of the final state particles as obtained from the solutions of the kinematic equations. We sum the parton distribution functions $F(x)$, evaluated at $Q^2 = \mtop^2$, over the possible LO initial parton states
($\PQu\PAQu$, $\PAQu\PQu$, $\PQd\PAQd$, $\PAQd\PQd$,
and $\Pg\Pg$); $x_1$ and $x_2$ are the Bjorken $x$ values for the initial-state partons which can be computed from the momenta of the final-state particles. The function $p(E|\mtop)$ is the probability density of observing a charged lepton of energy $E$ in the rest frame of a top quark of mass $\mtop$, given by \cite{DG}:
\begin{equation}
p(E|\mtop)=\frac{4\mtop E(\mtop^2-m_\PQb^2-2\mtop E)}{(\mtop^2-m_\PQb^2)^2+M_\PW^2(\mtop^2-m_\PQb^2)-2M_\PW^4},
\end{equation}
where the \PQb quark mass, $m_{\PQb}$,~is set to 4.8\GeV, and the \PW~boson mass, $M_{\PW}$, to 80.4\GeV. For each \mtop hypothesis, we find an overall weight by summing the weights of all solutions found.
To compensate for mismeasurements of the momenta due to finite detector resolution or the loss of correlation between the jet and quark momentum because of hard-gluon radiation, we account for the jet energy resolution during reconstruction. Every event in both the collision and simulated data is reconstructed 500 times, each time with jet momenta drawn randomly from a Gaussian distributions of widths given by the detector resolution and with means given by the measured momenta.
After this randomization procedure, approximately 96\% of all events in both the collision and simulated data have at least one solution, and hence a top quark mass estimator. The final weight curve of each event is given by the average of the weight distributions from each of the 500 randomizations, after excluding the cases for which there is no valid solution. This distribution serves as a measure of the relative probability that the observed event occurs for any given value of \mtop.

The estimator for \mtop is then the hypothesized mass with the highest average sum weight for each event, called the AMWT mass,
$m_{\PQt}^{\mathrm{AMWT}}$.

\section{Mass measurements\label{sec:mass}}
\label{mass}
\subsection{The lepton+jets channel}
\label{sec:ljets}

To check the compatibility of an event with the \ttbar hypothesis and improve the resolution of the reconstructed quantities, a kinematic fit~\cite{Abbott:1998dc} is applied to the events.
For each event, the inputs to the fitter are the four-momenta of the lepton and the four leading jets, the missing transverse energy, and their respective resolutions.
The fit constrains these to the hypothesis of the production of two heavy particles of equal mass, each one decaying to a \PW boson with an invariant mass of 80.4\GeV~\cite{PDG} and a bottom quark.
It minimizes
$\chi^{2}=\left(\mathbf{x}-\mathbf{x}^{m}\right)^{T}\textit{\textbf{E}}^{-1}\left(\mathbf{x}-\mathbf{x}^{m}\right)$
where $\mathbf{x}^{m}$ is the vector of measured observables, $\mathbf{x}$ is the vector of fitted observables, and $\textit{\textbf{E}}^{-1}$ is the inverse error matrix which is given by the resolutions of the observables.
The two \PQb-tagged jets are candidates for the bottom quarks in the \ttbar hypothesis, while the two untagged jets serve as candidates for the light quarks for one of the \PW boson decays.
This leads to two possible parton-jet assignments per event and two solutions for the $z$ component of the neutrino momentum.

For simulated \ttbar events, the parton-jet assignments are classified as \emph{correct permutations},  \emph{wrong permutations}, and \emph{unmatched permutations}. The correct permutation class includes those events for which all of the quarks from the \ttbar decay (after initial-state parton shower) are correctly matched to the selected jets within a distance $\Delta R<0.3$. The wrong permutations class covers the events for which the jets from the \ttbar decay are correctly matched to the selected jets, but where two or more of the jets are interchanged. Lastly, the unmatched permutations class includes the events for which at least one quark from the \ttbar decay is not matched unambiguously to any of the four selected jets.
To increase the fraction of correct permutations, we require $P_{\mathrm{gof}} > 0.2$ for the kinematic fit with two degrees of freedom.
This selects 28\,295 events for the mass measurement, with an estimated composition of 96.3\% \ttbar signal and 3.7\% non-\ttbar background, which is dominated by single top quark events.
In the mass extraction, the permutations are weighted by their $P_\mathrm{gof}$ values, and the effective fraction of correct permutations among the \ttbar signal improves from 13\% to 44\%, while the fractions of wrong and unmatched permutations change from 16\% to 21\% and 71\% to 35\%, respectively, determined in simulation.

\begin{figure*}
\centering
\includegraphics[width=0.49\textwidth]{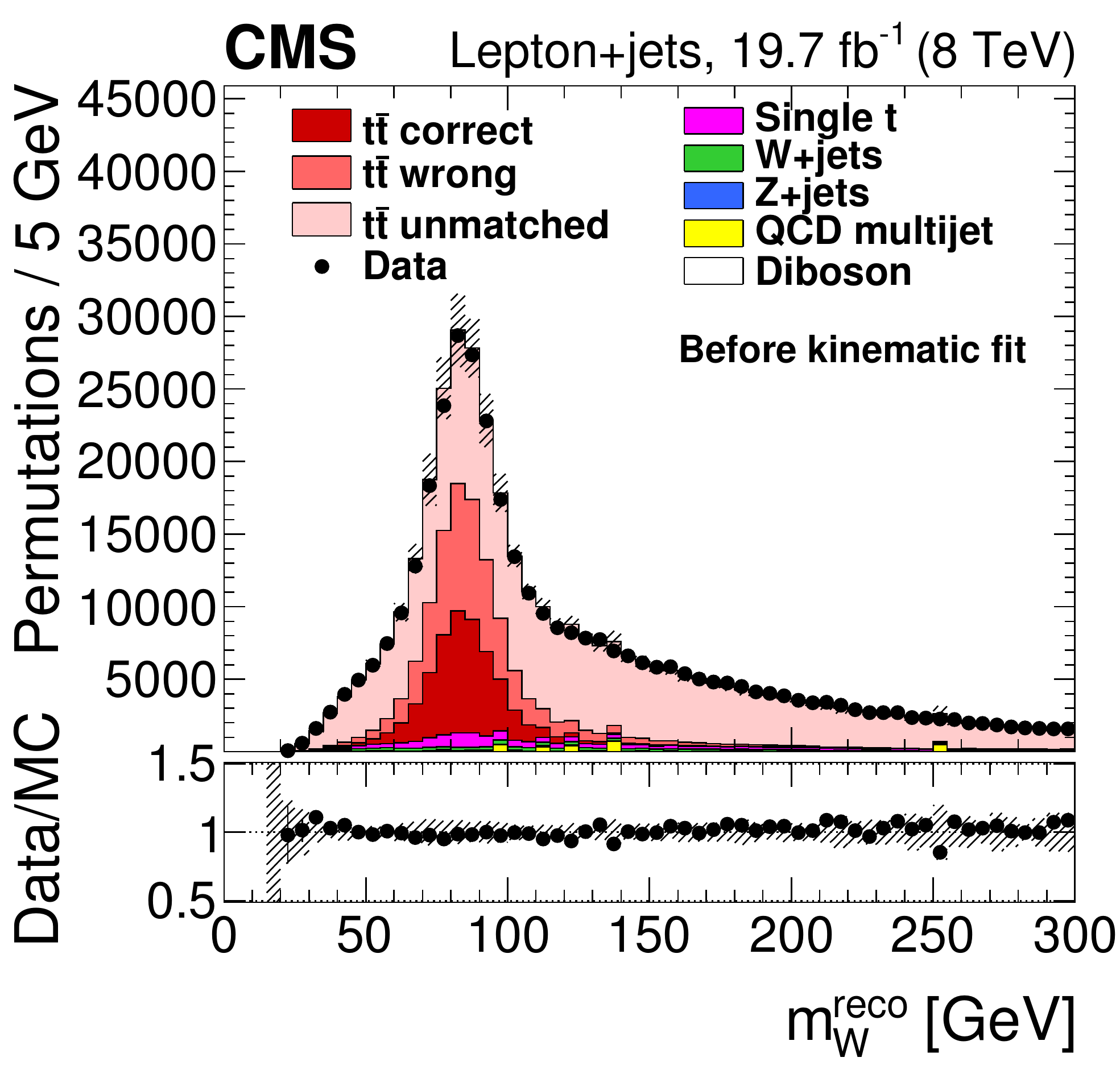}
\includegraphics[width=0.49\textwidth]{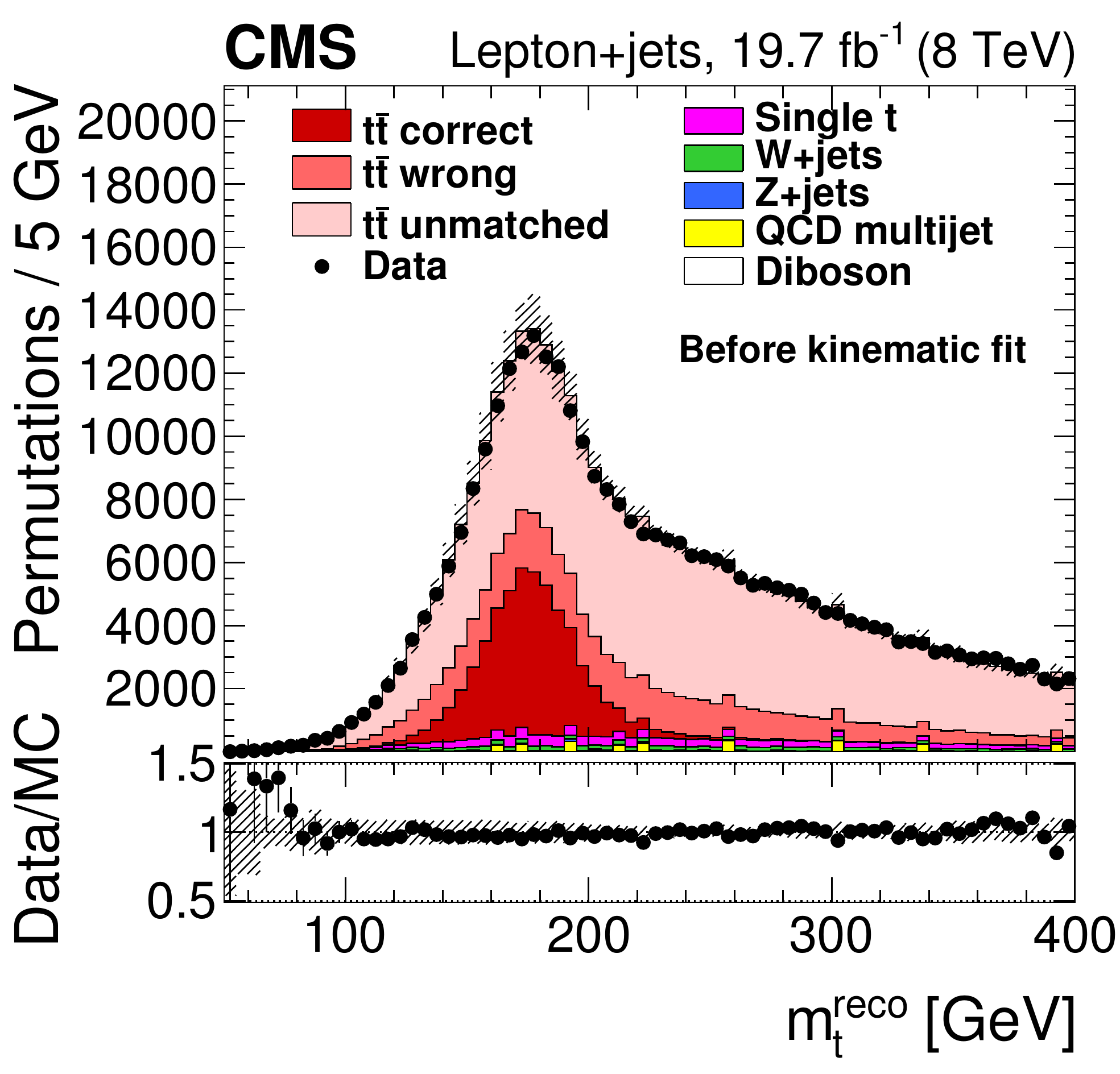}
\includegraphics[width=0.49\textwidth]{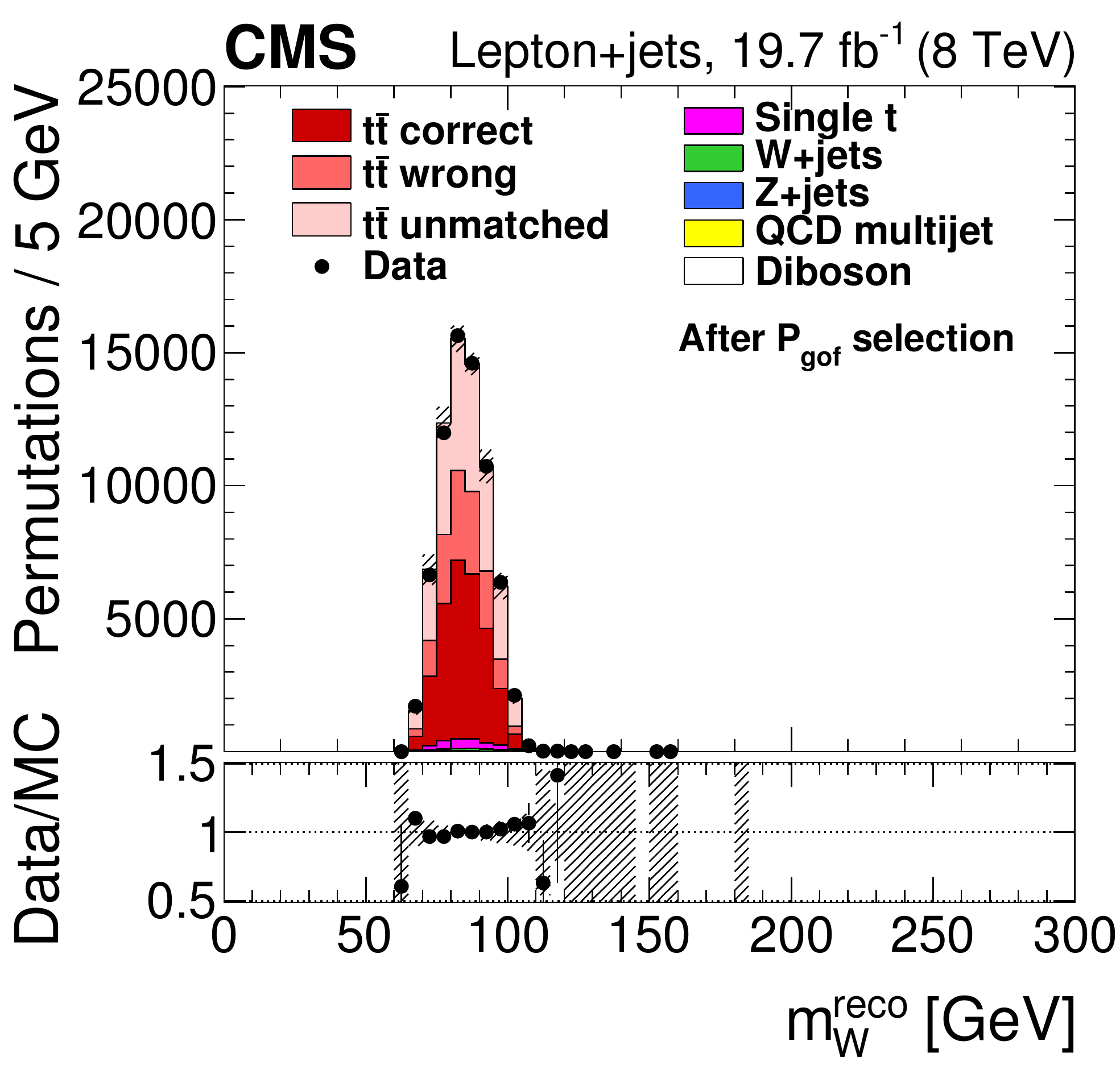}
\includegraphics[width=0.49\textwidth]{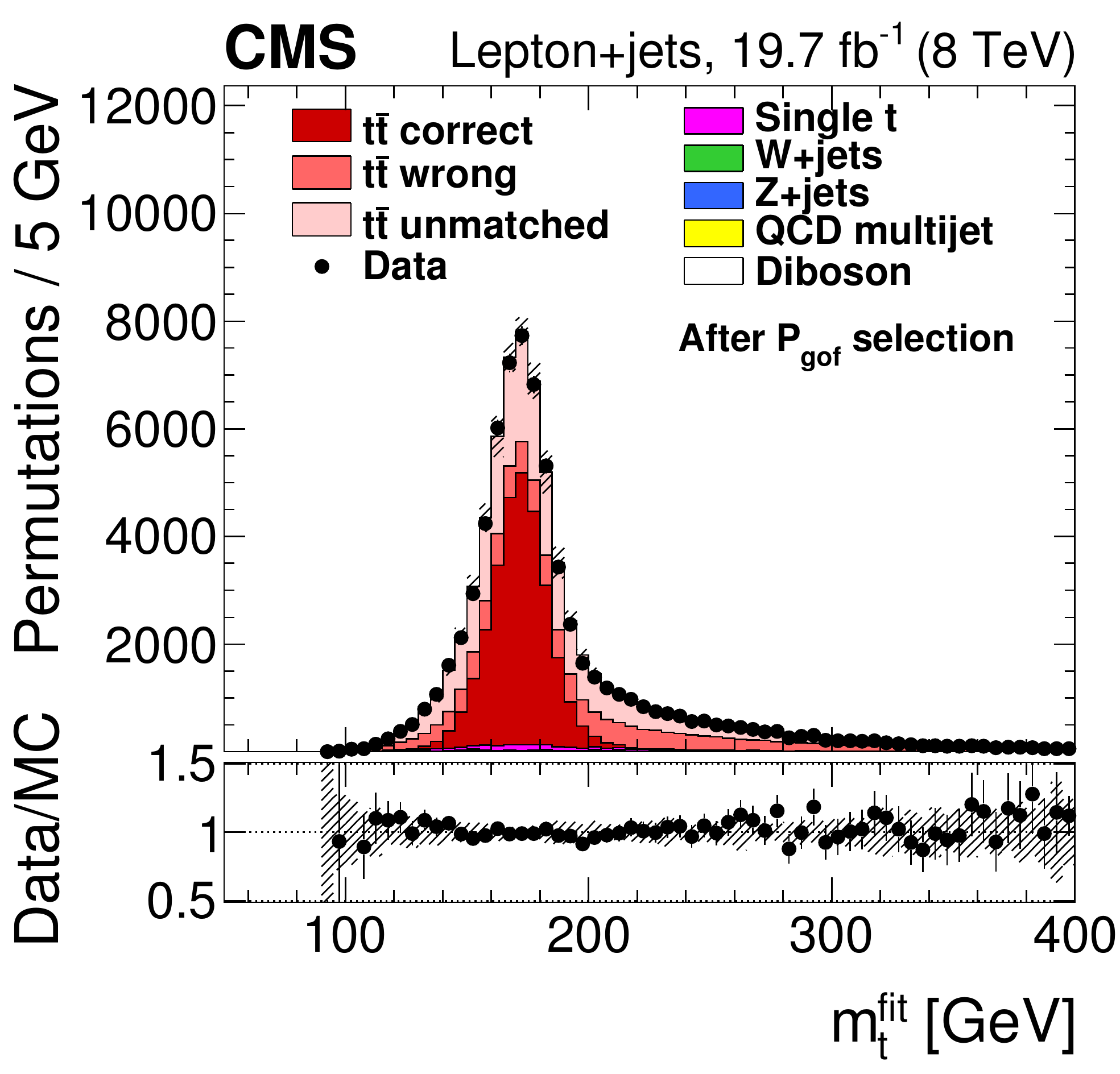}
\caption{\label{fig:controlplot-weighted-obs}
Reconstructed masses of (upper left) the W bosons decaying to $\PQq\PAQq$ pairs and (upper right) the corresponding top quarks, prior to the kinematic fitting to the \ttbar hypothesis. Panels (lower left) and (lower right) show, respectively, the reconstructed W boson masses and the fitted top quark masses after the goodness-of-fit selection. The total number of permutations found in simulation is normalized to be the same as the total number of permutations observed in data. The vertical bars show the statistical uncertainty and the hatched bands show the statistical and systematic uncertainties added in quadrature. The lower portion of each panel shown the ratio of the yields between the collision data and the simulation.}
\end{figure*}

Figure~\ref{fig:controlplot-weighted-obs} shows the distributions before and after the kinematic fit and $P_\mathrm{gof}$ selection of the reconstructed mass $m_\PW^\text{reco}$ of the \PW boson decaying to a $\qqbar$ pair and the mass $\mtop^\text{reco}$ of the corresponding top quark for all possible permutations.

The ideogram method (Section~\ref{sec:Ideogram}) is calibrated for each combination of the top quark mass hypothesis, $m_{\PQt}^{\text{gen}}$ and $\mathrm{JSF}$ values by conducting  10\,000 pseudo-experiments, separately for the muon and electron channels, using simulated \ttbar and background events.
The average deviations between extracted mass and JSF and their input values are obtained as a function of $m_{\PQt}^{\rm{gen}}$ and the bias is fit with a linear function for each generated JSF value.
From these fits, additional small corrections for calibrating the top quark mass and the jet energy scale are derived as linear functions of both the extracted top quark mass and JSF.
The corrections are approximately $-0.2$\GeV for $m_{\PQt}$ and $-0.4$\% for the JSF.
The statistical uncertainties of the method are also corrected by factors of approximately 1.04 that are derived from the widths of the corresponding pull distributions.

The 2D ideogram fit to the combined electron and muon channels yields:
\begin{equation*}\begin{split}
m_{\PQt}^{\text{2D}} & =  172.14\pm0.19\,\text{(stat+JSF)}\GeV,\\
\mathrm{JSF}^{\text{2D}} & =  1.005\pm0.002\stat.
\end{split}\end{equation*}

As \mtop and the JSF are measured simultaneously, the statistical uncertainty in $\mtop$ combines the statistical uncertainty arising from both components of the measurement.
The uncertainty of the measurement agrees with the expected precision obtained by performing pseudo-experiments.

\begin{figure}
\centering
\includegraphics[width=0.49\textwidth]{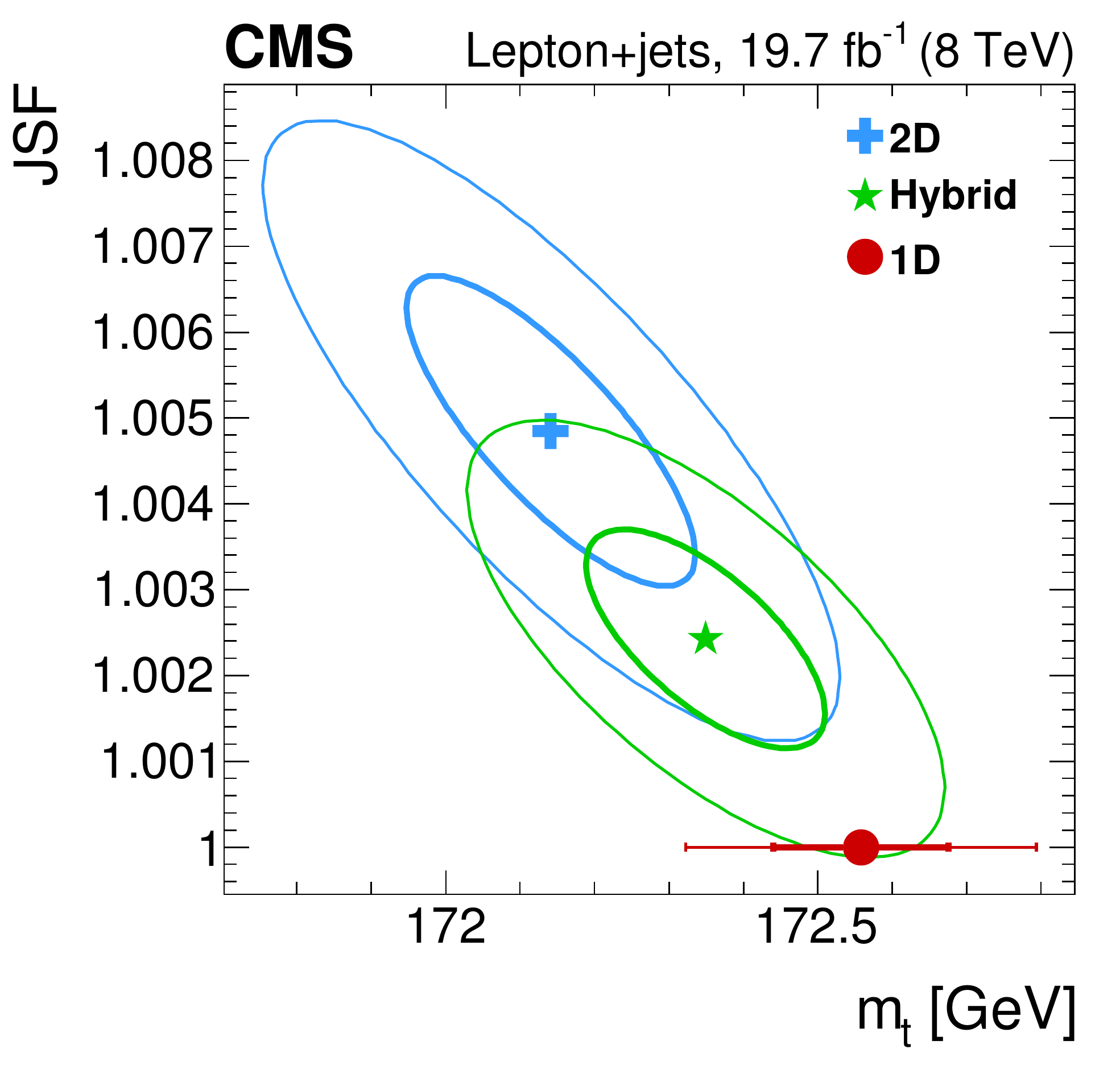}
\caption{\label{fig:result-ljets} The two-dimensional likelihood ($-2 \Delta \log \left(\mathcal{L}\right)$) for the lepton+jets channel for the 2D, hybrid, and 1D fits. The thick (thin) ellipses correspond to contours of $-2 \Delta \log \left(\mathcal{L}\right) = 1\,(4)$ allowing the construction of the one (two) $\sigma$ statistical intervals of $\mtop$.
For the 1D fit, the thick and thin lines correspond to the one and two $\sigma$ statistical uncertainties, respectively.
}
\end{figure}

The results in the individual muon and electron channels are compatible within their statistical uncertainties:
\ifthenelse{\boolean{cms@external}}{
\begin{equation*}\begin{split}
\mu\text{+jets: } \mtop^{\text{2D}} & =    172.03\pm0.27\,\text{(stat+JSF)}\GeV,\\&  \quad\mathrm{JSF}^{\text{2D}} = 1.007\pm0.003\stat,\\
\Pe\text{+jets: } \mtop^{\text{2D}} & =   172.26\pm0.28\,\text{(stat+JSF)}\GeV,\\&  \quad\mathrm{JSF}^{\text{2D}} = 1.003\pm0.003\stat.
\end{split}\end{equation*}
}{
\begin{equation*}\begin{split}
\mu\text{+jets: } \mtop^{\text{2D}} & =   172.03\pm0.27\,\text{(stat+JSF)}\GeV,\  \mathrm{JSF}^{\text{2D}} = 1.007\pm0.003\stat,\\
\Pe\text{+jets: } \mtop^{\text{2D}} & =   172.26\pm0.28\,\text{(stat+JSF)}\GeV,\  \mathrm{JSF}^{\text{2D}} = 1.003\pm0.003\stat.
\end{split}\end{equation*}
}
The 1D and hybrid analyses give results of
\begin{equation*}\begin{split}
m_{\PQt}^{\text{1D}} & =  172.56\pm0.12\stat\GeV,\\
m_{\PQt}^{\text{hyb}} & =  172.35\pm0.16\,\text{(stat+JSF)}\GeV,\\
\mathrm{JSF}^{\text{hyb}} & =  1.002\pm0.001\stat,
\end{split}\end{equation*}
respectively. Both the 2D and hybrid results for the JSF ($\mathrm{JSF}^{\text{2D}}$ and $\mathrm{JSF}^{\text{hyb}}$) are within 0.5\% of one. The results for $m_{\PQt}$ and the JSF are compared in Fig.~\ref{fig:result-ljets}, which shows the two-dimensional statistical likelihoods obtained from data in the 2D and hybrid cases and \mtop from the 1D analysis.

\subsection{The all-jets channel}

\label{alljets}
As in the lepton+jets channel, a kinematic fit~\cite{kinfitCMS} is used to improve the resolution of the reconstructed quantities and to check the compatibility of an event with the \ttbar hypothesis.
For each event, the inputs to the fit are the four-momenta of  the six leading jets.
The fit constrains these to the hypothesis of the production of two heavy particles of equal masses, each one decaying to a $\PW$~boson with its invariant mass constrained to 80.4\GeV~\cite{PDG} and a bottom quark.
The two $\PQb$-tagged jets are candidates for the bottom quarks in the \ttbar hypothesis, while the four untagged jets serve as candidates for the light quarks of the $\PW$~boson decays.
This leads to six possible parton-jet assignments per event and the assignment that fits best to the \ttbar hypothesis based on the $\chi^2$ of the kinematic fit is chosen.
As final selection criteria, we require $P_{\mathrm{gof}} > 0.1$ for the kinematic fit with three degrees of freedom, and the two \PQb quark jets be separated in $\eta$-$\phi$ space by $\Delta R_{\PQb\PAQb}>2.0$.
These requirements select 7049 events for the mass measurement in data and the fraction of signal events $f_{\text{sig}}$ increases from 14\% to 61\% based on the simulation.

For simulated \ttbar events, the parton-jet assignments are classified as \emph{correct permutations} and \emph{wrong permutations}.  The correct permutation class is defined in the same way as for the lepton+jets channel (Section~\ref{sec:ljets}). The wrong permutations class consists of permutations where at least one quark from the \ttbar decay is not unambiguously matched with a distance of  $\Delta R<0.3$  to any of the six selected jets.
For correct permutations, which compose 42\% of the selected \ttbar events, the kinematic fit improves the resolution of the fitted values of \mtop from 13.8 to 7.5\GeV.

The multijet background from QCD is modeled using a control sample that is obtained from data with the same event selection but without the \PQb tagging requirement. While this sample has a small contamination of a few percent coming from signal events, these have no influence on the results for the background model. For each event, the kinematic selection is applied to all possible assignments of the six jets to the six quarks from the \ttbar hypothesis.
The best fitting assignment is chosen and the event is used to model the background if it fulfills the  $P_\mathrm{gof}$ and $\Delta R_{\PQb\PAQb}$ criteria.
The modeled background  is compared to  the background predicted by an event mixing technique~\cite{AllJets2011} . Both predictions are found to agree within their uncertainties that are derived from the validation of the methods on simulated multijet events.

\begin{figure*}
\centering
\includegraphics[width=0.49\textwidth]{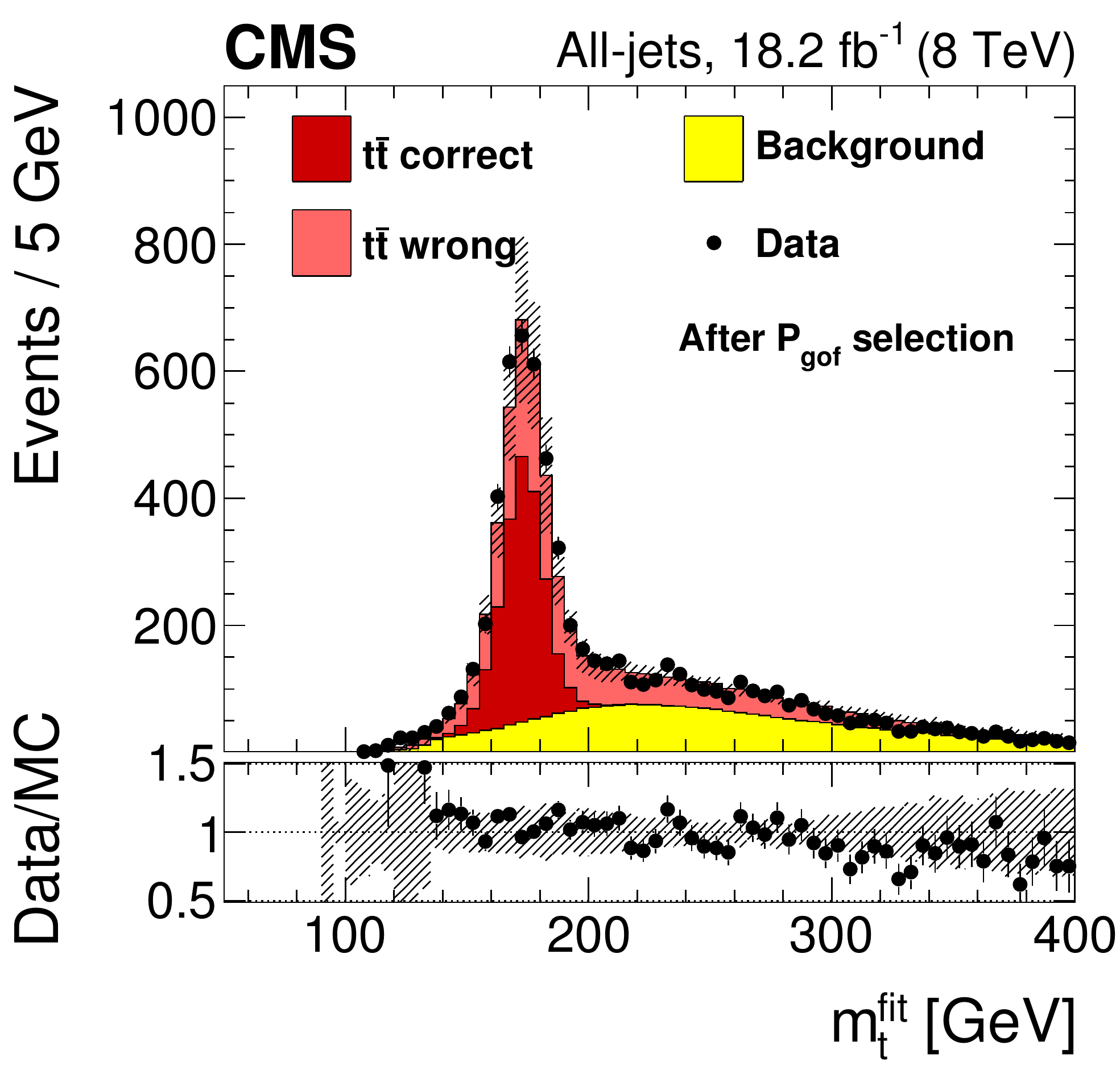}
\includegraphics[width=0.49\textwidth]{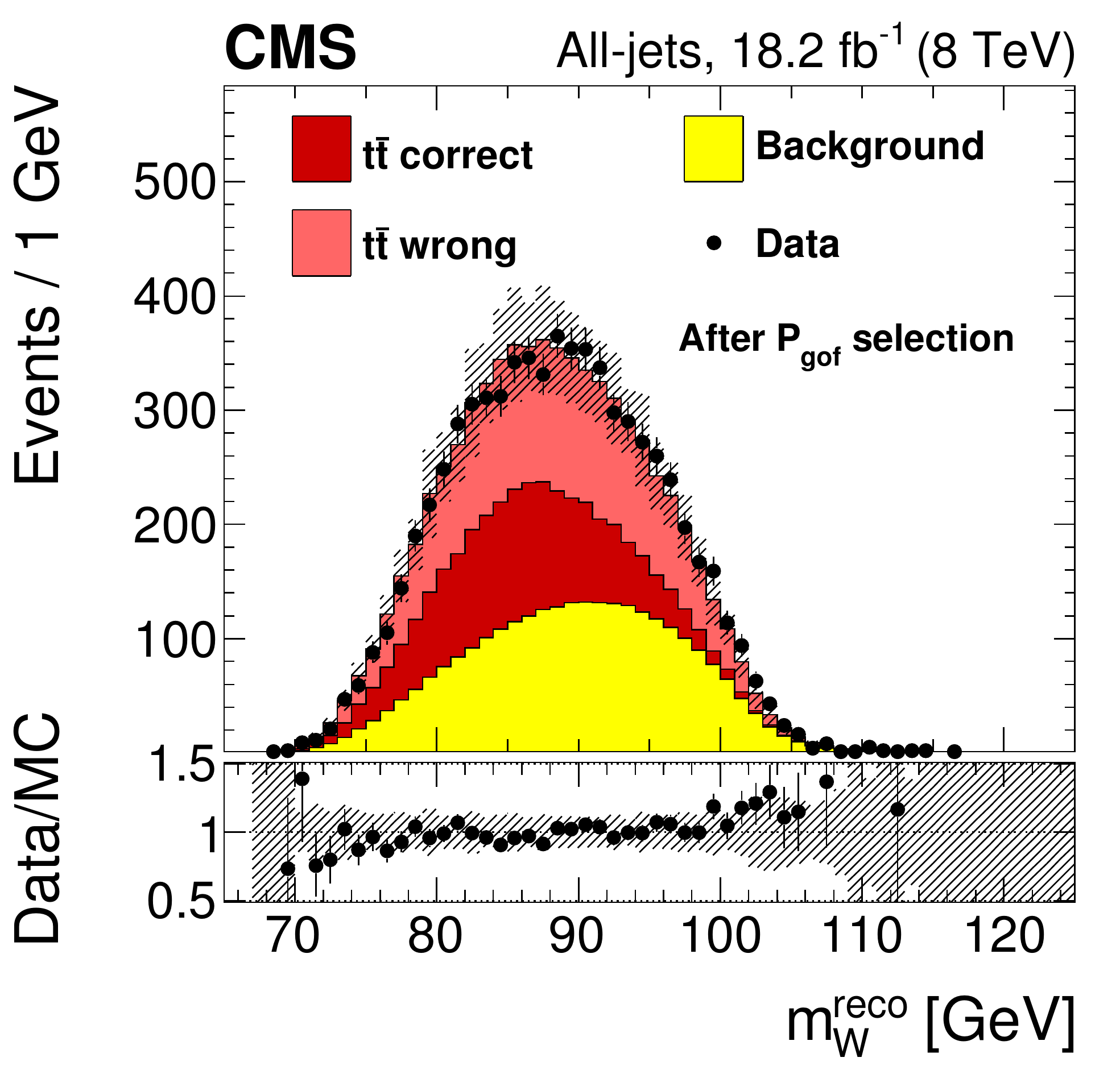}
\includegraphics[width=0.49\textwidth]{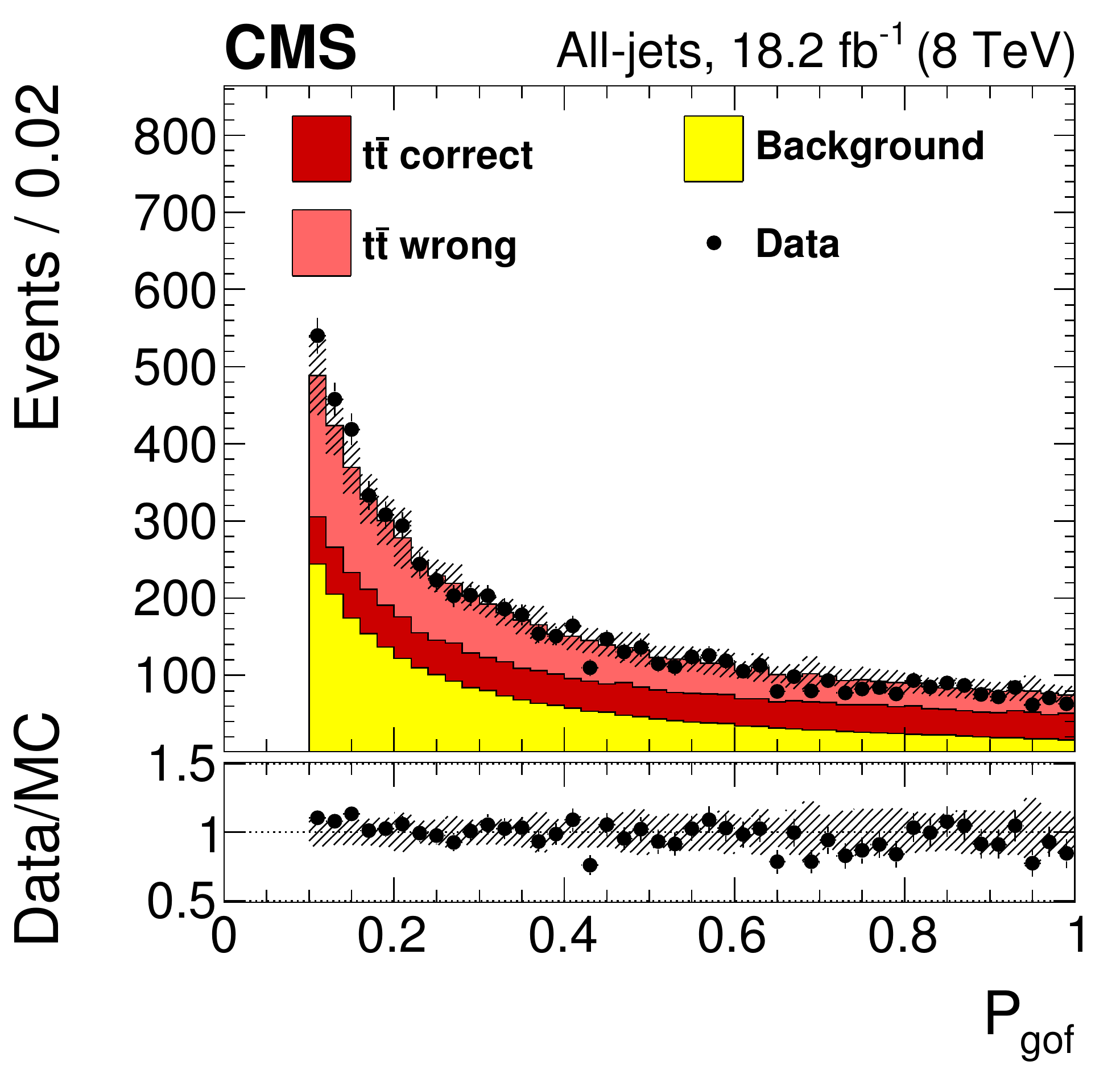}
\includegraphics[width=0.49\textwidth]{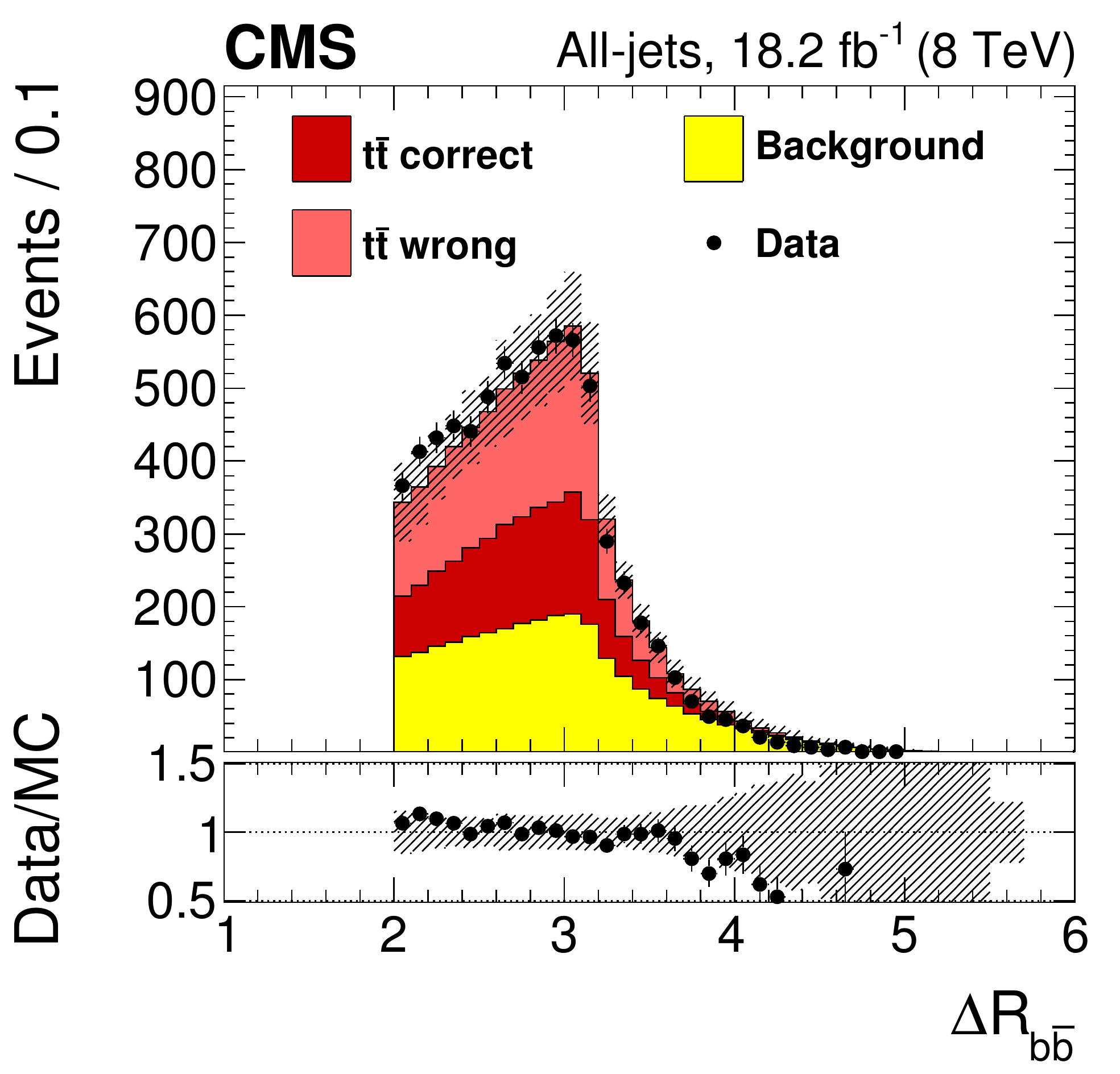}
\caption{\label{fig:controlplot-obs-alljets}
Distributions of (upper left) the reconstructed top quark mass from the kinematic fit, (upper right) the average reconstructed \PW boson mass, (lower left) the goodness-of-fit probability, and (loser right) the separation of the two \PQb quark jets for the all-jets channel.
The simulated \ttbar signal and the background from the control sample are normalized to data.
The value of \mtop used in the simulation is 172.5\GeV and the nominal jet energy scale is applied. The vertical bars show the statistical uncertainty and the hatched bands show the statistical and systematic uncertainties added in quadrature. The lower portion of each panel shows the ratio of the yields between the collision data and the simulation.}
\end{figure*}

Figure~\ref{fig:controlplot-obs-alljets}
compares data to the expectation from simulated \ttbar signal and background estimate from the data for $m_\PQt^\text{fit}$, $m_{\PW}^\text{reco}$, $P_\mathrm{gof}$, and $\Delta R_{\bbbar}$.

The ideogram method is calibrated for each combination of the $m_{\PQt}^{\rm{gen}}$  and $\mathrm{JSF}$ values by conducting  10\,000 pseudo-experiments.
The average deviations between extracted mass and JSF and their input values are obtained as a function of $m_{\PQt}^{\rm{gen}}$ and the bias is fit with a linear function for each generated JSF value.
From these fits, additional small corrections for calibrating the top quark mass and the jet energy scale are derived as linear functions of both the extracted top quark mass and JSF. The corrections are approximately $-0.6$\GeV for $m_{\PQt}$ and $+1.0$\% for the JSF. The statistical uncertainties of the method are corrected by factors of approximately 1.005 using values derived from the widths of the corresponding pull distributions.

Applying the ideogram method on data with no prior knowledge on the JSF (2D), yields the results:
\begin{equation*}\begin{split}
\mtop^{\text{2D}} & =  171.64 \pm 0.32\,\text{(stat+JSF)}\GeV,\\
\mathrm{JSF}^{\text{2D}} & =  1.011 \pm 0.003\stat.
\end{split}\end{equation*}

As \mtop and the JSF are measured simultaneously, the statistical uncertainty in $\mtop$ combines the statistical uncertainty arising from both components of the measurement.
The two additional free parameters in the fit, the signal fraction $f_\text{sig}$ and the fraction of correct permutations $f_\text{cp}$, are in agreement with the expectation from simulation.

Using the JEC determined from $\gamma/\Z$+jet events in combination with the JSF prior from the 2D fit yields the results in the 1D and hybrid approaches of
\begin{equation*}\begin{split}
\mtop^{\mathrm{1D} }     & =  172.46 \pm 0.23\stat\GeV, \\
\mtop^{\text{hyb} }     & =  172.32 \pm 0.25\,\text{(stat+JSF)}\GeV,\\
\mathrm{JSF}^{\text{hyb}} & =  1.002\pm0.001\stat.
\end{split}\end{equation*}

\begin{figure}
\centering
\includegraphics[width=0.49\textwidth]{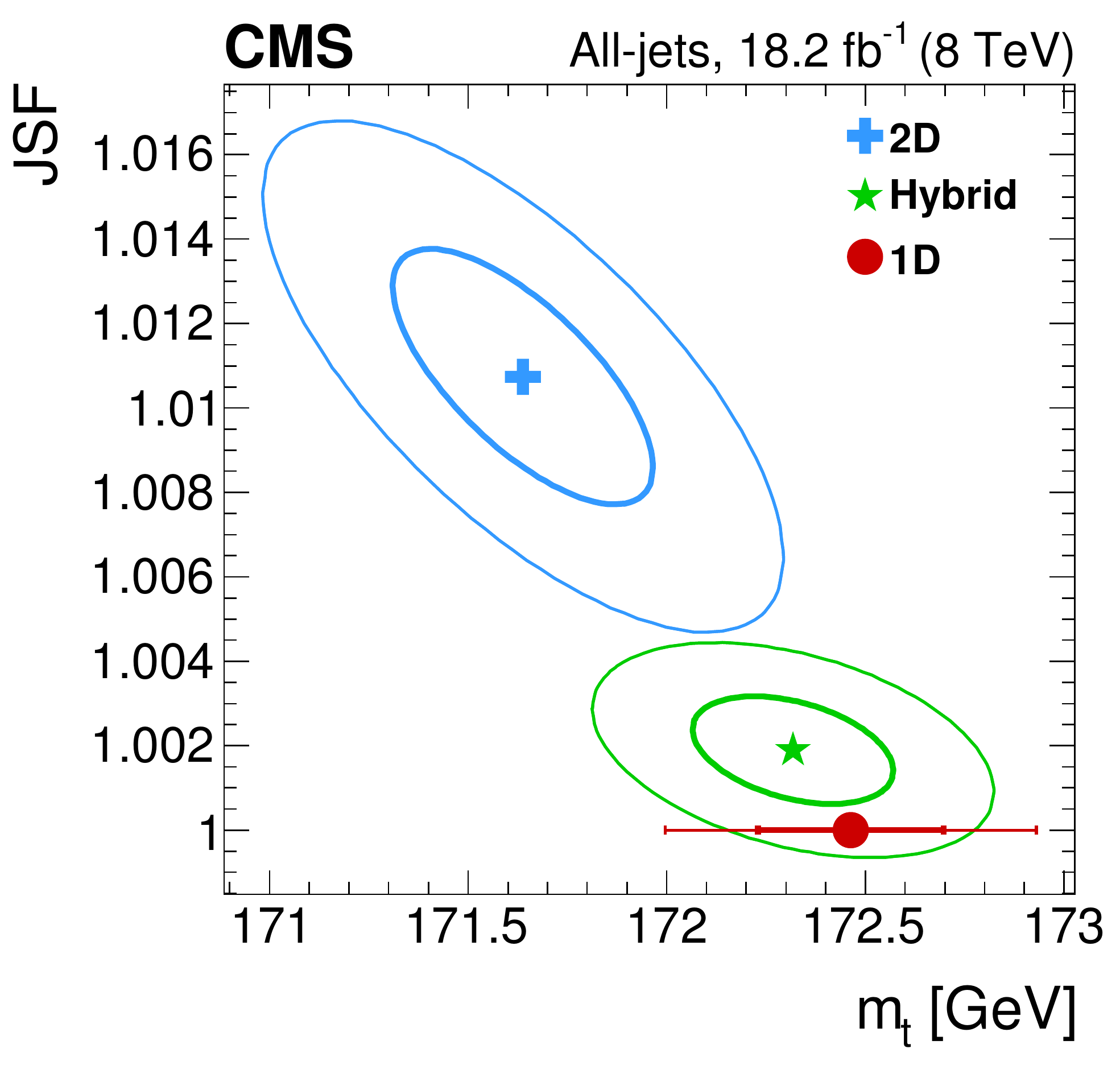}
\caption{\label{fig:result-alljets} The two-dimensional likelihood ($-2 \Delta \log \left(\mathcal{L}\right)$) for the all-jets channel for the 2D, hybrid, and 1D fits.  The thick (thin) ellipses correspond to contours of $-2 \Delta \log \left(\mathcal{L}\right) = 1\,(4)$ allowing the construction of the one (two) $\sigma$ statistical intervals of $\mtop$.
For the 1D fit, the thick and thin lines correspond to the one and two $\sigma$ statistical uncertainties, respectively.}
\end{figure}

For the all-jets channel, the fitted results for the JSF ($\mathrm{JSF}^{\mathrm{2D}}$ and $\mathrm{JSF}^{\text{hyb}}$) are both within 1.1\% of one. While the JSF results from the 2D analyses in the lepton+jets and all-jets channels differ by 0.6\%, the results from the hybrid analyses agree to within 0.2\%.
The all-jets results for $m_{\PQt}$ and the JSF are compared in Fig.~\ref{fig:result-alljets} which shows the two-dimensional statistical likelihoods obtained from data in the 2D and hybrid cases and \mtop from the 1D analysis.

\subsection{The dilepton channel}

\begin{figure}[htb]
\centering
\includegraphics[width=0.49\textwidth]{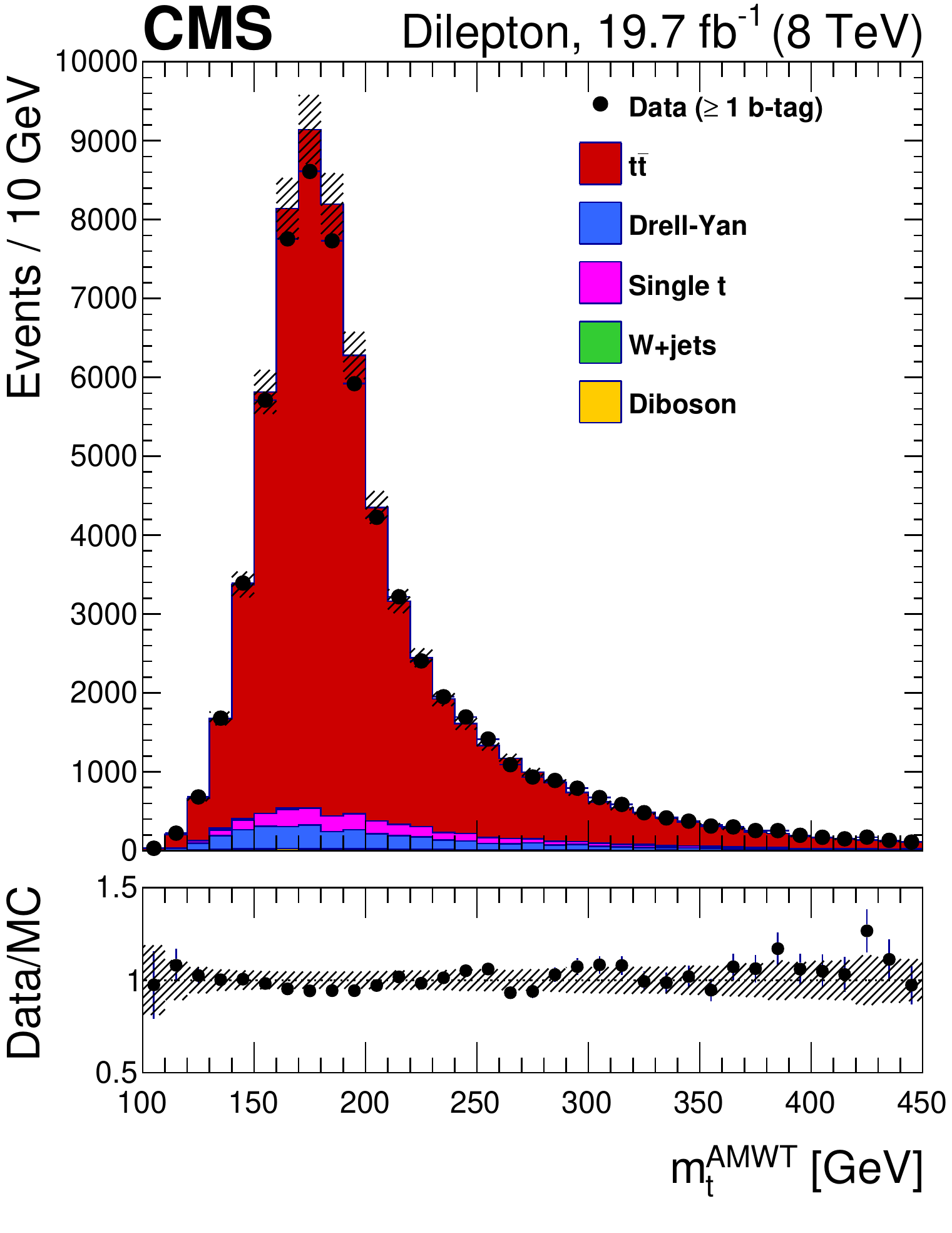}
        \caption{Distribution of $m_{\PQt}^{\mathrm{AMWT}}$ for the collision and simulated data with $\mtop = 172.5\GeV$. The vertical bars show the statistical uncertainty and the hatched bands show the statistical and systematic uncertainties added in quadrature. The lower section of the plot shows the ratio of the yields between the collision data and the simulation.}
\label{fig:amwtDistro}
\end{figure}

Figure~\ref{fig:amwtDistro} shows the distribution of $m_{\PQt}^{\mathrm{AMWT}}$ in data compared to a simulation with $\mtop = 172.5$\GeV for events containing at least one \PQb jet.
This channel is very clean with a negligible background from non \ttbar sources and the collision and simulated events are in good agreement.

AMWT masses are computed for all events in both the data and the simulations.
The $m_{\PQt}^{\mathrm{AMWT}}$ distributions computed for each of the seven simulated \ttbar mass samples are added to the distributions from the background samples, and these are treated as templates in a binned likelihood fit.
To minimize the effects of any bias from the poorly populated tails of the distribution, we only examine events with $m_{\PQt}^{\mathrm{AMWT}}$ between 100 and 400\GeV.
For each of the seven mass templates, a maximum likelihood fit is performed to the data distribution.
A parabola is fit to the negative logarithms of the maximum likelihoods returned by the fits, and the minimum of the parabola is taken as the measured mass value.

The fit is calibrated to correct for any biases induced by the reconstruction using pseudo-data. The calibration is performed by means of a test
using the simulated templates for the top quark masses between 169.5 and 175.5\GeV. We randomly draw 1000 samples of events, each selected such that the total number of events is the same as in the full data sample.
For each template, the 1000 measured masses are averaged together and subtracted from the input mass to obtain a numerical value for the bias induced by the fit.
The bias is then parametrized as a linear dependence on the generated value of \mtop, and the resulting calibration curve is used to correct for biases in the final result.

\begin{figure}[!htb]
\centering
\includegraphics[width=0.49\textwidth]{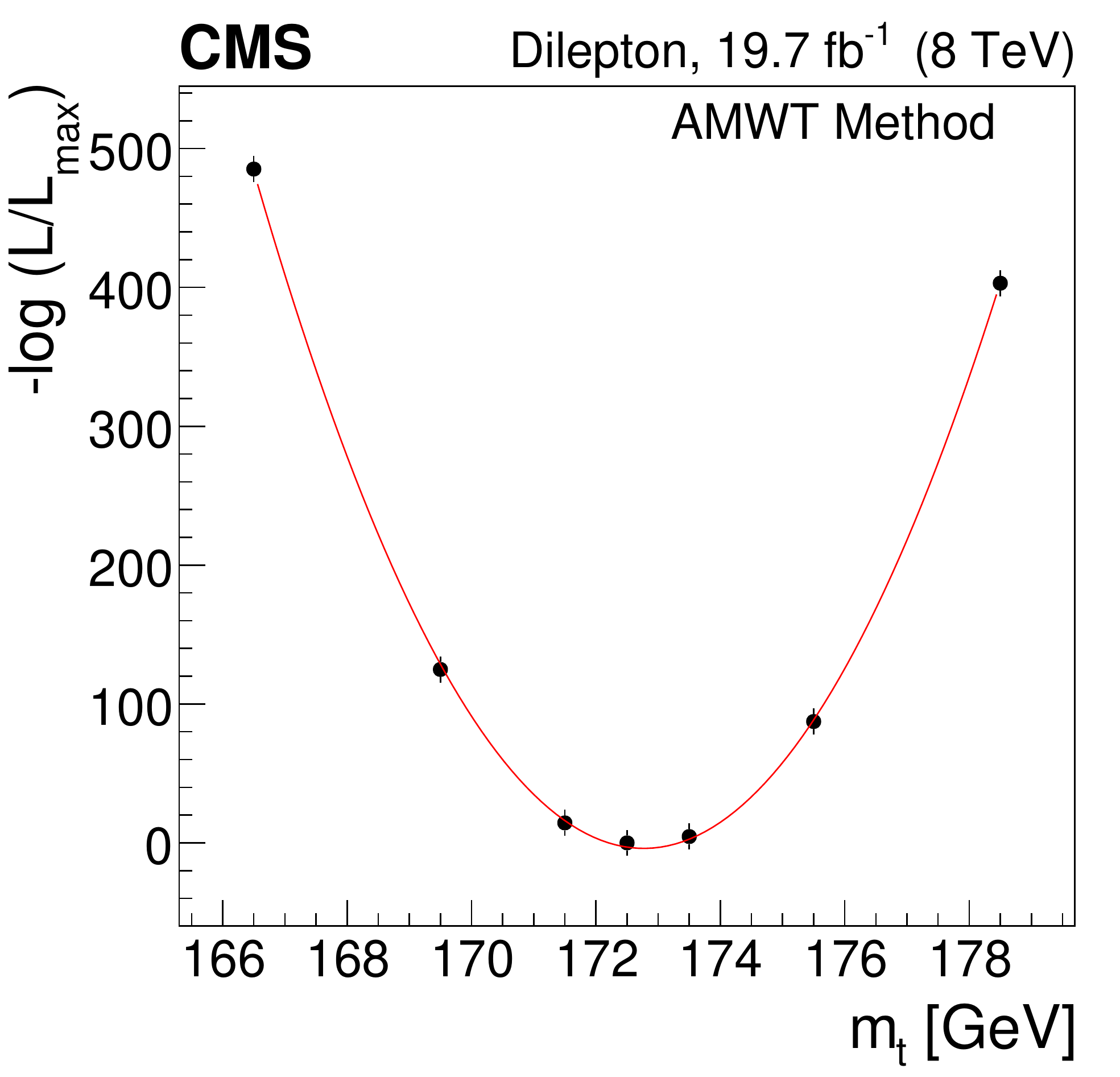}
\caption{Plot of the negative log-likelihood for data for the dilepton analysis. The continuous line represents a parabolic fit to the points.}
\label{fig:AMWTllplot}
\end{figure}

The likelihoods obtained from a fit of each of the seven simulated templates to data, as well as a second-order polynomial fit to these values, are shown in Fig. \ref{fig:AMWTllplot}. This yields an uncalibrated measured mass of $\mtop = 172.77\pm 0.19\stat\GeV$. After correcting for the fit bias, the result for the top quark mass is found to be
$ \mtop = 172.82 \pm 0.19\stat\GeV$.

The analysis was optimized with the value of \mtop blinded. The optimization of the event selection was done by minimizing the total expected (statistical+systematic) uncertainty. This resulted in the restriction of the analysis to events containing only two \PQb jets, rather than the requirement of at least one \PQb jet which was used initially.

\section{Systematic uncertainties\label{sec:uncert}}

\label{systcat}

The systematic uncertainties affecting each of the measurements can be grouped into four distinct categories: one experimental category and three theoretical categories that describe the modeling uncertainties. The experimental classification covers the uncertainties that arise from the precision of the calibration and resolution of the CMS detector and the effects coming from the backgrounds and pileup. The other three categories cover the modeling of the hard scattering process and the associated radiation; non-perturbative QCD effects, such as the simulation of the underlying event and color reconnection; and the modeling of the light- and \PQb-quark hadronization. Each of these is broken down into sub-categories leading to a total of 24 distinct systematic uncertainties. In each case the uncertainty is evaluated in terms of the largest shift that is observed in the value of \mtop that occurs when the parameter is varied by $\pm 1\sigma$, where $\sigma$ is the uncertainty assigned to that quantity. The only exception to this is if the statistical uncertainty in the observed shift is larger than the value of the calculated shift. In this case the statistical uncertainty is taken as the best estimate of the uncertainty in the parameter.

\subsection{Experimental effects}

\begin{itemize}
\item Intercalibration jet energy correction:
This is the part of the JES uncertainty originating from modeling of the radiation in the relative (\pt- and $\eta$-dependent) intercalibration procedure.
\item In situ jet energy calibration:
This is the part of the JES uncertainty coming from the uncertainties affecting the absolute JES determination using $\gamma/\Z$+jets events.
\item Uncorrelated jet energy correction:
This is the uncertainty source coming from the statistical uncertainty in the in situ jet energy calibration, the contributions stemming from the jet energy correction due to pileup effects, the uncertainties due to the variations in the calorimeter response versus time, and some detector specific effects.
To give a clear indication of the contribution to the JES uncertainty coming from pileup, we have sub-divided this uncertainty into non-pileup and pileup contributions.
\item Lepton energy scale (LES):
Analogous to the JES, the energy scale of the leptons may also induce a systematic bias. A typical variation of 0.6\% is taken for electrons in the barrel region and 1.5\% in the detector endcaps.
For the muons, the uncertainty is negligible.
\item \MET scale:
Measurement of the \MET is affected by the variation in LES and JES and by the uncertainty in scale of the unclustered energy. The unclustered energy scale is varied independently of LES and JES to obtain the \MET uncertainty.
\item Jet energy resolution:
The systematic uncertainty associated with the JER in the simulation is determined by increasing or decreasing the JER by 1$\sigma$.
\item \PQb~tagging:
The uncertainty in the \PQb tagging efficiency and misidentification probability of non-\PQb jets may lead to varying background and signal levels. This uncertainty is estimated by varying the \PQb~tagging discriminator requirements in simulations.
In the lepton+jets analysis, for example, the changes in the CSVM discriminator leads to an uncertainty in the  \PQb tagging efficiency of 1.2\% and the false tagging rate of 15\%, both of which correspond to a 1$\sigma$ variation in the value of the \PQb tagging scale factor. The all-jets and dilepton uncertainties are computed in a similar manner for the CSVT and CSVL discriminators, respectively.
Propagating the tagging efficiency uncertainty to the values of \mtop leads to the systematic uncertainty.
\item Trigger:
This systematic uncertainty captures the uncertainties related to the modeling of the trigger efficiency, and is only significant for the all-jets measurement.
\item Pileup:
During the data taking period, the instantaneous luminosity increased dramatically during the year, leading to an increased number of simultaneous proton-proton interactions overlapping with the primary hard scattering (in-time pileup) and possible effects due to the detector response to previous collision events (out-of-time pileup). These effects are evaluated by using pseudo-experiments in which the average number of pileup events was varied by $\pm$5\%.
\item Backgrounds:
The background contamination expected from simulation is $<$5\% in the lepton+jets and dilepton channels.
The effect of the background modeling on \mtop is estimated by varying the shape and normalization for each background within their uncertainty.
Uncertainties from simulated backgrounds are taken to be correlated across all the measurements.
The only channel for which there is a significant non-\ttbar background in the final fit sample is the all-jets channel.
For this, the shape of the QCD multijet background is estimated from a control sample in the data.
The method is validated using simulated QCD multijet events and with an alternative approach using event mixing in the data.
The predicted background shapes are varied to cover the residual differences found in the validation.
The uncertainties from the background estimation from control samples in the data are assumed to be uncorrelated.
\item Fit calibration:
For the calibration of the fits, the simulated samples are statistically limited. The uncertainty quoted is the statistical uncertainty in the residual bias in the fit calibration.
\end{itemize}

\subsection{Theoretical and modeling uncertainties}

\subsubsection{Hard scattering and radiation}

\begin{itemize}
\item Parton distribution functions:
PDFs are used in modeling the hard scattering in proton-proton collisions in the simulations. The uncertainties in the PDFs and their effect on the measured value of \mtop are studied by reweighting a \ttbar sample with different PDF eigenvectors using the PDF4LHC prescription~\cite{PDF4LHC}. The reweighted events are used to generate pseudo-experiments and the variation in the fitted mass is quoted as the uncertainty~\cite{Chatrchyan:2011ds}.
\item Renormalization and factorization scales:
This uncertainty is estimated using the simultaneous variation of the renormalization and factorization scales by factors of 2 and 0.5 in the matrix element calculation and the initial-state parton shower of the signal and the \PW{}+jets background.
\item ME-PS matching threshold:
In the \ttbar simulation, the matching thresholds used for interfacing the matrix elements (ME) generated with \MADGRAPH to the \PYTHIA parton showers (PS) are varied from the default of 40\GeV down to 30\GeV and up to 60\GeV and the uncertainty is taken as the maximal difference in \mtop induced by this variation.
\item ME generator:
The sensitivity to the parton-level modeling is estimated by comparing the reference samples (\MADGRAPH and \PYTHIA) to samples produced using \POWHEG and \PYTHIA.
The difference between the values of \mtop obtained with the two samples is taken as the uncertainty.
\item Top quark \pt uncertainty:
This term represents the uncertainty coming from the modeling of the top quark \pt distribution in the ME generator.
The uncertainty is estimated by taking the difference in shape between the parton level \pt spectrum from the ME generator and the unfolded \pt spectrum determined from the data~\cite{Top_ptunc}. The uncertainty is considered fully correlated across the measurements.
\end{itemize}

\subsubsection{Non-perturbative QCD}

\begin{itemize}
\item Underlying event:
This represents the uncertainty in modeling the soft underlying hadronic activity in the event, which affects the simulation of both signal and background. The uncertainty is estimated by comparing \PYTHIA tunes with increased and decreased underlying event activity relative to a central tune. For this we compare the results for the Perugia 2011 tune to the results obtained using the Perugia 2011 mpiHi and the Perugia 2011 Tevatron tunes~\cite{Perugia}.
\item Color reconnection:
The effects of possible mismodeling of color reconnection are estimated by comparing the mass calculated using underlying event tunes with and without the inclusion of these effects. For these simulations the Perugia 2011 and Perugia 2011 no CR tunes are used~\cite{Perugia}. The uncertainty is taken as the difference between the two computed values of \mtop.
\end{itemize}

\subsubsection{Hadronization}

\begin{itemize}
\item Flavor-dependent hadronization uncertainty:
This is the part of the JES uncertainty that comes from differences in the energy response for different jet flavors and flavor mixtures with respect to those used in the calibration procedures.
Four uncertainties are quoted that correspond to the uncertainties for light quarks (\PQu, \PQd, \PQs), charm quarks, bottom quarks and gluons.
These are evaluated by comparing Lund string fragmentation (\PYTHIA 6 ~\cite{Sjostrand:2006za}) and cluster fragmentation (\HERWIGpp~\cite{HERWIGPP}) for each category of jets. The models in \PYTHIA and \HERWIG allow for the differences between the jet types, and the uncertainty is determined by varying the jet energies within their respective flavor-dependent uncertainties. The full flavor-dependent uncertainty is obtained by taking a signed linear sum of these four contributions. For this we perform ${\pm}1\sigma$ shifts for each of the contributions and compute the total uncertainty from the sum of the $+1\sigma$ and $-1\sigma$ shifts separately. As these are symmetric, we quote the $+1\sigma$ shifts for the values of the uncertainties in Tables \ref{tab:ljetssyst}--\ref{tab:AMWTsyst}.
\item \PQb~quark fragmentation and \PQb hadron branching fraction uncertainties:
This term provides a description of the residual uncertainties not covered by the flavor-dependent hadronization term. It has two components: the uncertainty in the modeling of the \PQb~quark fragmentation function and the uncertainty from the measured \PQb hadron semileptonic branching fractions. The \PQb~quark fragmentation function in \PYTHIA is modeled using a Bowler--Lund model for the fragmentation into \PQb hadrons. The fragmentation uncertainty is determined from the difference between a version tuned to ALEPH~\cite{ALEPH} and DELPHI~\cite{DELPHI} data and the \PYTHIA Z2$^*$ tune. Lastly, the uncertainty from the semileptonic \PQb hadron branching fraction is obtained by varying by $-0.45$\% and $+0.77$\%, which is the range of the measurements from \PBz/\PBp decays and their uncertainties~\cite{PDG}.
\end{itemize}

\section{Individual channel results\label{sec:channels}}
\subsection{The lepton+jets channel}

After estimating the systematic uncertainties for the lepton+jets channel, the measurement of \mtop and the JSF from the 2D analysis gives
\begin{equation*}\begin{split}
m_{\PQt}^{\text{2D}} & =  172.14\pm0.19\,\text{(stat+JSF)}\pm0.59\syst\GeV,\\
\mathrm{JSF}^{\text{2D}} & =  1.005\pm0.002\stat\pm0.007\syst.
\end{split}\end{equation*}
The overall uncertainty in \mtop is 0.62\GeV and the measured JSF is compatible with the one obtained from events with \Z bosons and photons~\cite{Chatrchyan:2011ds} within the systematic uncertainties.

The measurements from the 1D and hybrid analyses are
\begin{equation*}\begin{split}
m_{\PQt}^{\text{1D}} & =  172.56\pm0.12\stat\pm0.62\syst\GeV,\\
m_{\PQt}^{\text{hyb}} & =  172.35\pm0.16\,\text{(stat+JSF)}\pm0.48\syst\GeV.
\end{split}\end{equation*}
Thus the hybrid approach delivers the most precise measurement of the methods studied for the lepton+jets channel with a total uncertainty of 0.51\GeV.

The breakdown of the systematic uncertainties for the three fits is shown in Table~\ref{tab:ljetssyst}. In the lepton+jets and all-jets measurements several uncertainty sources yield opposite signs in the 1D and 2D approaches. This arises because the untagged jets used for $m_\PW^\text{reco}$ have a softer \pt spectrum and larger gluon contamination compared to the \PQb jets.
As a consequence, the measurement of the JSF in the 2D measurement is more sensitive to low-\pt
effects and radiation uncertainties than the 1D measurement where the light-jet energies are bound to fulfill
the \PW~mass constraint. The net effect, when using a flat JSF, is that the uncertainties can be overcorrected
in the 2D fit and thus their signs reverse. The hybrid fit makes optimal use of the available information and leads to partial cancelation of these
uncertainties, resulting in the observed improvement of the precision of the mass measurement.

\begin{table*}[!htb]
\centering
\topcaption{\label{tab:ljetssyst}
Category breakdown of the systematic uncertainties for the 2D, 1D, and hybrid measurements in the lepton+jets channel. Each term has been estimated using the procedures described in Section~\ref{systcat}. The uncertainties are expressed in \GeV and the signs are taken from the $+1\sigma$ shift in the value of the quantity. Thus a
positive sign indicates an increase in the value of $m_{\PQt}$ or the JSF and a negative sign indicates a decrease.
For uncertainties determined on independent simulated samples the statistical precision of the shift is displayed.
With the exception of the flavor-dependent JEC terms (see Section~\ref{systcat}), the total systematic uncertainty is obtained from the sum in quadrature of the individual systematic uncertainties.}
\resizebox{\textwidth}{!}{
\begin{scotch}{l>{$}c<{$}>{$}c<{$}>{$}c<{$}>{$}c<{$}}
 & \multicolumn{4}{c}{\mtop fit type}\\ \cline{2-5}
 Lepton+jets channel  & \multicolumn{2}{c|}{2D} & \multicolumn{1}{c|}{1D} & \multicolumn{1}{c}{hybrid}\\
 & \delta m_{\PQt}^{\text{2D}}\,(\GeVns{})& \multicolumn{1}{c|}{$\delta\mathrm{JSF}$}  & \multicolumn{1}{c|}{$\delta m_{\PQt}^{\text{1D}}\,(\GeVns{})$} & \delta m_{\PQt}^{\text{hyb}}\,(\GeVns{})\\
\hline
\hline
Experimental uncertainties &  &  &  & \\
\hline
Method calibration  & 0.04  & 0.001  & 0.04  & 0.04\\
Jet energy corrections  &  &  &  & \\
-- JEC: Intercalibration  & {<}0.01  & {<}0.001  & +0.02  & +0.01\\
-- JEC: In situ calibration  & -0.01  & +0.003 & +0.24  & +0.12\\
-- JEC: Uncorrelated non-pileup & +0.09  & -0.004 & -0.26  & -0.10\\
-- JEC: Uncorrelated pileup & +0.06  & -0.002 & -0.11  & -0.04\\
Lepton energy scale  & +0.01  & {<}0.001  & +0.01  & +0.01\\
\MET scale & +0.04  & {<}0.001  & +0.03  & +0.04\\
Jet energy resolution  & -0.11  & +0.002  & +0.05  & -0.03\\
\PQb tagging  & +0.06  & <0.001  & +0.04  & +0.06\\
Pileup  & -0.12  & +0.002  & +0.05  & -0.04\\
Backgrounds  & +0.05  & <0.001  & +0.01  & +0.03\\
\hline
\hline
Modeling of hadronization &  &  &  & \\
\hline
JEC: Flavor-dependent &  &  &  & \\
-- light quarks (\PQu\PQd\PQs) & +0.11 & -0.002 & -0.02 & +0.05\\
-- charm & +0.03 & {<}0.001 & -0.01 & +0.01\\
-- bottom & -0.32 & {<}0.001 & -0.31 & -0.32\\
-- gluon & -0.22 & +0.003 & +0.05 & -0.08\\
b jet modeling &  &  &  & \\
-- \PQb fragmentation  & +0.06  & -0.001  & -0.06 & {<}0.01 \\
-- Semileptonic \PQb hadron decays  & -0.16  & {<}0.001  & -0.15 & -0.16\\
\hline
\hline
Modeling of perturbative QCD  &  &  &  & \\
\hline
PDF  & 0.09  & 0.001  & 0.06  & 0.04\\
Ren. and fact. scales & +0.17\pm0.08 & -0.004\pm0.001 & -0.24\pm0.06 & -0.09\pm0.07\\
ME-PS matching threshold  & +0.11\pm0.09  & -0.002\pm0.001  & -0.07\pm0.06  & +0.03\pm0.07\\
ME generator  & -0.07\pm0.11  & -0.001\pm0.001  & -0.16\pm0.07  & -0.12\pm0.08\\
Top quark \pt & +0.16 & -0.003 & -0.11 & +0.02\\
\hline
\hline
Modeling of soft QCD  &  &  &  & \\
\hline
Underlying event  & +0.15\pm0.15  & -0.002\pm0.001  & +0.07\pm0.09  & +0.08\pm0.11\\
Color reconnection modeling  & +0.11\pm0.13  & -0.002\pm0.001  & -0.09\pm0.08  & +0.01\pm0.09\\
\hline
\hline
Total systematic & 0.59 & 0.007 & 0.62 & 0.48\\
\hline
Statistical & 0.20 & 0.002 & 0.12 & 0.16\\
\hline
\hline
Total  & 0.62 & 0.007 & 0.63 & 0.51\\
\end{scotch}
}
\end{table*}

\subsection{The all-jets channel}

The 2D analysis in the all-jets channel yields a measurement of
\begin{equation*}\begin{split}
\mtop^{\text{2D}} & = 171.64 \pm 0.32\,\text{(stat+JSF)}\pm 0.95\syst\GeV,\\
\mathrm{JSF}^{\text{2D}} & =  1.011 \pm 0.003\stat\pm 0.011\syst,
\end{split}\end{equation*}
giving an overall uncertainty in the mass of 1.00\GeV.

The measurements from the 1D and hybrid analyses are
\begin{equation*}\begin{split}
\mtop^{\text{1D} }     & =  172.46 \pm 0.23\stat\pm 0.62\syst\GeV,\\
\mtop^{\text{hyb} }     & =  172.32 \pm 0.25\,\text{(stat+JSF)}\pm 0.59\syst\GeV,
\end{split}\end{equation*}
with overall uncertainties of 0.66 and 0.64\GeV  for the 1D and hybrid fits, respectively.

The breakdown of the systematic uncertainties for the three fits is shown in Table~\ref{tab:alljetssyst}.
\begin{table*}[!htb]
\centering
\topcaption{\label{tab:alljetssyst} Category breakdown of the systematic uncertainties for the 2D, 1D and hybrid measurements in the all-jets channel. Each term has been estimated using the procedures described in Section~\ref{systcat}. The uncertainties are expressed in \GeV and the signs are taken from the $+1\sigma$ shift in the value of the quantity. Thus a
positive sign indicates an increase in the value of $m_{\PQt}$ or the JSF and a negative sign indicates a decrease.
For uncertainties determined on independent simulated samples the statistical precision of the shift is displayed.
With the exception of the flavor-dependent JEC terms (see Section~\ref{systcat}), the total systematic uncertainty is obtained from the sum in quadrature of the individual systematic uncertainties.}
\resizebox{\textwidth}{!}{
\begin{scotch}{l>{$}c<{$}>{$}c<{$}>{$}c<{$}>{$}c<{$}}
 & \multicolumn{4}{c}{\mtop fit type}\\ \cline{2-5}
 All-jets channel  & \multicolumn{2}{c|}{2D} & \multicolumn{1}{c|}{1D} & \multicolumn{1}{c}{hybrid}\\
 & \delta{\mtop}^{\text{2D}}\,(\GeVns{})  & \multicolumn{1}{c|}{$\delta\mathrm{JSF}$}  & \multicolumn{1}{c|}{$\delta{\mtop}^{\text{1D}}\,(\GeVns{})$}& \delta m_{\PQt}^{\text{hyb}},(\GeVns{})\\
\hline
\hline
Experimental uncertainties &  &  & & \\
\hline
Method Calibration  & 0.06  & 0.001  & 0.06 & 0.06 \\
Jet energy corrections  &  &  &  &  \\
-- JEC: Intercalibration  & {<}0.01 & {<}0.001 & +0.02 & +0.02 \\
-- JEC: In situ calibration & -0.01  &  {<}0.001 & +0.23 & +0.19 \\
-- JEC: Uncorrelated non-pileup & +0.06  & -0.001 & -0.19 & -0.16 \\
-- JEC: Uncorrelated pileup & +0.04  & {<}0.001 & -0.08 & -0.06  \\
Jet energy resolution  & -0.10  & +0.001  & +0.03 & +0.02 \\
\PQb tagging  & +0.02  & {<}0.001 & +0.01 & +0.02 \\
Pileup  & -0.09  & +0.002  & +0.02 & {<}0.01 \\
Backgrounds  & -0.61  & -0.007 & -0.14 & -0.20 \\
Trigger & +0.04 &  {<}0.001 & -0.01 & {<}0.01 \\
\hline
\hline
Modeling of hadronization &  &  & \\
\hline
JEC: Flavor-dependent &  &  &  & \\
-- light quarks (\PQu\PQd\PQs) & +0.10 & -0.001 & -0.02 & +0.00\\
-- charm              & +0.03 & -0.001 & -0.01 & -0.01\\
-- bottom             & -0.30 & +0.000& -0.29 & -0.29 \\
-- gluon              & -0.17 & +0.002 & +0.02 & -0.02\\
\PQb jet modeling  &  &  &  &  \\
-- \PQb fragmentation  & +0.08  & -0.001  & +0.03 & +0.04\\
-- Semileptonic \PQb hadron decays  & -0.14  & {<}0.001 & -0.13 & -0.13 \\
\hline
\hline
Modeling of perturbative QCD  &  &  & \\
\hline
PDF  & 0.06  & {<}0.001  & 0.03 & 0.03  \\
Ren. and fact. scales & +0.29\pm0.16 & -0.005\pm0.001 & -0.19\pm0.11 & -0.12\pm0.12 \\
ME-PS matching threshold  & +0.18\pm0.16  & -0.002\pm0.001  & +0.12\pm0.11 & +0.13\pm0.12 \\
ME generator  & -0.04\pm0.20  & -0.002\pm0.002  & -0.18\pm0.14 & -0.16\pm0.14  \\
Top quark \pt & +0.04 & +0.001 & +0.08 & +0.06 \\
\hline
\hline
Modeling of soft QCD  &  &  & \\
\hline
Underlying event  & +0.27\pm0.25  & -0.002\pm0.002  & +0.13\pm0.18 &  +0.14\pm0.18\\
Color reconnection modeling  & +0.35\pm0.22  & -0.003\pm0.002  & +0.14\pm0.16 &  +0.16\pm0.16 \\
\hline
\hline
Total systematic  & 0.95 & 0.011  & 0.62 & 0.59 \\
\hline
Statistical & 0.32 & 0.003 & 0.23 & 0.25\\
\hline
\hline
Total  & 1.00 & 0.011 & 0.66 & 0.64\\
\end{scotch}
}
\end{table*}

\subsection{The dilepton channel}
For the dilepton channel the systematic uncertainties are defined as the difference between measurements of \mtop from pseudo-data events, selected at random from the MC events in the $\mtop = 172.5\GeV$ template. For each category of systematic uncertainty, modified templates were produced with a given systematic variable shifted, generically by $\pm1\sigma.$
The fit is repeated using the modified pseudo-data and the respective mean is subtracted from the mean of the default \ttbar MC simulation to calculate the final systematic uncertainty for each category. This yields a final mass measurement of
\begin{eqnarray*}
m_{\PQt} & = & 172.82\pm0.19\mbox{ (stat)}\pm1.22\mbox{ (syst)\GeV}.
\end{eqnarray*}
The breakdown of the systematic uncertainty for the dilepton mass measurement is shown in Table~\ref{tab:AMWTsyst}. In comparison with the lepton+jets (Table~\ref{tab:ljetssyst}) and the all-jets (Table~\ref{tab:alljetssyst}) channels, the systematic uncertainties are similar in size with the exception of the factorization and renormalization, and \PQb fragmentation terms, both of which are significantly larger. Studies of these indicate that this is probably the result of an increased boost of the visible decay products, coupled to the weak constraint of the \MET on the energies of the two neutrinos.
\noindent
\begin{table}[!htb]
\centering
\topcaption{\label{tab:AMWTsyst} Category breakdown of the systematic uncertainties for the AMWT measurement in the dilepton channel. Each term has been estimated using the procedures described in Section~\ref{systcat}. The uncertainties are expressed in \GeV and the signs are taken from the $+1\sigma$ shift in the value of the quantity. Thus a
positive sign indicates an increase in the value of $m_{\PQt}$ and a negative sign indicates a decrease.
For uncertainties determined on independent simulated samples the statistical precision of the shift is displayed.
With the exception of the flavor-dependent JEC terms (see Section~\ref{systcat}), the total systematic uncertainty is obtained from the sum in quadrature of the individual systematic uncertainties.}
\begin{scotch}{l>{$}c<{$}}
Dilepton channel
 & \delta m_{\PQt}\,(\GeVns{})\\
\hline
\hline
Experimental uncertainties & \\
\hline
Method calibration & 0.03\\
Jet energy corrections  & \\
-- JEC: Intercalibration  & +0.03\\
-- JEC: In situ calibration & +0.24\\
-- JEC: Uncorrelated non-pileup & -0.28\\
-- JEC: Uncorrelated pileup & -0.12\\
Lepton energy scale  & +0.12\\
\MET scale & +0.06\\
Jet energy resolution  & +0.06\\
\PQb tagging  & +0.04\\
Pileup  & +0.04\\
Backgrounds  & +0.02\\
\hline
\hline
Modeling of hadronization & \\
\hline
JEC: Flavor-dependent & \\
-- light quarks (\PQu\PQd\PQs) & +0.02\\
-- charm & +0.02\\
-- bottom & -0.34\\
-- gluon & +0.06\\
\PQb jet modeling & \\
-- \PQb fragmentation  & -0.69\\
-- Semileptonic \PQb hadron decays  & -0.17\\
\hline
\hline
Modeling of perturbative QCD  & \\
\hline
PDF  & 0.16\\
Ren. and fact. scales & -0.75\pm0.20\\
ME-PS matching threshold  & -0.12\pm0.20\\
ME generator  & -0.24\pm0.20\\
Top quark \pt & -0.25\\
\hline
\hline
Modeling of soft QCD  & \\
\hline
Underlying event  & +0.04\pm0.20\\
Color reconnection modeling  & -0.11\pm0.20\\
\hline
\hline
Total systematic & 1.22\\
\hline
Statistical & 0.19\\
\hline
\hline
Total  & 1.24\\
\end{scotch}
\end{table}

\subsection{The 2010 and 2011 measurements}
\begin{table*}[!htb]
\centering
\topcaption{\label{tab:2010_11} CMS measurements of the top quark mass using the data recorded at $\sqrt{s} = 7$\TeV.}
\begin{scotch}{clccccc}
\multicolumn{2}{c}{Analysis} &\multicolumn{1}{c}{Reference} &\multicolumn{1}{c}{\mtop} &\multicolumn{1}{c}{Stat. uncertainty} &\multicolumn{1}{c}{Syst. uncertainty}\\ \cline{1-2}
\multicolumn{2}{c}{} &\multicolumn{1}{c}{} &\multicolumn{1}{c}{$(\GeVns{})$} &\multicolumn{1}{c}{$(\GeVns{})$} &\multicolumn{1}{c}{$(\GeVns{})$}\\
\hline
2010 &dilepton (AMWT)& \cite{DiLep2010} & 175.50 & 4.60 & 4.52\\
\hline
2011 &lepton+jets (2D)& \cite{LepJets2011} & 173.49 & 0.27 & 1.03\\
2011 &all-jets  (1D)& \cite{AllJets2011} & 173.49 & 0.69 & 1.23\\
2011 &dilepton (AMWT)& \cite{DiLep2011} & 172.50 & 0.43 & 1.46\\
\end{scotch}
\end{table*}

The published CMS measurements are based on $\sqrt{s} = 7$\TeV data recorded during  2010 and 2011. Although much less precise than the new measurements, they come from independent data sets and have different sensitivities to the various systematic uncertainties. These are included in the combined mass analysis, which is discussed in Section~\ref{Combine}. For completeness we summarize these measurements in Table~\ref{tab:2010_11} below. The analysis techniques used for each of these are very similar to those used for the 2012 analyses. The dilepton results both use the AMWT method, which is described in Section~\ref{amwt}, and the lepton+jets (all-jets) result comes from the 2D (1D) ideogram technique, which is described in Section~\ref{ideogram}.

\section{Measured top quark mass as a function of kinematic observables\label{sec:kine}}
\label{sec:diff}
To search for possible biases in our measurements and the potential limitations of current event generators, a series of differential measurements of \mtop as a function of the kinematic properties of the \ttbar system is performed. To maximize the accuracy of the results, the study is performed in the lepton+jets channel using the hybrid fit technique.  The variables are chosen to probe potential effects from color reconnection, initial- and final-state radiation, and the kinematics of the jets coming from the top quark decays.

For each measurement, the hybrid analysis method is applied to subsets of events defined according to the value of a given kinematic event observable after the kinematic fit.
The contribution of the external JSF constraint is fixed to 50\% to ensure consistency of all bins with the inclusive result.
Constant shifts in the measured \mtop values may arise due to the systematic uncertainties of the inclusive measurement or from the use of different \mtop values in data and simulations. To search for kinematics-dependent biases the value of the mean measured top quark mass is subtracted and the results are expressed in the form $\mtop-\left<\mtop\right>$, where the mean comes from the inclusive measurement on the specific sample.
In each case, the event sample is divided into 3 to 5 bins as a function of the value of the kinematic observable and we populate each bin using
all permutations which lie within the bin boundaries. As some observables depend on the jet-quark assignment that cannot be resolved unambiguously, such as the \pt of a reconstructed top quark, a single event is allowed to contribute to multiple bins.

To aid in the interpretation of a difference between the value of $\mtop-\left<\mtop\right>$ and the prediction from a simulation in the same bin, a bin-by-bin calibration of the results is performed using the \MADGRAPH + \PYTHIA simulation. This is performed using the same technique as for the inclusive measurement~\cite{LepJets2011} except that it is performed on each bin separately. Thus, after calibration the value in each bin can be interpreted in terms of its agreement with respect to the inclusive measurement.

For eight kinematic variables the results for the calibrated mass difference, $\mtop-\left<\mtop\right>$, are shown as a function of the chosen variable, and we compare the results to the predictions of seven different simulations.
For each plotted point the statistical uncertainty and the dominant systematic uncertainties are combined in quadrature, where the latter include the JES (\pt-, $\eta$- and flavor-dependent), JER, pileup, \PQb fragmentation, renormalization and factorization scales, and ME-PS matching threshold.
The systematic uncertainties are assumed to be correlated among all bins, so that any constant shift is removed by subtracting $\left<\mtop\right>$. We note that this approximation may underestimate the uncertainties from the \pt/$\eta$-dependent JES.

For each plot we compare the data to simulations based on LO (\MADGRAPH and \SHERPA) and NLO (\POWHEG and \MCATNLO) matrix element calculations with both string (\PYTHIA) and cluster (\HERWIG) models for fragmentation. We also vary the choice of underlying event tune from Z2* to Perugia 2011 both with and without color reconnections, and the AUET2 tune. With the exception of the \MCATNLO and \SHERPA simulations, which are only used for this study, these are the same simulation as those discussed in Section~\ref{sec:simul}. The simulations used for this study are
\begin{itemize}
\item  \MADGRAPH with the \PYTHIA Z2* tune, which is the simulation used in the mass determinations~\cite{Alwall:2011uj,Artoisenet:2012st,Sjostrand:2006za,Chatrchyan:2011id};
\item  \MADGRAPH with the \PYTHIA Perugia 2011 tune~\cite{Alwall:2011uj,Sjostrand:2006za,Perugia};
\item  \MADGRAPH with the \PYTHIA Perugia 2011 noCR tune~\cite{Alwall:2011uj,Sjostrand:2006za,Perugia};
\item  \SHERPA 1.4.0 with up to 4 additional jets from the LO matrix element~\cite{Gleisberg:2008ta,Hoeche:2009rj};
\item  \POWHEG with the \PYTHIA Z2* tune~\cite{Alioli:2009je,Re:2010bp,Nason:2004rx,Frixione:2007vw,Alioli:2010xd,Sjostrand:2006za,Chatrchyan:2011id};
\item  \POWHEG with the \HERWIG 6.520 AUET2 tune~\cite{Alioli:2009je,Re:2010bp,Nason:2004rx,Frixione:2007vw,Alioli:2010xd,MCCMS3};
\item  \MCATNLO 3.41 with the \HERWIG 6.520 default tune~\cite{Frixione:2002ik,Frixione:2003ei,MCCMS3}.
\end {itemize}

\begin{figure*}[!htb]
\centering
  \includegraphics[width=0.48\textwidth]{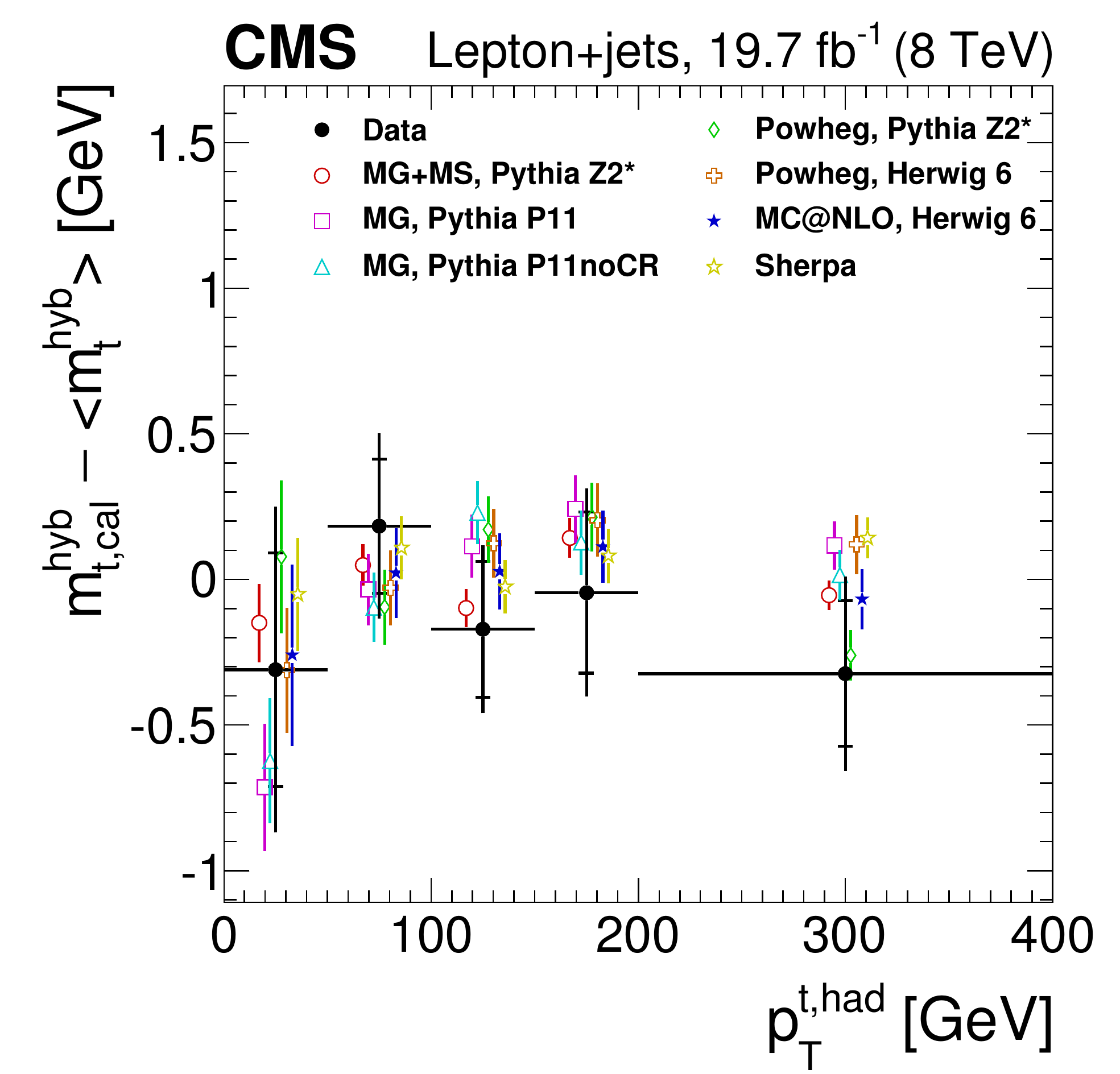}
  \includegraphics[width=0.48\textwidth]{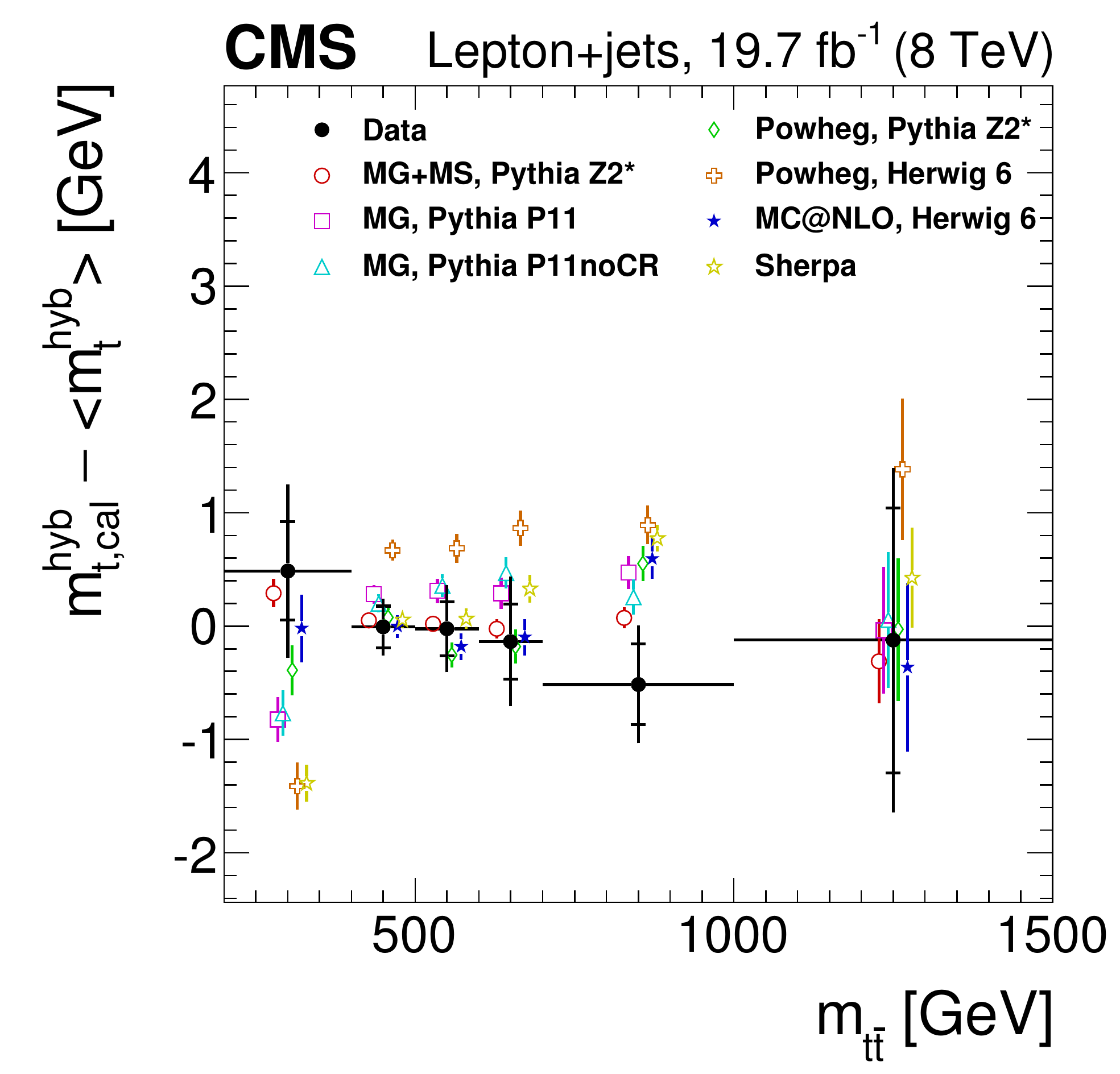}
  \includegraphics[width=0.48\textwidth]{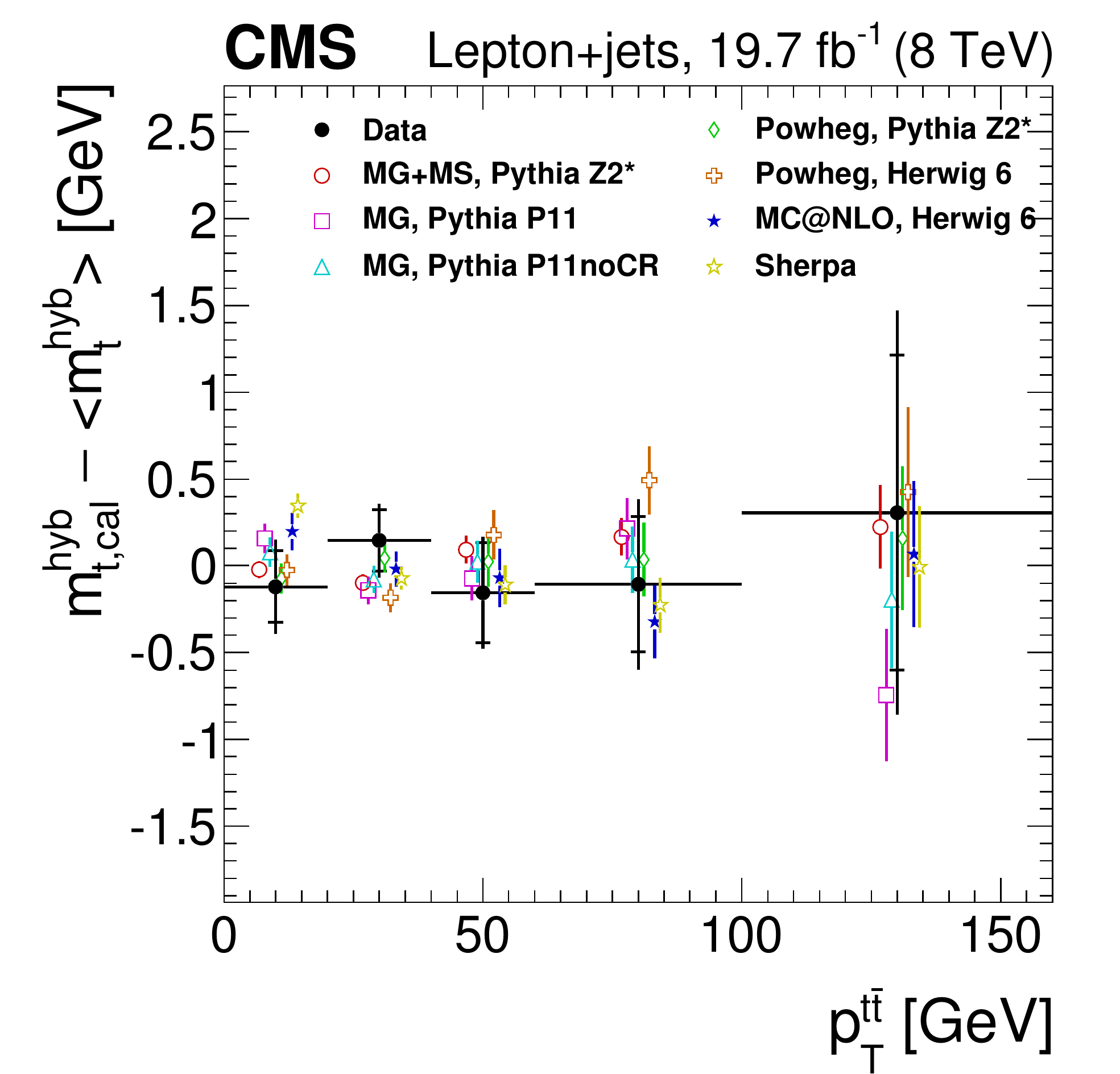}
  \includegraphics[width=0.48\textwidth]{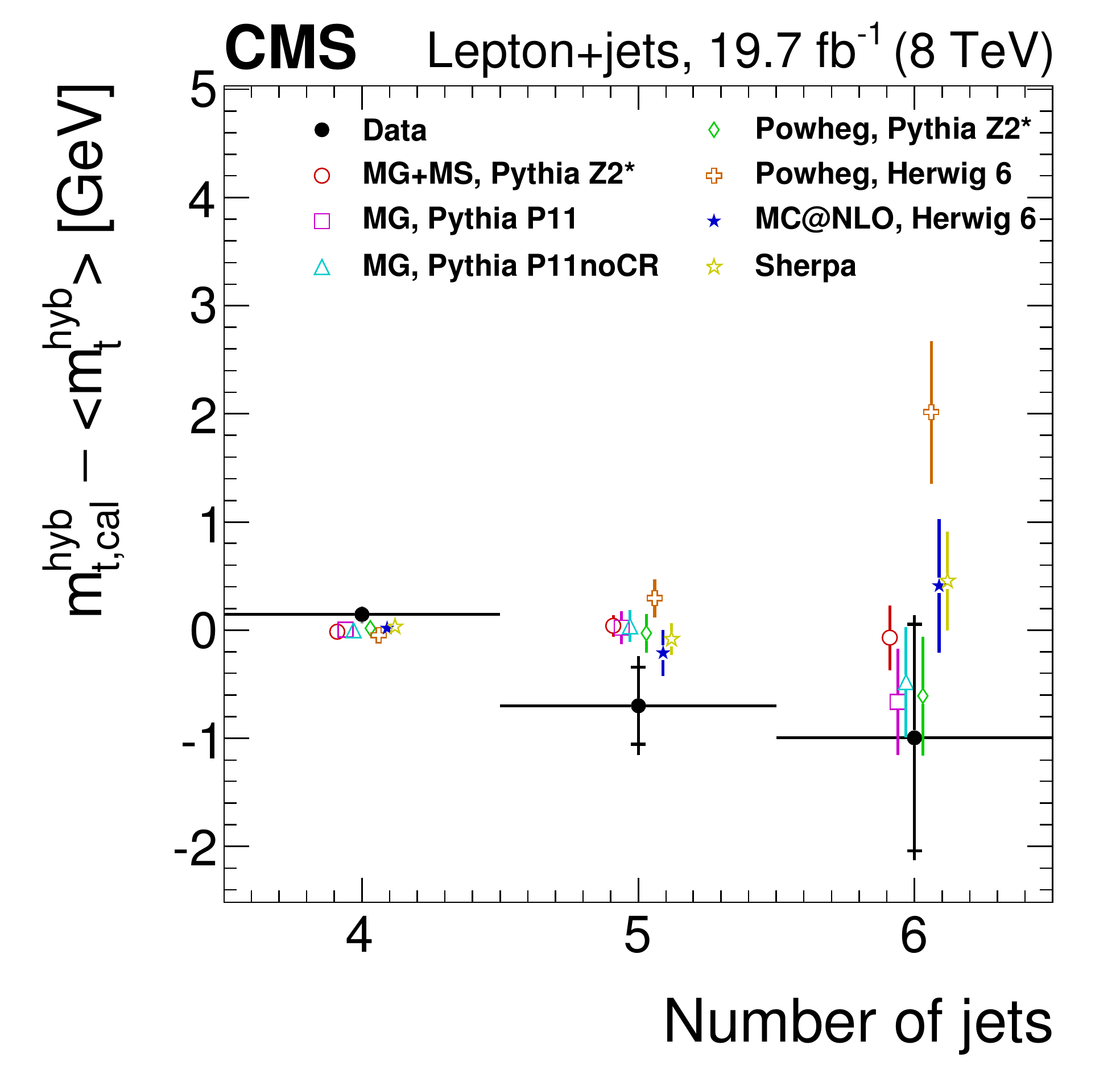}
  \caption{
Measurements of \mtop as a function of the transverse momentum of the hadronically decaying top quark ($\pt^{\rm{t,had}}$), the invariant mass of the \ttbar system ($m_{\ttbar}$), the transverse momentum of the \ttbar system ($\pt^{\ttbar}$), and the number of jets with $\pt > 30\GeV$.
The filled circles represent the data, and the other symbols are for the simulations. For reasons of clarity the horizontal error bars are shown only for the data points and each of the simulations is shown as a single offset point with a vertical error bar representing its statistical uncertainty.  \diffcaption ~The open circles correspond to \MADGRAPH with the \PYTHIA Z2* tune, the open squares to \MADGRAPH with the \PYTHIA Perugia 2011 tune, and the open triangles represent \MADGRAPH with the \PYTHIA Perugia 2011 noCR tune. The open diamonds correspond to \POWHEG with the \PYTHIA Z2* tune and the open crosses correspond to \POWHEG with \HERWIG6. The filled stars are for \MCATNLO with \HERWIG6 and the open stars are for \SHERPA.}
  \label{fig:diff_top}
\end{figure*}

\begin{figure*}[!htb]
\centering
  \includegraphics[width=0.48\textwidth]{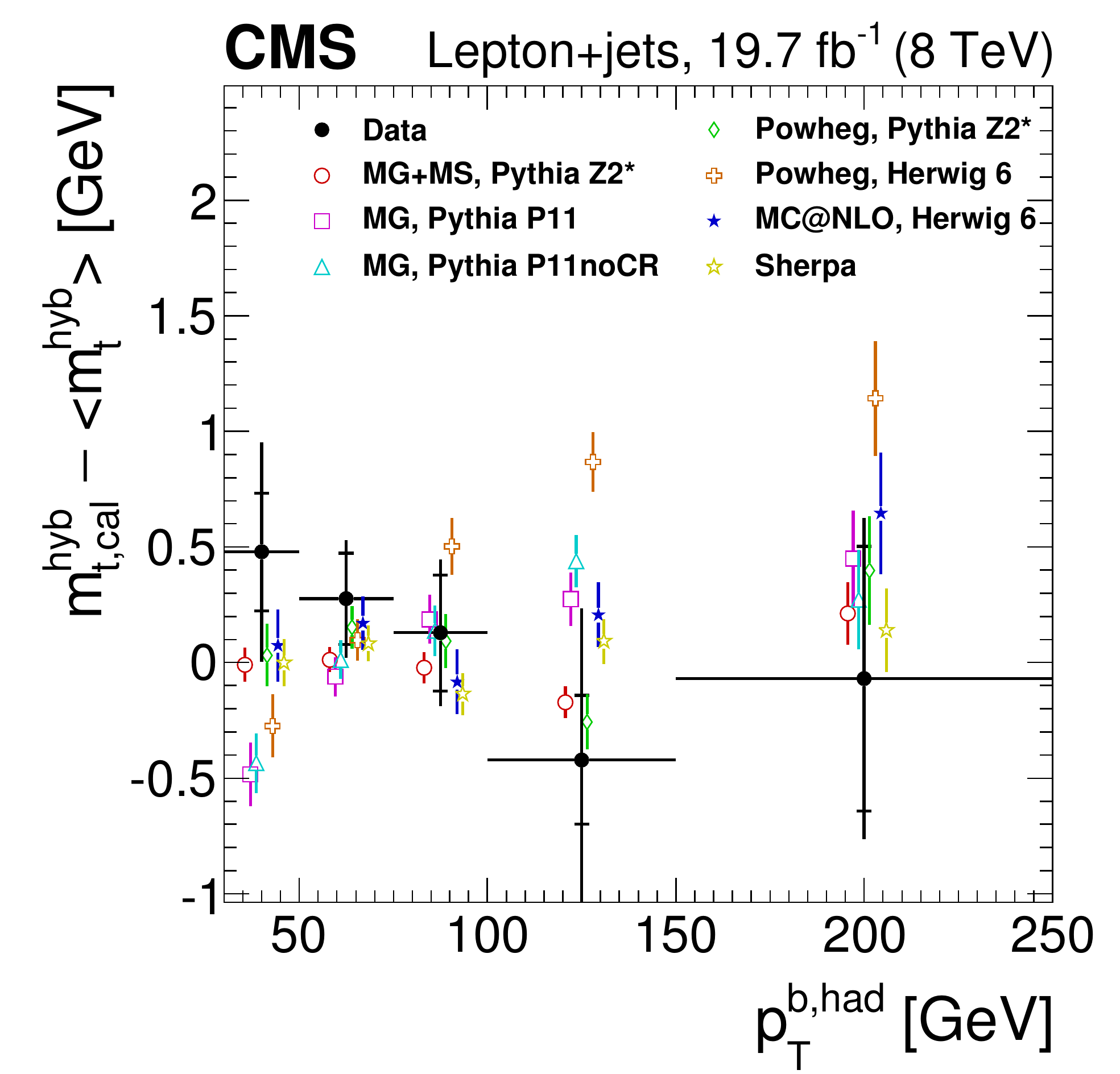}
  \includegraphics[width=0.48\textwidth]{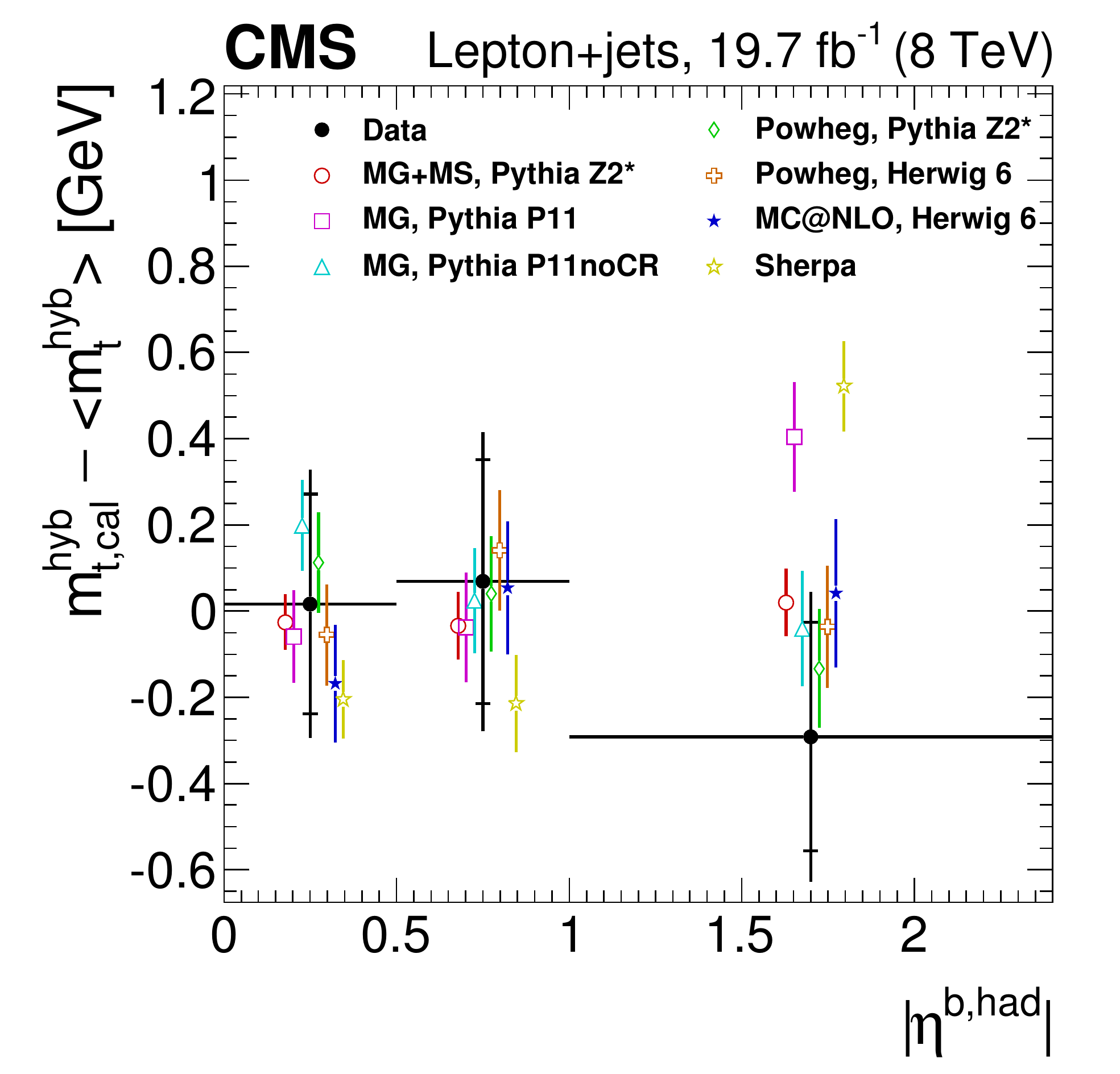}
  \includegraphics[width=0.48\textwidth]{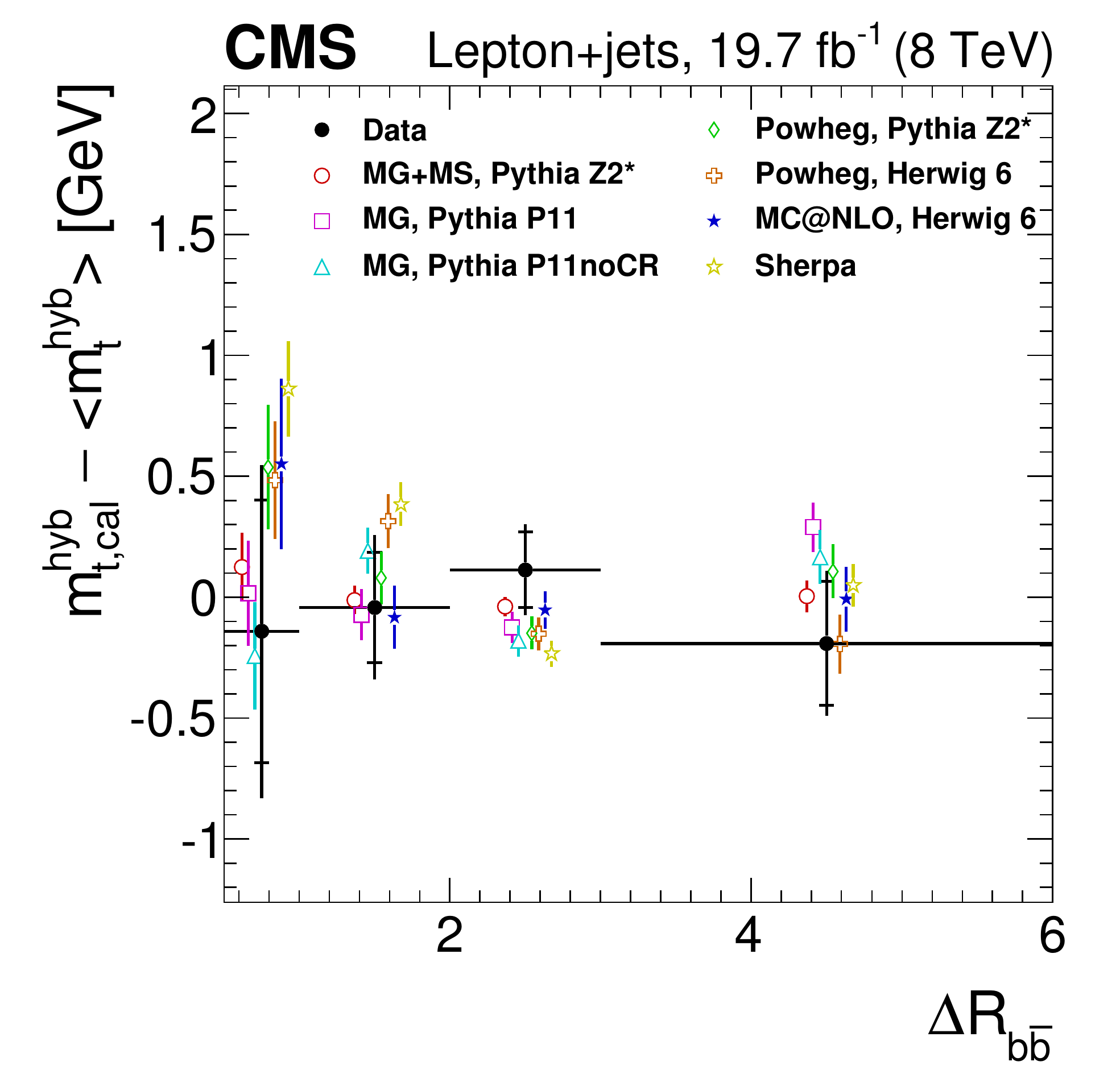}
  \includegraphics[width=0.48\textwidth]{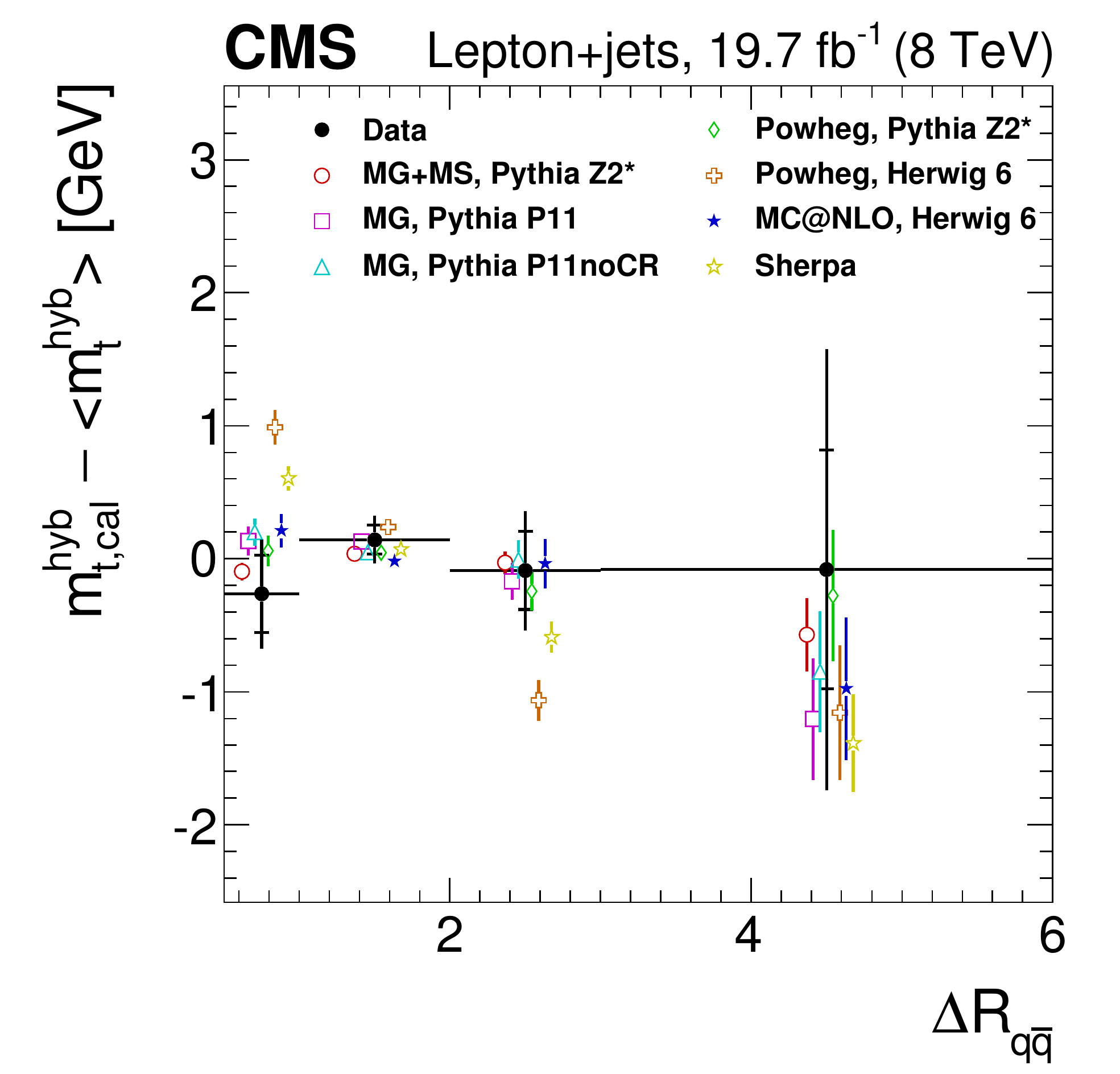}
  \caption{Measurements of \mtop as a function of the \pt of the \PQb jet assigned to the hadronic decay branch ($\pt^{\rm{b,had}}$), the pseudorapidity of the \PQb jet assigned to the hadronic decay branch ($\left|\eta^{\rm b,had}\right|$), the $\Delta R$ between the \PQb jets ($\Delta R_{\bbbar}$), and the $\Delta R$ between the light-quark jets ($\Delta R_{\qqbar}$).
The filled circles represent the data, and the other symbols are for the simulations. For reasons of clarity the horizontal error bars are shown only for the data points and each of the simulations is shown as a single offset point with a vertical error bar representing its statistical uncertainty.  \diffcaption ~The open circles correspond to \MADGRAPH with the \PYTHIA Z2* tune, the open squares to \MADGRAPH with the \PYTHIA Perugia 2011 tune, and the open triangles represent \MADGRAPH with the \PYTHIA Perugia 2011 noCR tune. The open diamonds correspond to \POWHEG with the \PYTHIA Z2* tune and the open crosses correspond to \POWHEG with \HERWIG6. The filled stars are for \MCATNLO with \HERWIG6 and the open stars are for \SHERPA.}
  \label{fig:diff_decay}
\end{figure*}

The variables were chosen for their potential sensitivity to modeling the kinematics of top quark production (Fig.~\ref{fig:diff_top}) and decay (Fig.~\ref{fig:diff_decay}).
No significant deviation in the value of the measured \mtop is observed, indicating that within the current precision, there is no evidence for a bias in the measurements. The agreement between the data and each of the simulations is quantified in Table~\ref{tab:chi2-all}. Here we show the cumulative $\chi^{2}$ for the 27 degrees of freedom represented by the eight distributions studied (Figs.~\ref{fig:diff_top} and \ref{fig:diff_decay}) and the corresponding number of standard deviations between the data and the simulation, where we have assumed two-sided Gaussian confidence intervals for each simulation. In all cases, with the possible exception of \POWHEG + \HERWIG6 simulation, the data is well described by the models.

\begin{table}[!tfb]
\topcaption{Comparison of different simulations and the data. The summed $\chi^{2}$ values and number of standard deviations are computed for the 27 points entering Figs.~\ref{fig:diff_top}~and~\ref{fig:diff_decay} assuming two-sided Gaussian statistics.
}
\label{tab:chi2-all}
\centering
\begin{scotch}{lcc}
Simulation & $\chi^{2}$ & Standard deviations\\
\hline
MG + \PYTHIA~6 Z2{*}        & 17.55 & 0.10 \\
MG + \PYTHIA~6 P11          & 37.68 & 1.73 \\
MG + \PYTHIA~6 P11noCR      & 31.57 & 1.15 \\
\POWHEG + \PYTHIA~6 Z2{*}   & 19.70 & 0.20 \\
\POWHEG + \HERWIG~6         & 76.48 & 4.84 \\
\MCATNLO + \HERWIG~6        & 20.47 & 0.24 \\
\SHERPA                     & 46.79 & 2.56 \\
\end{scotch}
\end{table}

\section{Combining the mass measurements\label{sec:combo}}
\label{Combine}
In this section, results for the combined top quark mass measurement are presented. As inputs we use the new results presented in this paper and the published CMS measurements from the 2010 \cite{DiLep2010} and 2011 \cite{LepJets2011,DiLep2011,AllJets2011} analyses.
To combine the results, the best linear unbiased estimate method (BLUE) \cite{BLUE} is used.
This determines a linear combination of the input measurements which takes into account statistical and systematic uncertainties by minimizing the total uncertainty of the combined result. The procedure takes account of the correlations that exist between the different uncertainty sources through the use of correlation coefficients. These are chosen to reflect the current knowledge of the uncertainties for both the correlations between measurements in a given decay channel from different years ($\rho_{\text{chan}}$) and between the measurements in different decay channels from the same year ($\rho_{\text{year}}$). The nominal values are set to either zero for uncorrelated or unity for fully correlated (see Table~\ref{tab:combination}).
Because the measurements from the 2012 analyses are significantly more precise, both statistically and systematically, than those from the 2010 and 2011 analyses, the use of unity coefficients for $\rho_{\text{chan}}$ and $\rho_{\text{year}}$ is problematic. To mitigate this,
we have chosen to perform combinations in which the correlation coefficients are limited to value of less than unity. This has been done by setting the correlation coefficients for each pair of measurements in the fully correlated cases
to $\rho = \sigma_i/\sigma_j$, where $\sigma_{i}$ and $\sigma_{j}$ are the uncorrelated components of the uncertainties in measurements $i$ and $j$, respectively, and $\sigma_{i}<\sigma_{j}$.
For all of the measurements, the statistical uncertainties are assumed to be uncorrelated.
\begin{table}[!htb]
\centering
\topcaption{\label{tab:combination} Nominal correlation coefficients for the systematic uncertainties,
 The term $\rho_{\rm chan}$ is the correlation factor for measurements in the same top quark decay channel, but different years and the term $\rho_{\rm year}$ is the correlation between measurements in different channels from the same year.
}
\begin{scotch}{lcc}
&\multicolumn{2}{c}{Correlations}\\ \cline{2-3}
&\multicolumn{1}{c}{$\rho_\text{chan}$} &\multicolumn{1}{c}{$\rho_\text{year}$}\\
\hline
Experimental uncertainties & &\\
\hline
Method calibration                            & 0 & 0 \\
JEC: Intercalibration                        & 1 & 1 \\
JEC: In situ calibration                     & 1 & 1 \\
JEC: Uncorrelated non-pileup       & 0 & 1 \\
Lepton energy scale                       & 1 & 1 \\
\MET scale                                         & 1 & 1 \\
Jet energy resolution                       & 1 & 1 \\
\PQb tagging                                         & 1 & 1 \\
Pileup                                                 & 0 & 1 \\
Non-\ttbar background (data)         & 0 & 0 \\
Non-\ttbar background (simulation)   & 1 & 1 \\
Trigger                                               & 0 & 0 \\
\hline
Modeling of hadronization& & \\
\hline
JEC: Flavor-dependent                 & 1 & 1 \\
\PQb jet modeling                                 & 1 & 1 \\
\hline
Modeling of perturbative QCD && \\
\hline
PDF                                                     & 1 & 1 \\
Ren. and fact. scales                       & 1 & 1 \\
ME-PS matching threshold             & 1 & 1 \\
ME generator                                     & 1 & 1 \\
Top quark \pt  & 1 & 1 \\
\hline
Modeling of soft QCD&& \\
\hline
Underlying event                         & 1 & 1 \\
Color reconnection modeling    & 1 & 1 \\
\end{scotch}
\end{table}

\subsection{Measurement permutations}
The precision of any combination of the measurements will be dominated by the set of new measurements, derived from the 2012 data. To investigate the effect of the choice of fit method on the result, we perform a series of combinations in which the 2012 inputs from the lepton+jets and all-jets decay channels are varied. For simplicity of discussion, these are classified according to the type of fit used for each channel. They are labeled as follows: 2 for a 2D fit, 1 for a 1D or AMWT fit, and h for a hybrid fit. Thus a lepton+jets:all-jets:dilepton fit is denoted 211 in the case of a 2D fit for the lepton+jets channel a 1D fit for the all-jets channel and the AMWT fit in the dilepton channel.

The most precise set of non-hybrid measurements corresponds to the set 211, which gives a result of
\ifthenelse{\boolean{cms@external}}{
\begin{multline*}
\mtop = 172.40 \pm 0.13\,\text{(stat+JSF)}\pm 0.54\syst\GeV\\ \text{(211 combination)}.
\end{multline*}
}{
\begin{equation*}
\mtop = 172.40 \pm 0.13\,\text{(stat+JSF)}\pm 0.54\syst\GeV\quad\text{(211 combination)}.
\end{equation*}
}

To verify that this gives the most precise combination, combinations are performed using the other permutations of the 2012 measurements. The results, listed in Table~\ref{tab:permute}, are in good agreement with the 211 result but have less precision, as expected.

\begin{table}[!h]
\centering
\topcaption{\label{tab:permute} Combination results for the permutations of the 2D, 1D,  and hybrid measurements. The permutation order is defined to be lepton+jets:all-jets:dilepton, thus 211 corresponds to the 2D lepton+jets:1D all-jets:AMWT dilepton combination.}
\begin{scotch}{cccc}
\multirow{2}{*}{Combination} &\mtop &Stat+JSF uncertainty &Syst uncertainty\\
 &(\GeVns{}) &(\GeVns{}) &(\GeVns{})\\
\hline
211 & 172.40 & 0.13 & 0.54\\
\hline
121 & 172.61 & 0.11 & 0.57\\
221 & 172.30 & 0.15 & 0.58\\
111 & 172.66 & 0.12 & 0.56\\
\hline
h11 & 172.45 & 0.13 & 0.47\\
hh1 & 172.44 & 0.13 & 0.47\\
\hline
2h1 & 172.35 & 0.14 & 0.53\\
\end{scotch}
\end{table}

For the hybrid results, the effect of constraining the JSF factor in the mass fits can be examined. There are three significant new permutations to consider, the h11, hh1, and 2h1 combinations. The results, shown in Table~\ref{tab:permute}, are in good agreement with the 211 result, with the h11 and hh1 combinations giving the most precise measurements, as expected. For these the results are
\ifthenelse{\boolean{cms@external}}{
\begin{equation*}\begin{split}
\mtop &= 172.45 \pm 0.13\,\text{(stat+JSF)}\pm 0.47\syst\GeV\\
&\quad\text{(h11 combination)},\\
\mtop &= 172.44 \pm 0.13\,\text{(stat+JSF)}\pm 0.47\syst\GeV\\
&\quad\text{(hh1 combination)},
\end{split}\end{equation*}
}{
\begin{equation*}\begin{split}
\mtop &= 172.45 \pm 0.13\,\text{(stat+JSF)}\pm 0.47\syst\GeV\quad\text{(h11 combination)},\\
\mtop &= 172.44 \pm 0.13\,\text{(stat+JSF)}\pm 0.47\syst\GeV\quad\text{(hh1 combination)},
\end{split}\end{equation*}
}
both with an overall improvement in precision of 0.07\GeV with respect to the 211 analysis, and a total uncertainty of 0.48\GeV.

\subsection{Anticorrelation effects}
\begin{figure}[!htb]
\centering
\includegraphics[width=0.45\textwidth]{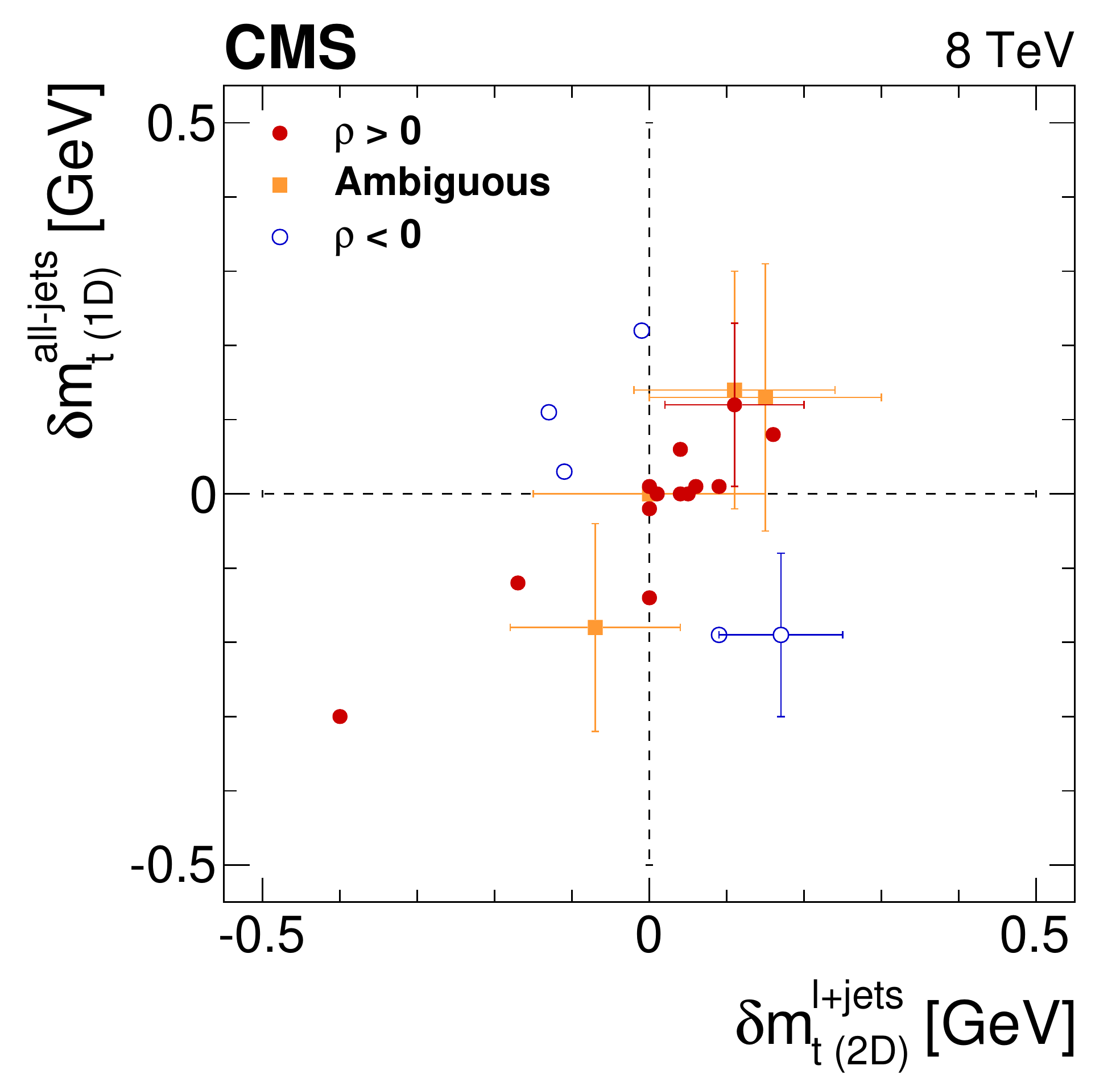}
\includegraphics[width=0.45\textwidth]{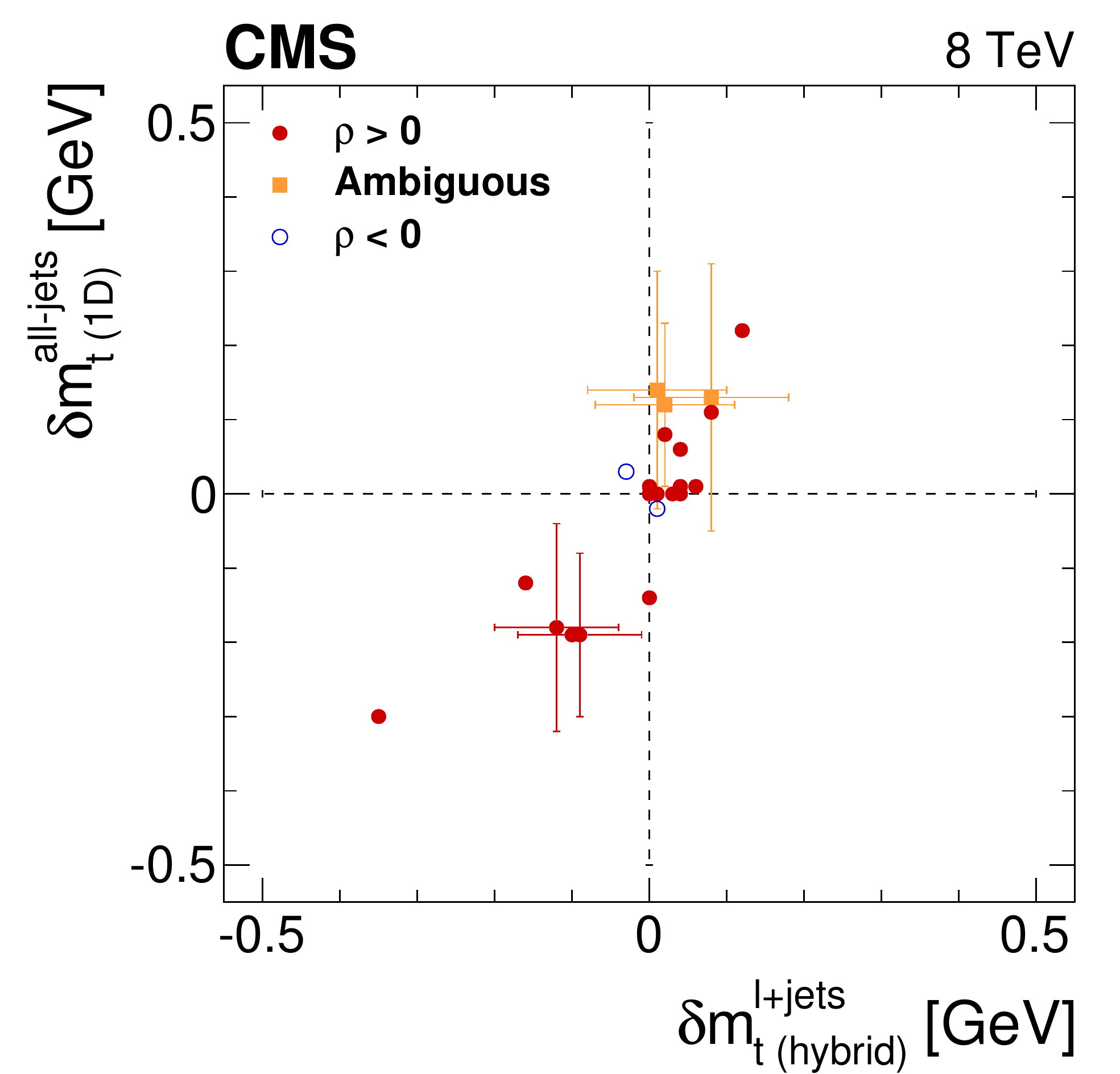}
\caption{Systematic uncertainty correlations for mass measurements in the lepton+jets and all-jets channels. Each point represents a single systematic uncertainty taken from Tables~\ref{tab:ljetssyst} and~\ref{tab:alljetssyst}. Panel (\cmsLeft) for the 2D lepton+jets and 1D all-jets measurements, and
 Panel (\cmsRight) for the hybrid lepton+jets and the 1D all-jets measurements. The filled circles correspond to the systematic uncertainties which show a positive correlation between the two fit methods and the open circles to the systematic terms which show a negative correlation. The points shown as filled squares are those for which the systematic estimation is dominated by a statistical uncertainty, so no clear categorization is possible. The vertical and horizontal error bars correspond to the statistical uncertainties in the systematic uncertainties.}
\label{fig:syst_correl}
\end{figure}

For the results presented here, the signs of most of the uncertainty contributions are well defined (i.e. for a 1$\sigma$ shift in a given quantity, the statistical component of the estimated systematic uncertainty is significantly smaller than the value of the uncertainty).
This allows a comparison of the signs of the systematic uncertainties for the different channels and for the different fitting techniques.
An anticorrelation (i.e. opposite signs) is observed between several of the terms when comparing the results from a 2D and a 1D (or AMWT) fit. However, if the 2D fit is replaced by the corresponding hybrid result, the anticorrelations are removed. This is illustrated in Fig.~\ref{fig:syst_correl}, which shows the uncertainty correlations between the lepton+jets and all-jets channels for the 2D vs. 1D and the hybrid vs. 1D cases. In the 2D vs. 1D plot (Fig.~\ref{fig:syst_correl} \cmsLeft) we observe a significant number of anticorrelated terms (coming primarily from the JES and pileup terms), whereas in the hybrid vs. 1D plot (Fig.~\ref{fig:syst_correl} \cmsRight) we see no significant anti-correlations. Given the uncertainty terms that vary between the 2D and hybrid treatments, it is believed that the observed effect arises from the variation in the JSF factors between the 2D, 1D, and hybrid results (Sections~\ref{sec:ljets} and~\ref{alljets}).

These effects are not considered in the standard 211 combination as the input correlation coefficients are positive for all of the correlated cases (see Table~\ref{tab:combination}). To estimate the effect of including anticorrelations, the correlation coefficients are set to negative values for the cases where an anticorrelation (opposite sign) is observed and positive values where a positive (same sign) or neutral (statistically limited) correlation is observed and the 211 combination analysis is repeated. This gives a result of $172.40 \pm 0.13\,\text{(stat+JSF)}\pm 0.47\syst\GeV$. Thus, while the result for the mass is unchanged, the systematic uncertainty is decreased and becomes comparable to that achieved in the hybrid combinations.

\section{Results\label{sec:results}}
\begin{figure}[!thb]
   \centering
   \includegraphics[width=\cmsFigWidth]{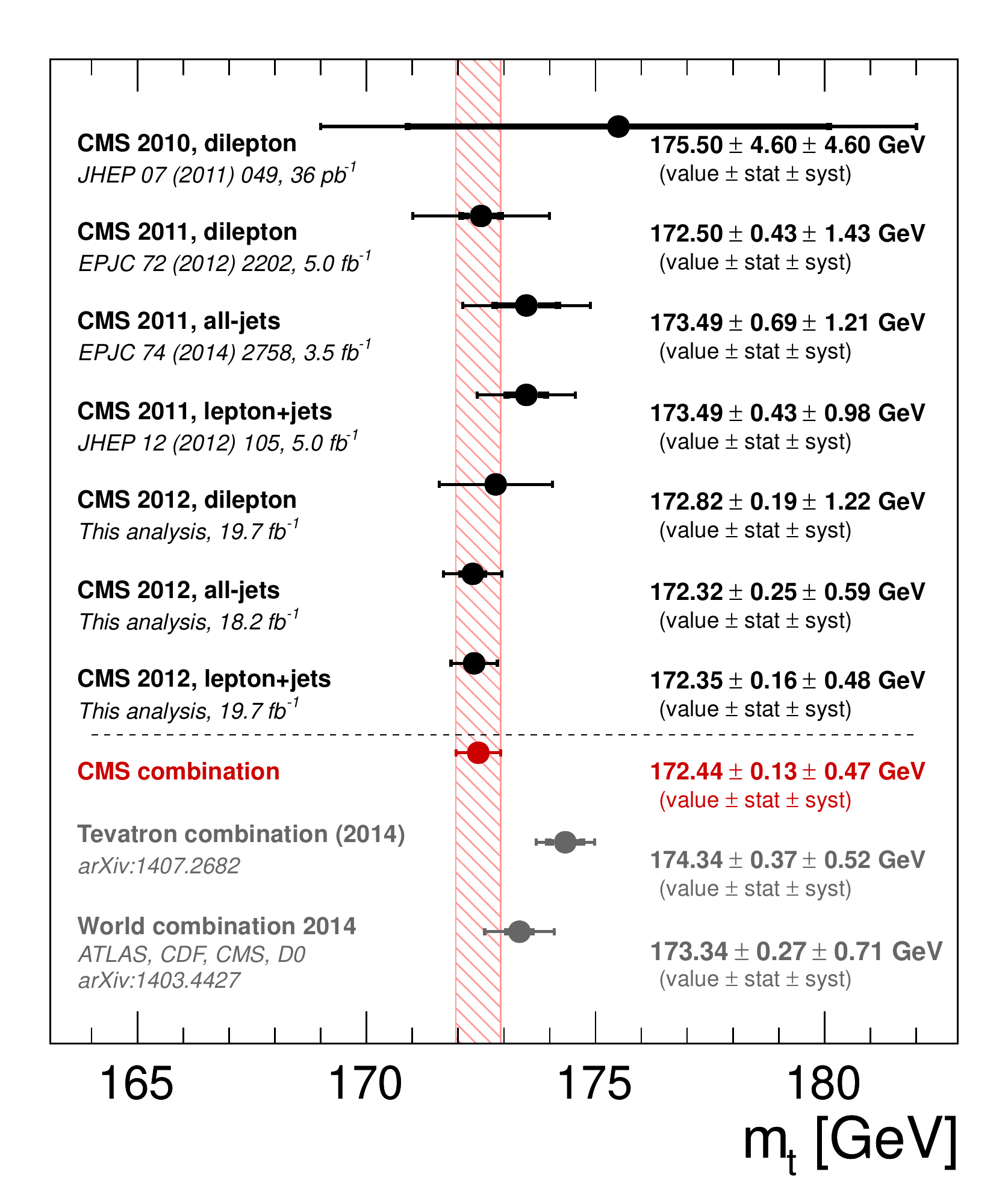}
   \caption{\label{fig:summary}
   Summary of the CMS \mtop measurements and their combination. The thick error bars show the statistical uncertainty and the thin error bars show the total uncertainty. Also shown are the current Tevatron~\cite{Tevatron2014} and world average~\cite{World2013} combinations.}
\end{figure}

\begin{figure}[!htb]
\centering
 \includegraphics[width=0.48\textwidth]{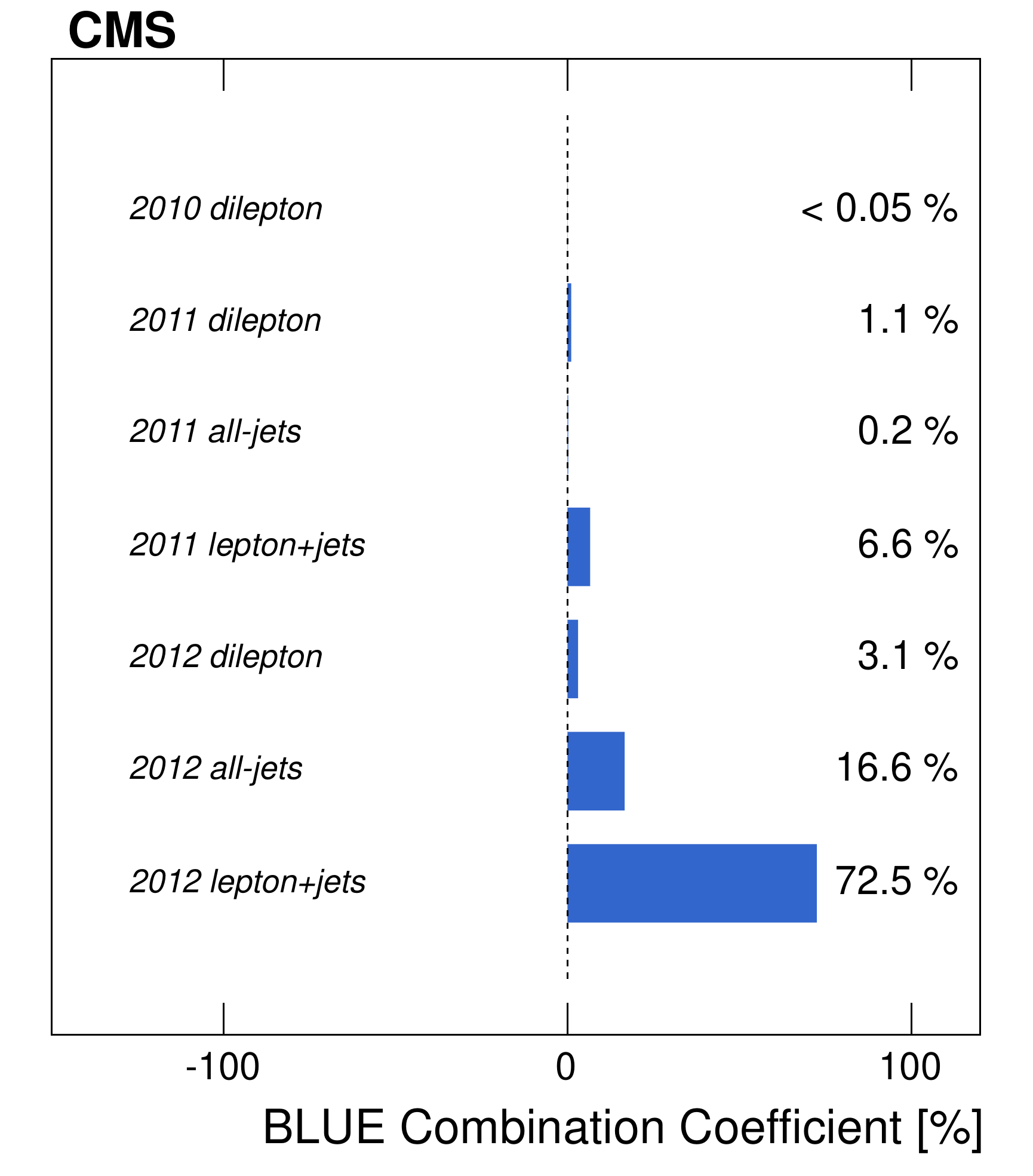}
 \includegraphics[width=0.48\textwidth]{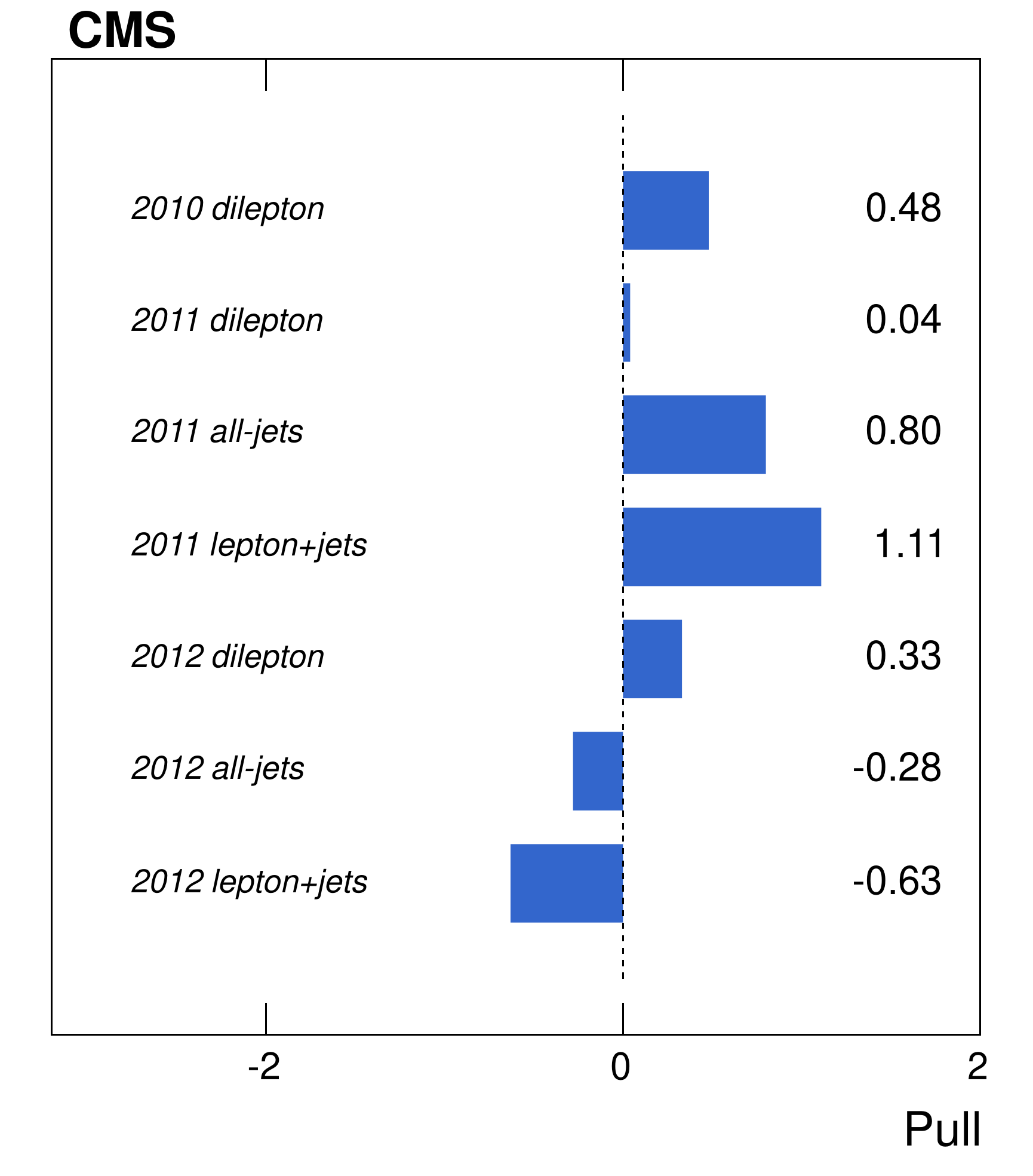}
   \caption{\label{fig:coepul}
   Results of the BLUE combining procedure on the CMS measurements showing (\cmsLeft) the combination coefficients, and (\cmsRight) the pulls for each contribution.}
\end{figure}

Based on the expected uncertainties for each of the individual measurements (Tables \ref{tab:ljetssyst}--\ref{tab:AMWTsyst}) and the consistency of the hybrid and 1D results for the JSF (Sections~\ref{sec:ljets},~\ref{alljets}), the hh1 combination is chosen as the preferred result. Combining the seven input measurements
(four from $\sqrt{s} = 7\TeV$ and three from this analysis)
gives a combined top quark mass measurement of
\begin{equation*}
\mtop = 172.44 \pm 0.13\,\text{(stat+JSF)}\pm 0.47\syst\GeV,
\end{equation*}
for which the combination $\chi^{2}$ is 2.5 for six degrees of freedom, corresponding to a probability of 87\%. This is compared to the full set of  Run 1 measurements in Fig.~\ref{fig:summary} where the current world average {\cite{World2013} and Tevatron \cite{Tevatron2014} combinations are also shown. The result is consistent with all of the published LHC measurements and is the most precise measurement to date with a precision of 0.3\%.

\begin{table*}[htb]
\centering
\topcaption{\label{tab:inputcorr}Correlations between input measurements. The elements in the table are labelled according to the analysis they correspond to (rows and columns read  as 2010, 2011, 2012 followed by the \ttbar decay channel name).}
\resizebox{\textwidth}{!}{
\begin{scotch}{ccccccccc}
&&\multicolumn{1}{c|}{2010} &\multicolumn{3}{c|}{2011} &\multicolumn{3}{c}{2012}\\ \cline{3-9}
&& \multicolumn{1}{c|}{dilepton} & dilepton & lepton+jets & \multicolumn{1}{c|}{all-jets} & dilepton & lepton+jets & all-jets \\
\hline
  2010 & dilepton  &  1.00 &             &            &        &        &         &     \\

\hline
            & dilepton  &  0.15 &  1.00  &        &        &         &     &   \\
  2011 & lepton+jets  &  0.09 &  0.37  &  1.00  &        &         &&  \\
            & all-jets &  0.10 &  0.62  &  0.31  &  1.00  &         &&  \\
\hline
            & dilepton &  0.09 &  0.26  &  0.17  &  0.17  &   1.00  &&  \\
  2012 & lepton+jets  &  0.05 &  0.21  &  0.30  &  0.26  &   0.26  &   1.00  &  \\
            & all-jets &  0.06   &  0.20  &  0.27  &   0.28  &   0.32  &   0.61 &    1.00\\
\end{scotch}
}
\end{table*}

The correlations between each of the measurements is shown in Table~\ref{tab:inputcorr}. Figure~\ref{fig:coepul}
shows the combination coefficients and pulls, where the pull is defined as
($m_{\text{top}}^{\text{comb}}-m_{\text{top}}^{\text{meas}}$) / $\sqrt{\smash[b]{ \sigma_{\text{meas}}^{2} - \sigma_{\text{comb}}^{2}}}$
where $m_{\text{top}}^{\text{comb}}$ and $m_{\text{top}}^{\text{meas}}$ are the combined and the individual measurements of \mtop, respectively, and $\sigma_{\text{comb}}$ and $\sigma_{\text{meas}}$ are the corresponding total uncertainties.
The 2010 measurement contributes very little to the overall result. As the treatment of the systematic uncertainty for this analysis is the least sophisticated of the seven measurements, the final combination is repeated to verify that it does not influence the final result. Excluding this measurement
produces negligible changes in the values of \mtop or its total uncertainty, $\delta m_{\PQt}$.
For the combination of the remaining six measurements the $\chi^{2}$ is 2.3 for five degrees of freedom, corresponding to a probability of 80\%.

The breakdown of the systematic uncertainties for the combination is shown in Table~\ref{tab:h11syst}. The dominant uncertainty in the measurement arises from the modeling of the hadronization, with 0.33\GeV coming from the flavor-dependent jet energy corrections and a further 0.14\GeV coming from the \PQb jets. There are a further six terms with uncertainties in the range of 0.11--0.12\GeV. Of these, four are coming from theory and only two, the JEC in situ (0.12\GeV) and the JEC Uncorrelated non-pileup (0.10\GeV) are experimental. The theoretical uncertainties are computed using the same models so they should be fully correlated. For the two experimental terms, the strength of the assumed correlations is varied by 50\% of their nominal values to check the sensitivity to the assumed correlation strength. In both cases this produces changes of less than 0.01\GeV in
\mtop and $\delta m_{\PQt}$. We therefore conclude that the result is quite stable against reasonable changes in the assumed correlation strength.

Although we do not believe that the use of 100\% correlation strengths is appropriate to use for the correlated systematic uncertainties, for completeness we have re-run the final combination without the constraint on the correlation strengths. In this case we observe shifts of $-0.28$\GeV in \mtop and $-0.03$\GeV in $\delta m_{\PQt}$. For this combination, four of the seven measurements have negative combination coefficients and the central mass lies outside of the boundaries of the measurements. This corresponds to the result obtained using the standard BLUE method.

\begin{table}[!htb]
\centering
\topcaption{\label{tab:h11syst} Category breakdown of systematic uncertainties for the combined mass result. The uncertainties are expressed in \GeVns.}
\begin{scotch}{lc}
Combined \mtop result
 & $\delta \mtop (\GeVns{})$\\
\hline
\hline
Experimental uncertainties &  \\
\hline
Method calibration  & 0.03  \\
Jet energy corrections &  \\
-- JEC: Intercalibration  & 0.01 \\
-- JEC: In situ calibration  & 0.12 \\
-- JEC: Uncorrelated non-pileup & 0.10 \\
Lepton energy scale  & 0.01 \\
\MET scale & 0.03  \\
Jet energy resolution  & 0.03 \\
b tagging  & 0.05  \\
Pileup  & 0.06 \\
Backgrounds  & 0.04  \\
Trigger & $<$0.01\\
\hline
\hline
Modeling of hadronization &  \\
\hline
JEC: Flavor & 0.33 \\
b jet modeling & 0.14 \\
\hline
\hline
Modeling of perturbative QCD  &  \\
\hline
PDF  & 0.04  \\
Ren. and fact. scales & 0.10 \\
ME-PS matching threshold  & 0.08 \\
ME generator  & 0.11 \\
Top quark \pt & 0.02 \\
\hline
\hline
Modeling of soft QCD  &  \\
\hline
Underlying event  & 0.11 \\
Color reconnection modeling  & 0.10 \\
\hline
\hline
Total systematic & 0.47 \\
\hline
Statistical & 0.13 \\
\hline
\hline
Total Uncertainty & 0.48 \\
\end{scotch}
\end{table}

Figure~\ref{fig:channels} shows the mass values obtained from each of the three channels separately. These correspond to combinations h2 (2012, 2011) for the lepton+jets channel, 111 (2012, 2011, 2010) for the dilepton channel, and h1 (2012, 2011) for the all-jets channel, respectively. The results are all in good agreement with the combined measurement.
\begin{figure}[!thb]
\centering
   \includegraphics[width=\cmsFigWidth]{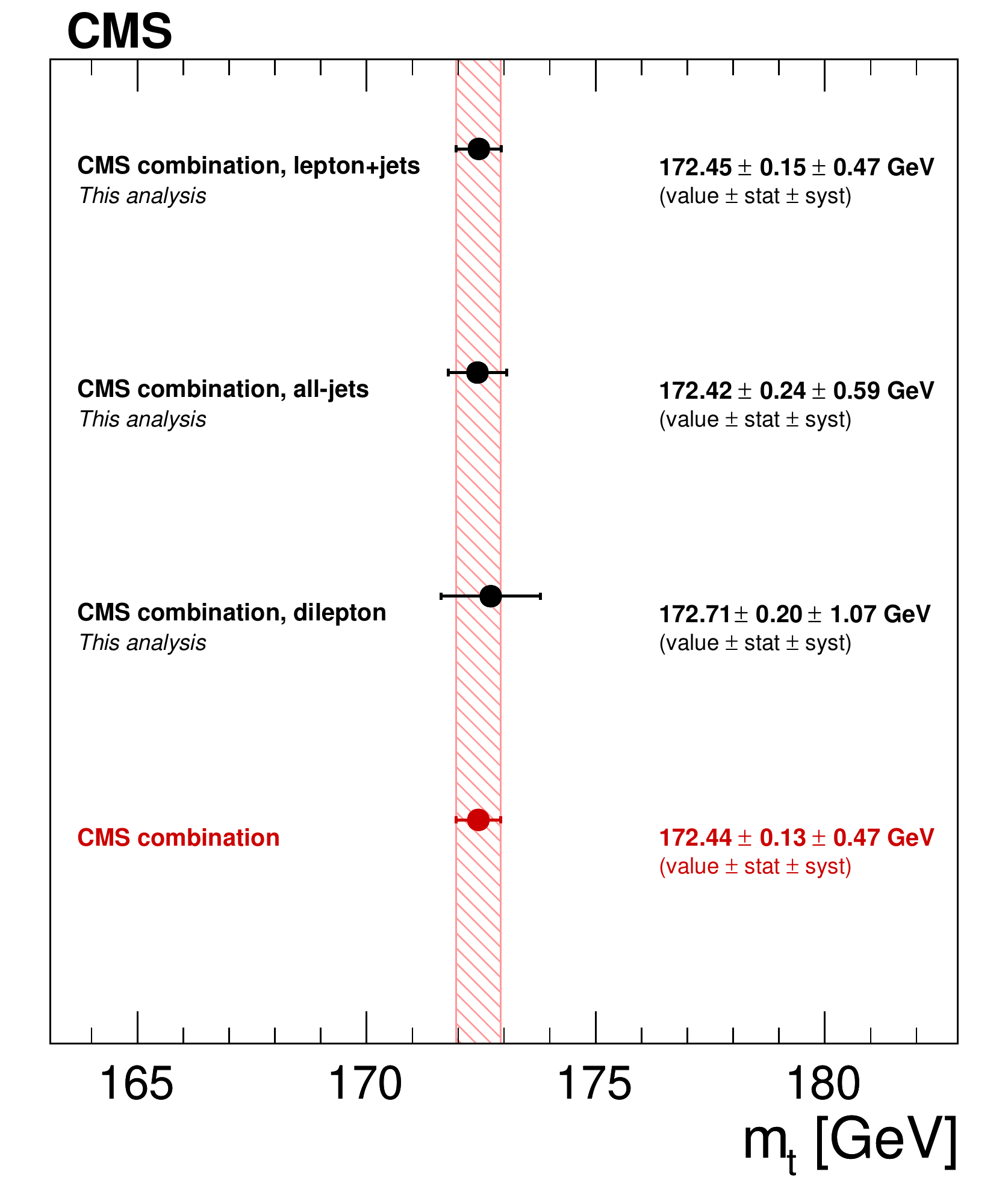}
   \caption{\label{fig:channels}
    The combined $\sqrt{s} = 7$ and 8\TeV measurements of \mtop for each of the $\ttbar$ decay channels. }
\end{figure}

\section{Summary}
A new set of measurements of the top quark mass has been presented, based on the data recorded by the CMS experiment at the LHC at $\sqrt{s} = 8\TeV$ during 2012, and corresponding to a luminosity of 19.7\fbinv. The top quark mass has been measured in the lepton+jets, all-jets and dilepton decay channels, giving values of
$172.35 \pm 0.16\stat\pm 0.48\syst\GeV$,
$172.32 \pm 0.25\stat\pm 0.59\syst\GeV$,
and 1$72.82 \pm 0.19\stat\pm 1.22\syst\GeV$, respectively.
Individually, these constitute the most precise measurements in each of the decay channels studied. When combined with the published CMS results at $\sqrt{s} = 7\TeV$, a top quark mass measurement of $172.44 \pm 0.13\stat\pm 0.47\syst\GeV$ is obtained. This is the most precise measurement of \mtop to date, with a total uncertainty of 0.48\GeV, and it supersedes all of the previous CMS measurements of the top quark mass.

The top quark mass has also been studied as a function of the event kinematical properties in the lepton+jets channel. No indications of a kinematical bias in the measurements is observed and the data are consistent with a range of predictions from current theoretical models of \ttbar production.
\begin{acknowledgments}
\hyphenation{Bundes-ministerium Forschungs-gemeinschaft Forschungs-zentren} We congratulate our colleagues in the CERN accelerator departments for the excellent performance of the LHC and thank the technical and administrative staffs at CERN and at other CMS institutes for their contributions to the success of the CMS effort. In addition, we gratefully acknowledge the computing centers and personnel of the Worldwide LHC Computing Grid for delivering so effectively the computing infrastructure essential to our analyses. Finally, we acknowledge the enduring support for the construction and operation of the LHC and the CMS detector provided by the following funding agencies: the Austrian Federal Ministry of Science, Research and Economy and the Austrian Science Fund; the Belgian Fonds de la Recherche Scientifique, and Fonds voor Wetenschappelijk Onderzoek; the Brazilian Funding Agencies (CNPq, CAPES, FAPERJ, and FAPESP); the Bulgarian Ministry of Education and Science; CERN; the Chinese Academy of Sciences, Ministry of Science and Technology, and National Natural Science Foundation of China; the Colombian Funding Agency (COLCIENCIAS); the Croatian Ministry of Science, Education and Sport, and the Croatian Science Foundation; the Research Promotion Foundation, Cyprus; the Ministry of Education and Research, Estonian Research Council via IUT23-4 and IUT23-6 and European Regional Development Fund, Estonia; the Academy of Finland, Finnish Ministry of Education and Culture, and Helsinki Institute of Physics; the Institut National de Physique Nucl\'eaire et de Physique des Particules~/~CNRS, and Commissariat \`a l'\'Energie Atomique et aux \'Energies Alternatives~/~CEA, France; the Bundesministerium f\"ur Bildung und Forschung, Deutsche Forschungsgemeinschaft, and Helmholtz-Gemeinschaft Deutscher Forschungszentren, Germany; the General Secretariat for Research and Technology, Greece; the National Scientific Research Foundation, and National Innovation Office, Hungary; the Department of Atomic Energy and the Department of Science and Technology, India; the Institute for Studies in Theoretical Physics and Mathematics, Iran; the Science Foundation, Ireland; the Istituto Nazionale di Fisica Nucleare, Italy; the Ministry of Science, ICT and Future Planning, and National Research Foundation (NRF), Republic of Korea; the Lithuanian Academy of Sciences; the Ministry of Education, and University of Malaya (Malaysia); the Mexican Funding Agencies (CINVESTAV, CONACYT, SEP, and UASLP-FAI); the Ministry of Business, Innovation and Employment, New Zealand; the Pakistan Atomic Energy Commission; the Ministry of Science and Higher Education and the National Science Centre, Poland; the Funda\c{c}\~ao para a Ci\^encia e a Tecnologia, Portugal; JINR, Dubna; the Ministry of Education and Science of the Russian Federation, the Federal Agency of Atomic Energy of the Russian Federation, Russian Academy of Sciences, and the Russian Foundation for Basic Research; the Ministry of Education, Science and Technological Development of Serbia; the Secretar\'{\i}a de Estado de Investigaci\'on, Desarrollo e Innovaci\'on and Programa Consolider-Ingenio 2010, Spain; the Swiss Funding Agencies (ETH Board, ETH Zurich, PSI, SNF, UniZH, Canton Zurich, and SER); the Ministry of Science and Technology, Taipei; the Thailand Center of Excellence in Physics, the Institute for the Promotion of Teaching Science and Technology of Thailand, Special Task Force for Activating Research and the National Science and Technology Development Agency of Thailand; the Scientific and Technical Research Council of Turkey, and Turkish Atomic Energy Authority; the National Academy of Sciences of Ukraine, and State Fund for Fundamental Researches, Ukraine; the Science and Technology Facilities Council, UK; the US Department of Energy, and the US National Science Foundation.

Individuals have received support from the Marie-Curie program and the European Research Council and EPLANET (European Union); the Leventis Foundation; the A. P. Sloan Foundation; the Alexander von Humboldt Foundation; the Belgian Federal Science Policy Office; the Fonds pour la Formation \`a la Recherche dans l'Industrie et dans l'Agriculture (FRIA-Belgium); the Agentschap voor Innovatie door Wetenschap en Technologie (IWT-Belgium); the Ministry of Education, Youth and Sports (MEYS) of the Czech Republic; the Council of Science and Industrial Research, India; the HOMING PLUS program of the Foundation for Polish Science, cofinanced from European Union, Regional Development Fund; the OPUS program of the National Science Center (Poland); the Compagnia di San Paolo (Torino); the Consorzio per la Fisica (Trieste); MIUR project 20108T4XTM (Italy); the Thalis and Aristeia programs cofinanced by EU-ESF and the Greek NSRF; the National Priorities Research Program by Qatar National Research Fund; the Rachadapisek Sompot Fund for Postdoctoral Fellowship, Chulalongkorn University (Thailand); and the Welch Foundation, contract C-1845.
\end{acknowledgments}

\bibliography{auto_generated}

\cleardoublepage \appendix\section{The CMS Collaboration \label{app:collab}}\begin{sloppypar}\hyphenpenalty=5000\widowpenalty=500\clubpenalty=5000\textbf{Yerevan Physics Institute,  Yerevan,  Armenia}\\*[0pt]
V.~Khachatryan, A.M.~Sirunyan, A.~Tumasyan
\vskip\cmsinstskip
\textbf{Institut f\"{u}r Hochenergiephysik der OeAW,  Wien,  Austria}\\*[0pt]
W.~Adam, E.~Asilar, T.~Bergauer, J.~Brandstetter, E.~Brondolin, M.~Dragicevic, J.~Er\"{o}, M.~Flechl, M.~Friedl, R.~Fr\"{u}hwirth\cmsAuthorMark{1}, V.M.~Ghete, C.~Hartl, N.~H\"{o}rmann, J.~Hrubec, M.~Jeitler\cmsAuthorMark{1}, V.~Kn\"{u}nz, A.~K\"{o}nig, M.~Krammer\cmsAuthorMark{1}, I.~Kr\"{a}tschmer, D.~Liko, T.~Matsushita, I.~Mikulec, D.~Rabady\cmsAuthorMark{2}, B.~Rahbaran, H.~Rohringer, J.~Schieck\cmsAuthorMark{1}, R.~Sch\"{o}fbeck, J.~Strauss, W.~Treberer-Treberspurg, W.~Waltenberger, C.-E.~Wulz\cmsAuthorMark{1}
% ...
\vskip\cmsinstskip
% ...
1:~~Also at Vienna University of Technology, Vienna, Austria\\
2:~~Also at CERN, European Organization for Nuclear Research, Geneva, Switzerland\\
% ...

\end{sloppypar}
\end{document}